\documentclass[11pt,a4paper,oneside,dvipsnames]{article}

\usepackage{latexsym}
\usepackage{graphicx}
\usepackage{multicol,multirow}
\usepackage{amsmath,amssymb,amsfonts}
\usepackage{mathrsfs}
\usepackage{amsthm}
\usepackage{apacite}
\usepackage{rotating}
\usepackage{appendix}
\usepackage[authoryear]{natbib}
\usepackage{ifpdf}
\usepackage[T1]{fontenc}
\usepackage{times}
\usepackage{sourcesanspro}
\usepackage{newtxmath}
\usepackage{textcomp}%
\usepackage{xcolor}%
\usepackage{csvsimple}
\usepackage{adjustbox}
\usepackage{xr}

\usepackage{xr-hyper}
\usepackage{hyperref}
\usepackage{siunitx}
\usepackage{authblk}

\usepackage{subcaption}
\usepackage[utf8]{inputenc}

\DeclareUnicodeCharacter{2080}{\textsubscript{0}}
\DeclareUnicodeCharacter{2081}{\textsubscript{1}}
\DeclareUnicodeCharacter{2082}{\textsubscript{2}}
\DeclareUnicodeCharacter{2083}{\textsubscript{3}}
\DeclareUnicodeCharacter{2084}{\textsubscript{4}}
\DeclareUnicodeCharacter{2085}{\textsubscript{5}}
\DeclareUnicodeCharacter{2093}{\textsubscript{x}}

\usepackage{geometry}

\newgeometry{vmargin={15mm}, hmargin={12mm,17mm}}   

\usepackage{booktabs}
\usepackage{pgfplotstable} 

\definecolor{reddish}{HTML}{FBB4AE}
\definecolor{blueish}{HTML}{B3CDE3}
\definecolor{magentish}{HTML}{FF00AA}
\definecolor{greenish}{HTML}{a1d99b}

\makeatletter
\newcommand*{\addFileDependency}[1]{
  \typeout{(#1)}
  \@addtofilelist{#1}
  \IfFileExists{#1}{}{\typeout{No file #1.}}
}
\makeatother

\newcommand*{\myexternaldocument}[2]{%
    \externaldocument[#2]{#1}%
    \addFileDependency{#2.tex}%
    \addFileDependency{#2.aux}%
}

\myexternaldocument{supplementary}{S-}

\begin{document}

\title{Estimating Ambient Air Pollution Using Structural Properties of Road Networks}

\author{
  Liam J. Berrisford$^{1,2,3}$\\
  \texttt{l.berrisford@exeter.ac.uk}
  \and
  Ronaldo Menezes$^{1}$\\
  \texttt{r.menezes@exeter.ac.uk}
  \and
  Eraldo Ribeiro$^{4}$\\
  \texttt{eribeiro@fit.edu}
}

\date{%
    $^1$ BioComplex Laboratory, Department of Computer Science, University of Exeter, England\\%
    $^2$ Department of Mathematics, University of Exeter, England\\
    $^3$ UKRI Centre for Doctoral Training in Environmental Intelligence, University of Exeter, England\\%
    $^4$ Department of Computer Science, Florida Institute of Technology, USA\\[2ex]%
    \today
}


\maketitle

\abstract{In recent years, the world has become increasingly concerned with air pollution. Particularly in the global north, countries are implementing systems to monitor air pollution on a large scale to aid decision-making.
Such efforts are essential but they have at least three shortcomings: {\em (1)} they are costly and are difficult to implement expediently; {\em (2)} they focus on urban areas, which is where most people live, but this choice is prone to inequalities; and {\em (3)} the process of estimating air pollution lacks transparency. In this paper, we demonstrate that we can estimate air pollution using open-source information about the structural properties of roads; we focus on England and Wales in the United Kingdom (UK) in this paper although the methods here described are not dependent on specific datasets. Our approach makes it possible to implement an inexpensive method of estimating air pollution concentrations to an accuracy level that can underpin policymakers' decisions while providing an estimate in all districts, not just urban areas, and in a process that is transparent and explainable. \\

\textbf{Impact Statement}. We show that a linear regression model using a single structural property---length of the track and unclassified road network within 0.36\% of districts within England and Wales (in the UK)---can accurately estimate which districts are the most polluted. The model presents a transparent and low-cost, yet effective, alternative to more expensive models such as the one currently used by DEFRA in the UK. The model has apparent practical uses for policymakers who want to pursue clean-air initiatives but lack the capital to invest in comprehensive monitoring networks. Its low implementation cost, accessible model design, and worldwide coverage of the dataset provide a basis for implementing systems to estimate air pollution concentrations in low-income countries.}

\section{Introduction}
\label{sec:introduction}

The world is becoming increasingly urbanised. Urban environments bring efficiencies and opportunities unavailable in rural settings, such as job opportunities, education and healthcare access. However, urbanisation comes with several problems, such as air pollution \citep{qing2018urbanization}. However, air pollution affects not only those who live within the urban setting themselves, but those in surrounding rural areas. Although the urban-rural gap has increased since 1970s (33\% decrease in rural population and a 100\% increase in urban population) and the air pollution has affected more urban areas (by a factor of more than 5) \citep{deng2021effect}, there is plenty of evidence of rural areas continuing to suffer consequences of air pollution mostly generated in the large urban concentrations \citep{agrawal2005effects,du2018household,aunan2019hidden}.

Wealthy nations, particularly in the global north, have the resources to install monitoring stations and sensors to monitor the situation and aid decision-makers in minimising the threat posed by air pollution. However, these monitoring stations are predominately placed within urban settings; as such, there is an evident lack of information available in some of the rural or even outskirts of central urban areas. \footnote{\url{https://waqi.info}. World's Air Pollution: Real-time Air Quality Index. (Last accessed: 24 June 2022).} Clearly this leads to possible inequalities in exposure monitoring which, coupled with the cost of stations and estimations, further disadvantage rural dwellers. The urban-rural inequalities in the global north are akin to the inequalities between the global north and global south, where resources are scarce. Hence, a simple and inexpensive mechanism for estimating air pollution, as the one proposed in this paper, is highly desirable.

Given the potential impact air pollution can have on human health and well-being \citep{kelly2015air,mannucci2017health,landrigan2017air}, it is vital to have a complete spatial picture of air pollution across an area of interest. Therefore, it is crucial to look at alternative ways of estimating air pollution without monitoring stations to help augment real-world measurements of air pollution concentrations. Hence, the purpose of this work is to {\it determine how well ambient air pollution concentrations can be estimated in a district by the structural properties of its road transport infrastructure}. We have selected road structural properties due to the abundance of data related to roads across the United Kingdom (UK) and worldwide. In this work we focus on England and Wales within the UK. This choice is based on the fact that we use MSOA (Middle Layer Super Output Areas) as the district of aggregation within the study for the benefit of having sociodemographic data attached as part of its use within the England and Wales census. For a full discussion of the choice, see Section \ref{sec:motivation} and Section \ref{S-sec:MAUP}. Henceforth, references prefixed with S refer to elements in the supplementary material.

Our contributions therefore are 3-fold: {\em (1)} our approach is inexpensive to produce and hence accessible to a wider range of districts; {\em (2)} our process is transparent, and the model proposed can be easily explained and hence be more useful to numerous stakeholders; and, indirectly, {\em (3)} we demonstrate that lack of data about air pollution may not be a problem if the information about air pollution can be extracted from datasets that have not been primarily proposed to be used as air pollution estimators (in our case, road networks).

This work is organised as follows. In Section \ref{sec:relatedwork} related work to this study is discussed. In Section \ref{sec:motivation} we motivate and formulate the problem this work tackles and justify the study area of choice. Section \ref{sec:data} discusses the data used in this work. The three model variants (\ref{sec:methodology:models}), feature selection process (\ref{sec:methodology:featureselection}) and policymaker focused missing districts model (\ref{sec:methodology:singleyeartests}) are detailed in Section \ref{sec:methodology}. We then discuss in Section \ref{sec:usecases} the use cases in which the proposed approach excels. Finally, the work ends with a discussion in Section \ref{sec:discussion}, setting apart this work's approach and use from existing studies.

\section{Related Work}
\label{sec:relatedwork}

As the effect of climate change becomes more prevalent, pressure mounts on governments and policymakers to act to counter the detrimental effect on society. Amongst the many issues, air pollution is quite prevalent for many reasons including its effect on health, its incriminatory effects (it affects everyone regardless of possible differences), and its global effect (no location in the world is immune to its effects despite their own contribution to the issue) \citep{shaddick2020half}. In fact, the issue is so globalised that certain nations are asking for restitution (loss and damage)\footnote{\url{https://www.un.org/en/climatechange/adelle-thomas-loss-and-damage}. Loss and damage: A moral imperative to act. (Last accessed: 14 November 2022).} from polluting countries \citep{johnson2017holding}.

In order to have a global view of the issues, we need estimators because of the lack of real-time monitoring infrastructures. Recent works have attempted to estimate air pollution caused by road traffic in a transport network \citep{Gualtieri:1998:Predicting,Karppinen:2000:Modelling}. These works have taken the approach of estimating the use of the road infrastructure by vehicles alongside a dispersion model for the pollution that results in a fine-grained pollution map for the study period over the study area. However, the issue present with this approach is the data and computation needed to underpin the model. The constraints presented by the model have recently improved with modern techniques such as Artificial Neural Networks (ANNs) \citep{Catalano:2016:Improving}. However, the task of collecting road traffic data is still prohibitively expensive and time-consuming to collect, resulting in spatially sparse datasets.\footnote{\url{https://roadtraffic.dft.gov.uk/downloads}. Road traffic bulk downloads. (Last accessed: 14 November 2022).} The prospect of using road traffic data for air pollution estimation over a sizeable spatial extent is further compounded due to the changing traffic dynamics on roads over small distances. These constraints make it challenging to expand the method to a national level. While there are models to assess air pollution at a national level, such as the UK DEFRA Modelling of Ambient Air Quality (MAAQ) contract model, they are not open source, convoluted and expensive \citep{Ricardo:2021:Technical}, with the 2021 renewal contract valued at £3,800,000 \citep{DEFRA:2021:Contract}.

In the scientific literature, one will encounter works that are based on deep learning and others that are not. The non-deep-learning methods are further separated into deterministic and statistical methods \citep{liu2021intelligent}. Deterministic models are argued to have limited predictive performance due to various factors including parameter estimation \citep{stern2008model,pak2020deep}. Statistical methods are further subdivided into works using classical statistics and ones based on machine learning (ML). Although statistical methods can capture interesting features such as non-linearity in the data, they are expensive to run and not able to extract complex features such as spatial correlations in historic data \citep{yan2021multi}.

This work explores the possibility of estimating ambient air pollution at a coarse but adequate level to inform policy-makers decisions using open source, low cost and non-invasive data extracted from the structural properties of road networks. Our work's final output helps free resources for direct investment into clean air initiatives rather than monitoring infrastructure.

\subsection{Literature Gap}
\label{sec:motivation}

A 2019 review by Public Health England outlined air pollution as the most significant environmental threat to health in the UK, with 28,000-36,000 deaths attributable yearly to long-term exposure \citep{PHE:2018}; unfortunately, this is also true worldwide \citep{shaddick2020half}. 

\begin{figure}
  \begin{subfigure}{0.31\textwidth}
    \includegraphics[width=\linewidth]{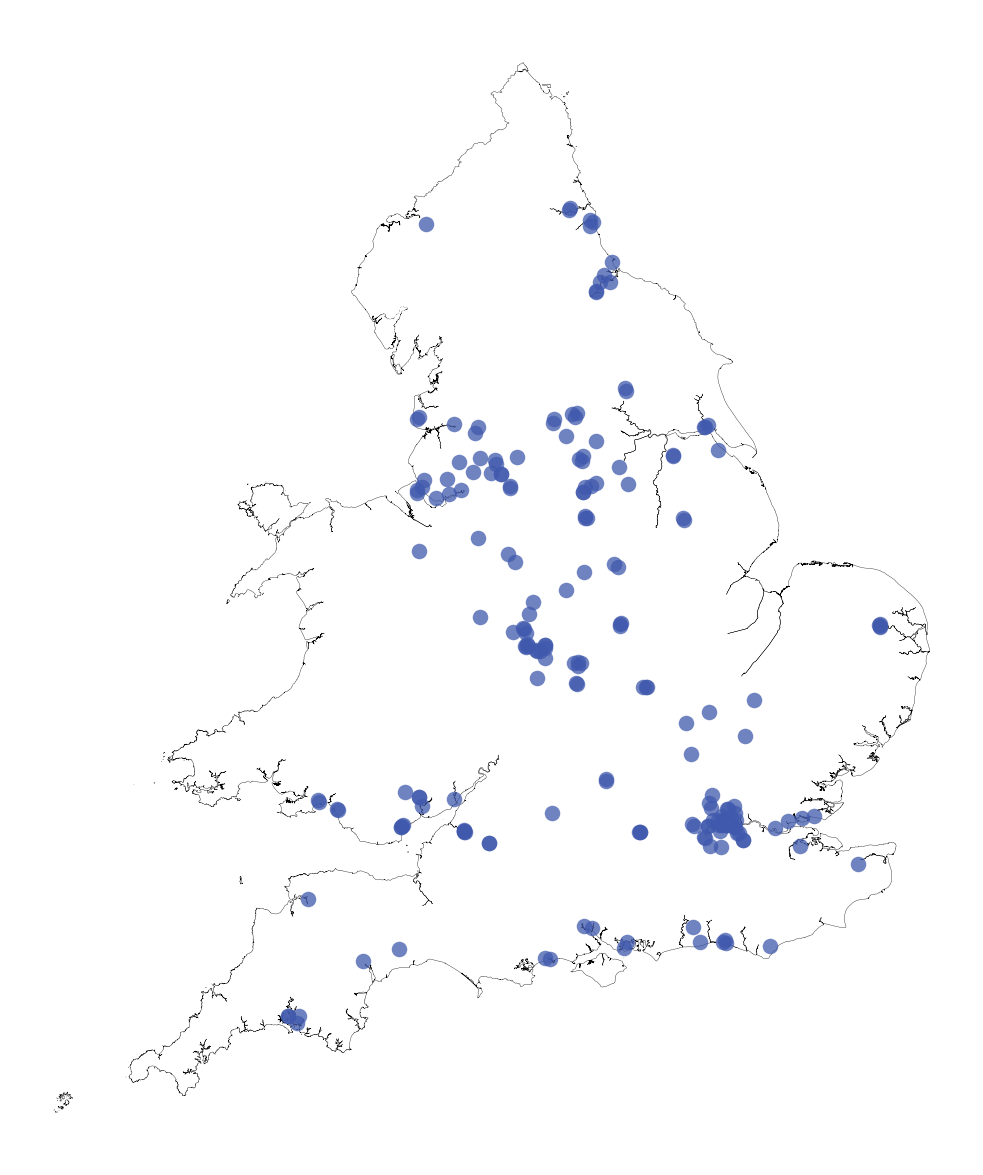}
    \caption{202 Urban Stations} \label{fig:ukaurnUrban}
  \end{subfigure}%
  \hspace*{\fill}   
  \begin{subfigure}{0.31\textwidth}
    \includegraphics[width=\linewidth]{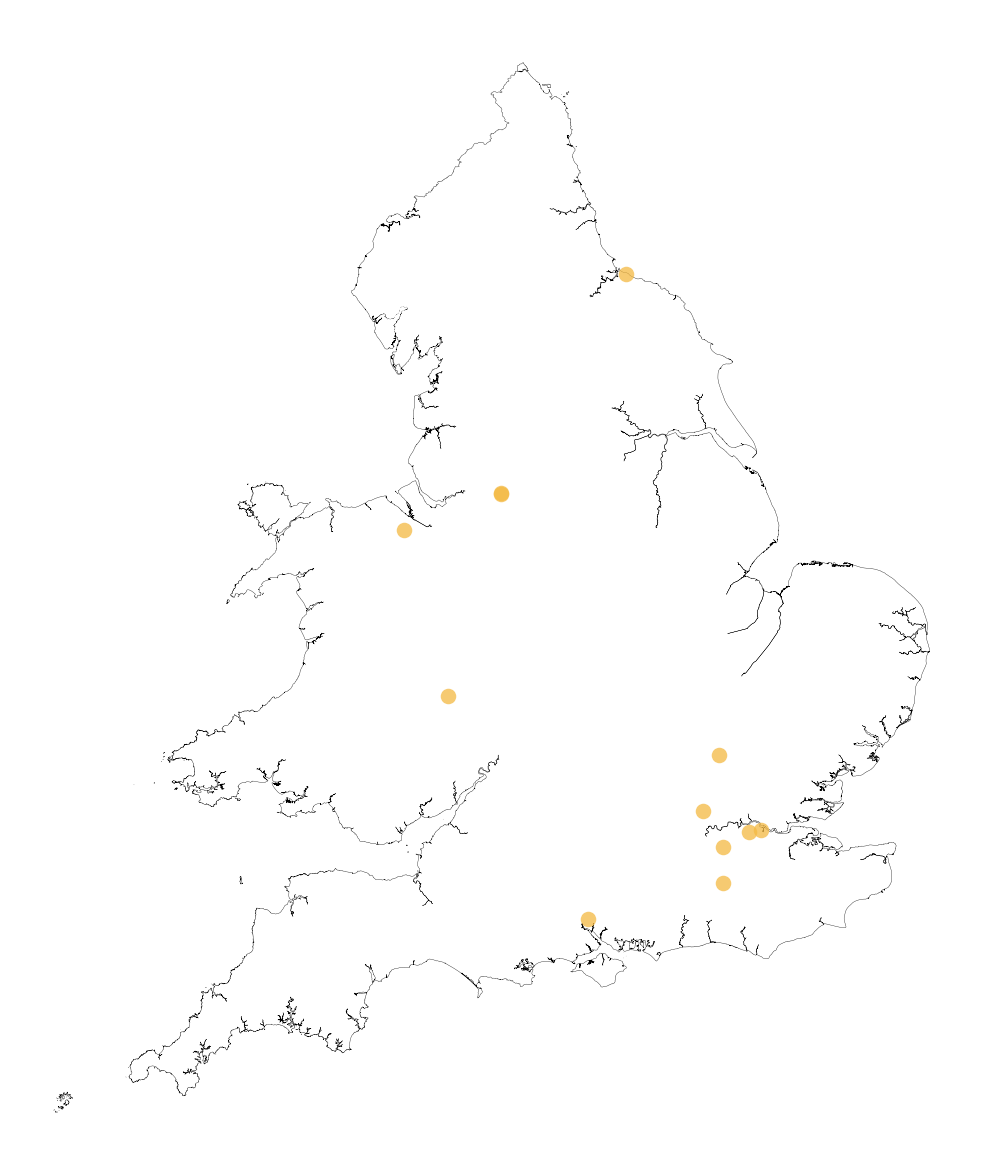}
    \caption{12 Suburban Stations} \label{fig:ukaurnSuburban}
  \end{subfigure}%
  \hspace*{\fill}   
  \begin{subfigure}{0.31\textwidth}
    \includegraphics[width=\linewidth]{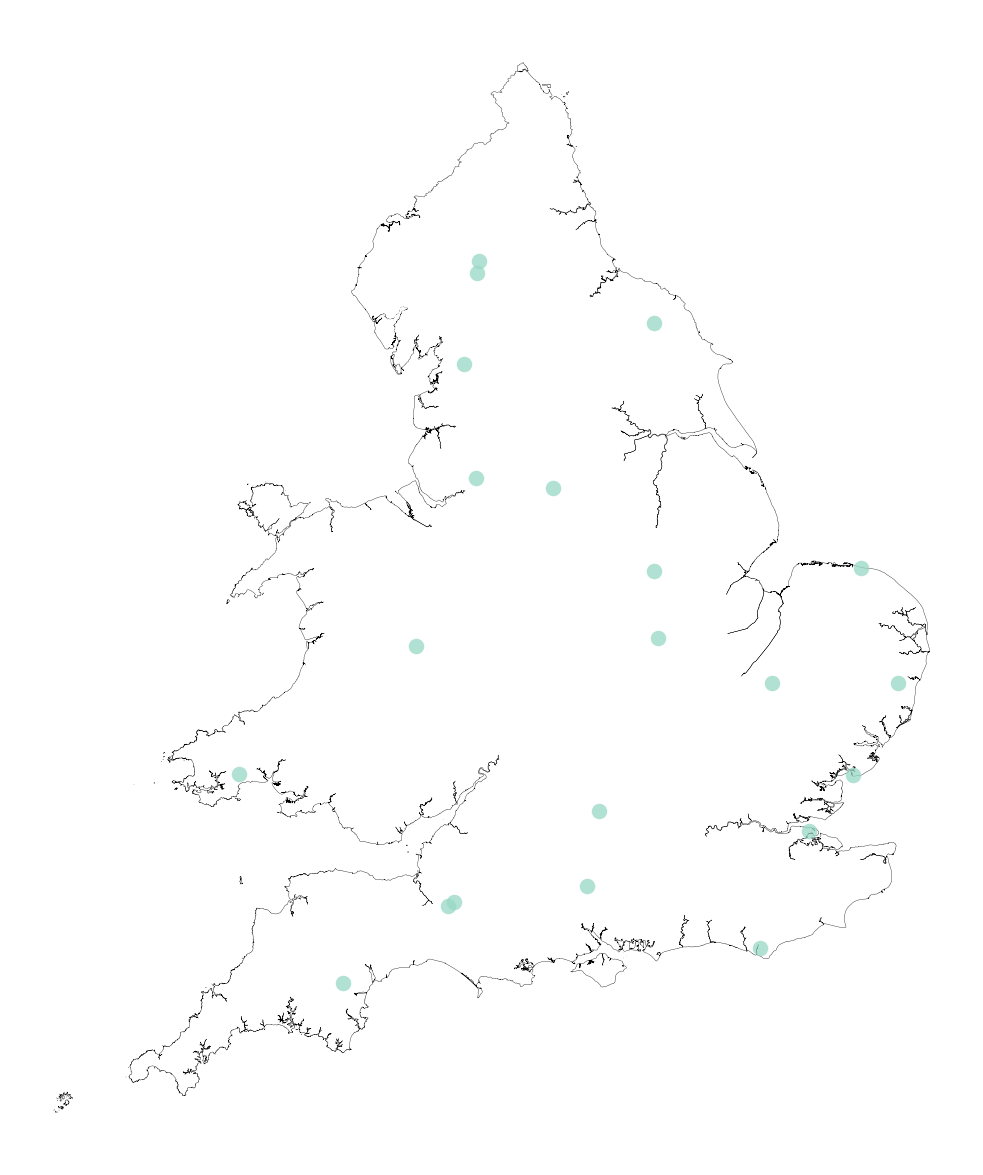}
    \caption{21 Rural Stations} \label{fig:ukaurnRural}
  \end{subfigure}

\caption{{\bfseries Spatial distribution and classification of monitoring stations within England and Wales.} The Automatic Urban and Rural Network (AURN) stations are divided into 3 classes: urban (202), suburban (12), and rural (21). Note the inequality of station numbers; most of the stations are in urban settings. The AURN has 274 stations total out of which 235 are in England and Wales} \label{fig:ukaurn}
\end{figure}

The first step to tackling air pollution is knowing where it is polluted; mapping the levels of pollution around the country. The most robust method of determining air pollution within an area is to perform ground measurements with specialised equipment. However, the cost of this equipment can be prohibitive, especially if there is a need for covering large areas. Costly equipment limits the scope of the air pollution monitoring networks. Even in wealthier countries such as the UK, which has reduced air pollution as a core policy goal \citep{Parliament:2021:Environment}, the scope of its air pollution monitoring network is limited. Figure \ref{fig:ukaurn} shows the scope of the England and Wales automatic air pollution monitoring network a part of the UK Automatic Urban and Rural Network (AURN). Note that despite the name of the network, the majority of the stations are in urban environments, which leads to poor performance in rural settings and inequalities between urban and rural residents.

Given the gaps that exist in ground monitoring, model estimates for air pollution can supplement the ground observation network to ensure compliance with policy targets across the whole of the UK~\citep{DEFRA:2019:AirPollutionReport}. The UK's air pollution datasets are augmented with outputs from the UK DEFRA Modelling of Ambient Air Quality (MAAQ) model, which like other aforementioned models suffer from shortcomings such as lack of transparency; we do not actually know the details of how this model works. 

The goal of this work is threefold: {\em (i)} we want to estimate pollution concentrations based on datasets that are already available in most countries (road infrastructure properties); {\em (ii)} we aim to achieve the same or better results of current models using open source input data while also making the model itself open source, ensuring no barrier to implementing the approach based on cost alongside making the approach accessible; {\em (iii)} the process is made as streamlined as possible using minimal data to inform data acquisition for policymakers constrained by a lack of data availability, which is becoming an increasing divide between countries \citep{Karlsson:2002:North}.

The amount of data collected around the world and the fact that information about a phenomenon can be embedded in datasets collected for other purposes, mean that lack of data to be used in air pollution models may not necessarily be an excuse for not having a model. This work shows that we can effectively use secondary datasets containing information about the phenomenon we want to model; this can be important in locations where data is collected by institutions such as non-profits, foundations, the United Nations, etc. but no curated pollution datasets exist. 

We chose the UK (England and Wales) as the study area to build a proof of concept model, as its air pollution monitoring data are open source and freely available. However, we see no reason for not being able to use a similar approach in other locations around the world, as long as we have similar data used here. 

When dealing with spatial data, we often have to choose the aggregation level for the model. We chose Middle Layer Super Output Areas (MSOA) as the districts under which we would aggregate the road and air pollution data. The three district designs considered can be seen in Section \ref{S-sec:MAUP}. The OS National Grid district design was ruled out because many districts had no roads, which created districts where we could not create a feature vector. A feature vector is an ordered list of features about an observed phenomenon. Each feature describes a measurable property of the observed phenomena, in this case, the road network within a given district. A corresponding target vector is then created, in which the goal is for the model to learn the relationship between the corresponding features and targets. This study's target vector is related to observed ambient air pollution concentrations within a district. The Local Authority District (LAD) design was a viable option but had very inconsistent geographical sizes for the different districts across the study area, ranging from 2.8km$^2$ area in the City of London LAD to 26,159.6km$^2$ area in the Highland LAD, alongside only 381 districts. The MSOA design offered a higher number of districts (at 7,201) while also providing a measure of population density with the population for each MSOA readily available; the population density then allowed for an estimation of how urbanised an area was in a continuous spectrum from least to most urbanised. 

\section{Data and Methods}
\label{sec:data}

In this work, we used OpenStreetMaps (OSM) data from 2014-2019. We selected 2014-2017 for training the models, 2018 for the test set, and 2019 for data exploration and parameter searching. Data prior to 2014 is quite sparse and post 2019 was avoided in this first instance because of the COVID-19 pandemic, which have resulted in changes in the way people behave spatially \citep{santana2022changes} and also because 
the COVID-19 pandemic had significant implications on air pollution across the world \citep{Brown:2021:Traffic}.

The OpenAir package \citep{Carslaw:2012:OpenAir} provides access to data from the UK Automatic Urban and Rural Network (AURN) \citep{Stevenson:2009:QAQC} with measurements taken every 15 minutes.  Data starts from early 1973. 
Our study focused on 12 pollutants: Carbon Monoxide (CO), Nitrogen Oxide (NO), Nitrogen Dioxide (NO$_\text{2}$) Nitrogen Oxides (NO$_{x}$), Particles < 10$\mu$m (PM$_{10}$), Particles < 2.5$\mu$m (PM$_{25}$), Non-volatile PM$_{10}$ (NV$_{10}$), Non-volatile PM$_{25}$ (NV$_{25}$), Volatile PM$_{10}$ (V$_{10}$), Volatile PM$_{25}$ (V$_{25}$), Ozone (O$_\text{3}$), and Sulphur Dioxide (SO$_\text{2}$). The number of active AURN stations per pollutant by year is shown in Figure \ref{S-fig:ActiveAURNStations}

Inverse Distance Weighting (IDW) Interpolation \citep{shepard:1968:two} was used to achieve full spatial coverage of the study area from the AURN ground observations by creating a raster from which air pollution values at specific locations could be sampled; IDW interpolation uses a power parameter to determine the effect of neighbouring sample points on a location's value which can be tuned for specific air pollutants that are more or less susceptible to neighbouring pollution. We experimented with a range of possible values from 0 to 3.5 to determine the power parameter value for each pollutant, with the value that minimised the root-mean-square error (RMSE) \citep{Oxford:2009:PhysicsDictionary} from leave-one-out-validation process used; Figure \ref{S-fig:RMSEInterpolationPowerScatterPlots} shows the  plots for various power parameter values which were used for the choice of the power parameter. Figure \ref{S-fig:InterpolationIDWSurfaces} shows the raster produced by the IDW process with the best performing power parameter for each of the 12 pollutants.

We compared the resulting raster to the UK Government DEFRA air pollution model output for the same year, 2019 \citep{DEFRA:2019:PM25}, to verify the suitability of using interpolation to create a raster from ground observations. The dataset from the DEFRA model gives point estimates with a 1km resolution. We sampled the raster produced at the exact location of the 1km point, which we then aggregated to the district level, giving a value for the pollution in a given area. 

\begin{figure}[ht]
\begin{center}
\includegraphics[width=0.5\textwidth]{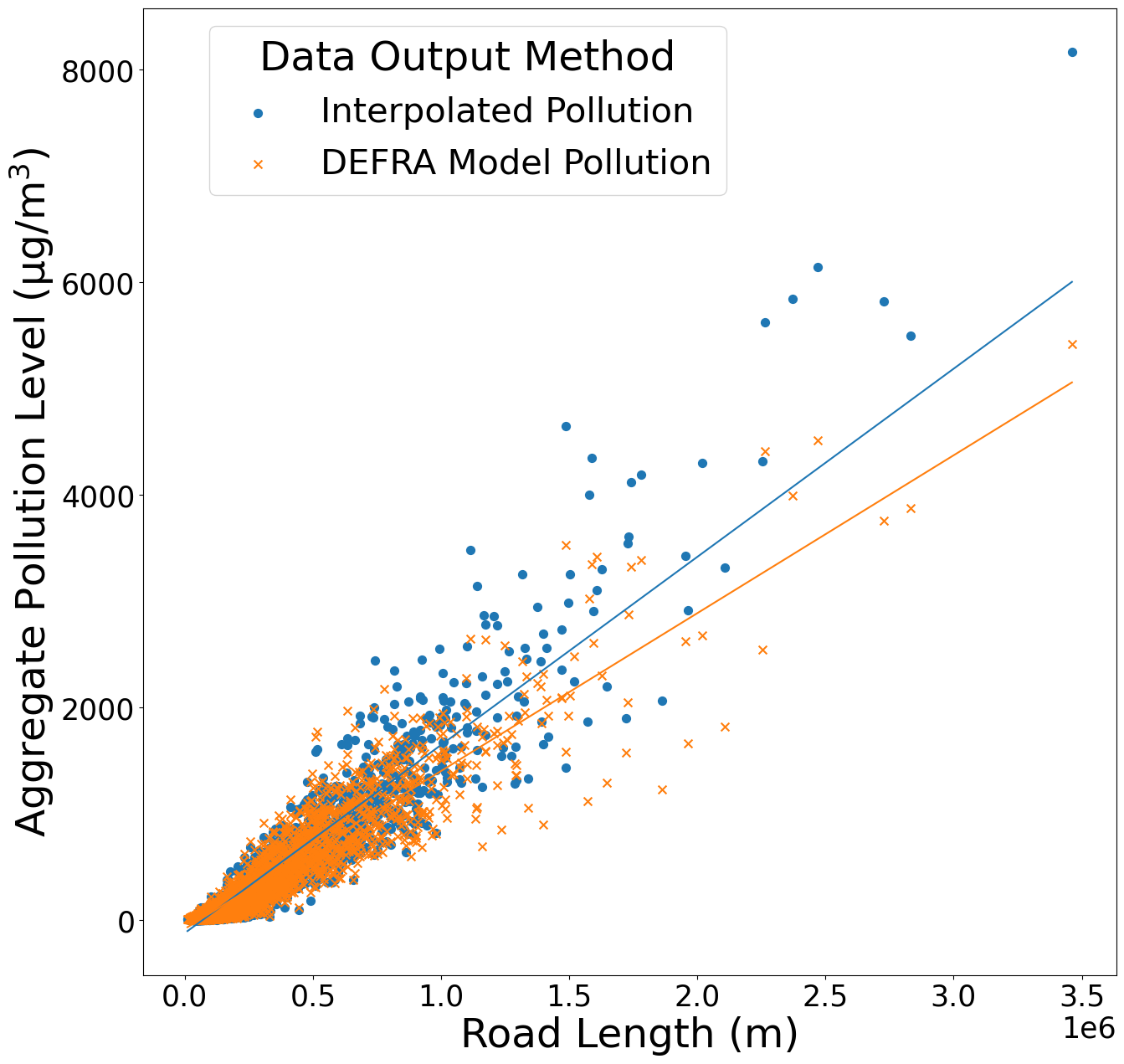}
\caption{{\bfseries AURN ground observations interpolation vs. DEFRA air pollution model.} Comparison between the two approaches. The blue points and line of the best fit depict the relationship between total air pollution in an MSOA (acquired via interpolating ground observations). In contrast, the orange points and line of the best fit represent air pollution data acquired via the UK DEFRA Ambient Air Quality Contract Model. The chart shows the two measurements of pollution, ordered by road length (x-asis). A similar relationship between the two lines of best fit indicates that the methods for generating the air pollution data are similar}
\label{fig:InterpolatedDEFRAPollutionComparison}
\end{center}
\end{figure}

We conducted a Pearson correlation coefficient \citep{benesty2009pearson} analysis was calculated between the DEFRA and interpolated output shown for various road length per district to ensure that the relationship between datasets was similar. We achieved a correlation value of 0.95 and 0.94 indicating that the datasets describe the same phenomena and the IDW process is able to replicate the point sample annual observation on a complete spatial scale. The spearman rank-order correlation coefficient \citep{spearman1961proof} was also calculated between the DEFRA and interpolated districts to ensure they had a similar ordering of the most / least polluted districts. The correlation value was high at 0.98 indicating that there is also agreement between the two datasets as to which districts are the most and least polluted across the study area. 
\figurename~\ref{fig:InterpolatedDEFRAPollutionComparison} shows the pearson correlation plot between the DEFRA and interpolated output and the road length within a given district.

OpenStreetMaps is an open-source collaborative project that contains road data for most parts of the world, and it is quite comprehensive for the UK \citep{OpenStreetMap:2021:OpenStreetMap}. The method used to create historical road infrastructure datasets was to use the OpenStreetMaps history files (.osh.pbf) to revert an OpenStreetMaps file (.osm.pbf) to its state at a given point in time. We could then use the historical OpenStreetMaps file to extract the network's structural properties at that time, such as the total road length per district. 

The feature vector within our study is based on the length of the road network within a given area, determined by summing the length of the roads within the district under the WGS 84 / World Mercator EPSG: 3395 coordinate reference system (CRS), giving the length of the line segments of roads in meters. This process is then repeated for each of the MSOA districts. Each MSOA district then gives one element to the final feature vector for a given time. 

The target vector is created by summing point estimates across a uniform point grid with 250m intervals. We chose the 250m interval to ensure that every MSOA had a point estimate. Details of the uniform point grid used can be seen in Table \ref{S-tab:Table 2 Target Vector Example}, alongside a visualisation for a single MSOA in Figure \ref{S-fig:ToyExampleOfRasterSampleWithUniformPointGrid}.

\begin{figure}[ht]
\begin{center}
\includegraphics[width=0.5\textwidth]{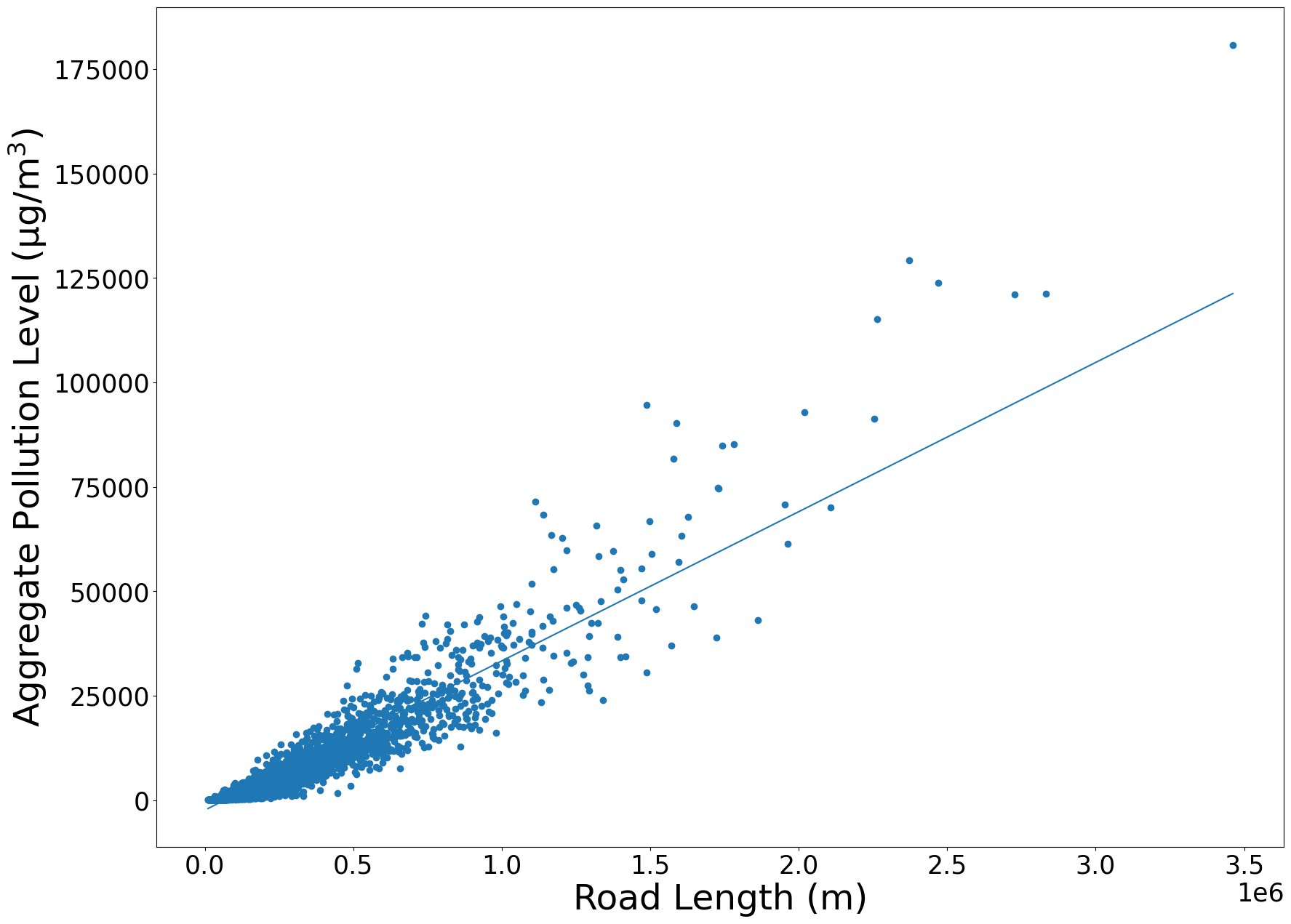}
\caption{{\bfseries MSOA district road network total length vs aggregate PM2.5 pollution in 2019.} Line of best fit for a set of points depicting the relationship between the aggregate PM2.5 pollution within an MSOA and the total road length within the corresponding MSOA. The linear relationship led to our decision to use a linear regression model}
\label{fig:RoadLengthPM25ScatterPlot}
\end{center}
\end{figure}

We observed a positive linear correlation between the total road network length and the annual aggregate air pollution within a district. This correlation is shown in Figure \ref{fig:RoadLengthPM25ScatterPlot} for the pollutant PM2.5 in 2019. This observation led us to use a linear model to predict the annual level of air pollution. The Pearson correlation coefficient between the aggregated interpolation pollution and road length was 0.943.

\section{Methodology}
\label{sec:methodology}

\subsection{Models}
\label{sec:methodology:models}

We used three variations of the model using different feature vectors. Example feature vectors and predicted pollution values for the models can be seen in Section \ref{S-sec:modelDetails}. 

\begin{figure}[ht]
\begin{center}
\includegraphics[width=0.9\textwidth]{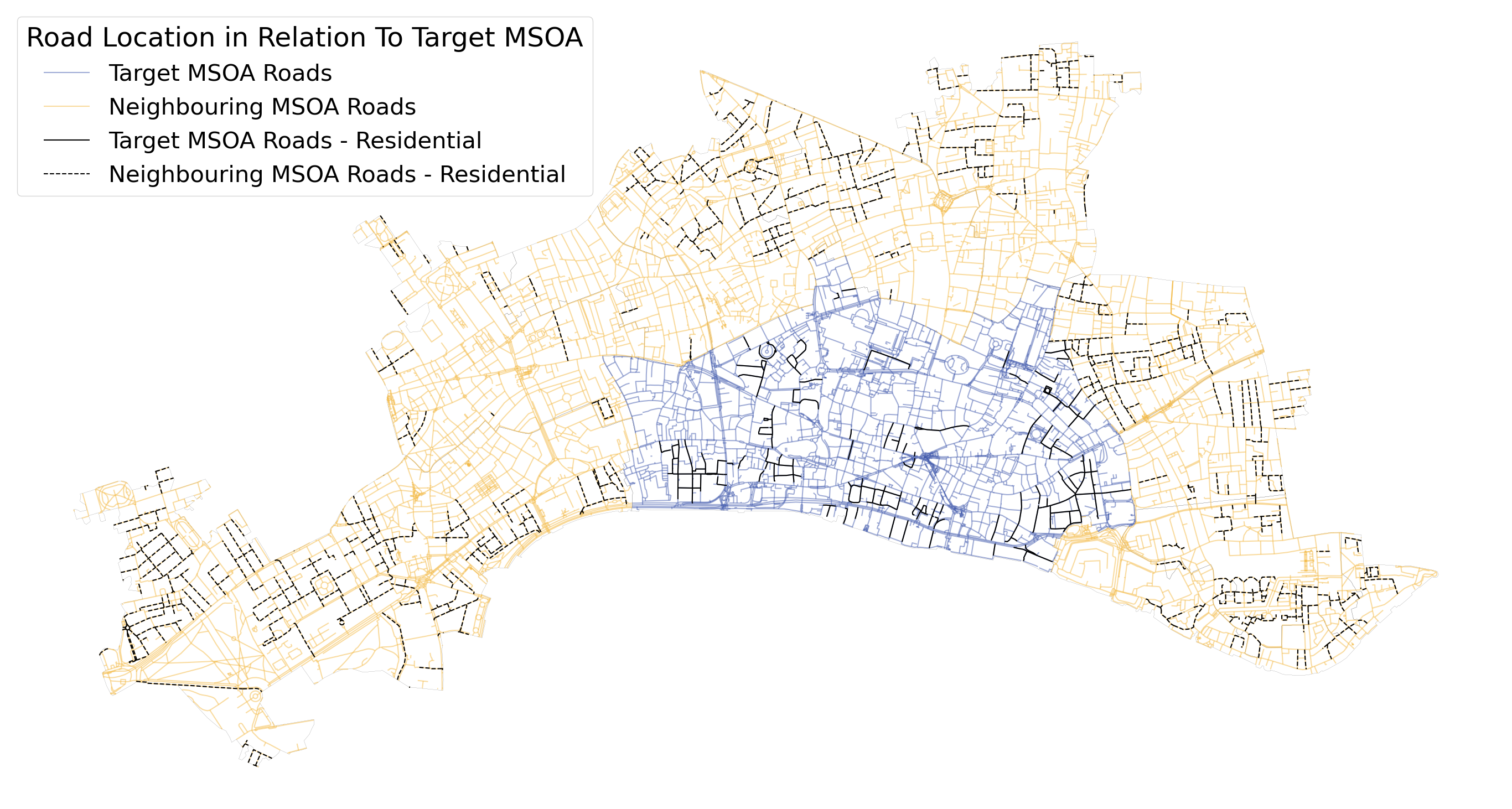}
\caption{{\bfseries Road network structure within MSOA City of London 001 (blue) and surrounding MSOAs (yellow).} The solid black line represents the residential roads within MSOA City of London 001 and the dashed black line the residential roads within the neighbouring MSOAs. The road network will contribute differently to the feature vector values depending on the framework used. In the case of the length model, the simple overall length of the road network is used, which for MSOA City of London 001 is 223,092m (in meters), visualised with the blue and black solid lines. In the case of the composition model, each road type contributes to a different single feature vector element, with the black solid residential road contributing a single element of the feature vector with the value 23,235m. The spatial variant of the model takes into account the total road length for different road types in adjacent MSOAs alongside the considerations of the composition model, meaning the total for the dashed black line compromises a single element of the feature vector used, in this case, 82,500m}
\label{fig:SpatialModelVisualisation}
\end{center}
\end{figure}

The first developed model aimed to predict a district's pollution from the total road network length; a simple linear regression \citep{freedman:2009:statistical} was used. The results from this model can be seen in Table \ref{tab:Table 3 All highway types results length}, detailing the mean absolute error, mean squared error, and the R\textsuperscript{2} (coefficient of determination) \citep{draper:1998:applied}. This table should be seen in conjunction to the other tables described later to allow a fair comparison between the models. An example feature vector for the length model can be seen in Table \ref{S-tab:Table 4 Curated Example Length Feature Vector}, with predictions made by the model shown in Table \ref{S-tab:Lenght Model Example Outputs}. In relation to Figure \ref{fig:SpatialModelVisualisation} the model only considers the total length of the blue road network. 

\begin{table}
\resizebox{\linewidth}{!}{
\begin{filecontents*}{CSVFiles/AllHighwayTypesResultsLengthRounded.csv}
Pollution Name,MAE Length,MSE Length,R2 Length
CO,57.808,11254.374,0.544
NO,3596.192,42036395.288,0.711
NO₂,4748.005,71646196.869,0.819
NOₓ,10277.881,337089243.351,0.773
NV₁₀,2458.049,19646741.387,0.861
NV₂₅,1484.307,7014357.411,0.842
O₃,7956.092,221380110.422,0.871
PM₁₀,3005.557,29519093.175,0.859
PM₂₅,1942.964,12019157.647,0.842
SO₂,546.223,1006964.606,0.533
V₁₀,559.122,1027117.426,0.858
V₂₅,477.524,733041.363,0.86
\end{filecontents*}
\pgfplotstabletypeset[
    multicolumn names=l, 
    col sep=comma, 
    string type, 
    header = has colnames, 
    columns={Pollution Name, MAE Length, MSE Length, R2 Length},
    columns/Pollution Name/.style={column type=l, column name=Pollutant Name},
    columns/MAE Length/.style={column type={S[round-precision=2, table-format=1.2e1,scientific-notation = true, table-number-alignment=center]}, column name=Mean Absolute Error (µg/m$^3$)},
    columns/MSE Length/.style={column type={S[round-precision=2, table-format=1.2e1,scientific-notation = true, table-number-alignment=center]}, column name=Mean Squared Error (µg/m$^3$)},
    columns/R2 Length/.style={column type={S[round-precision=2, table-format=1.2, table-number-alignment=center]}, column name=$R^2$ Coefficient of Determination},
    every head row/.style={before row=\toprule, after row=\midrule},
    every last row/.style={after row=\bottomrule}
    ]{CSVFiles/AllHighwayTypesResultsLengthRounded.csv}}
    \smallskip
    \caption{{\bfseries Results from the length feature vector model using all road types.} The results show that model performance achieves a high of 0.87 in $R^2$ score across the 12 pollutants. Of note is that CO and SO$_{2}$ have a lower score of around 0.535. As shown in Figure \ref{S-fig:RMSEInterpolationPowerScatterPlots} the interpolation process was unable to estimate a power parameter for the sample locations for CO, SO$_{2}$, V$_{10}$, indicating that there are not enough stations within the UK to pick up the variability over the distances separating the stations at the annual temporal level. However, V$_{10}$ achieved a 0.86 R2 score. The reason for this discrepancy between the three pollutants is the change in air pollution between the training (2014-2017) and test year (2018). For the V$_{10}$interpolated raster, a change of concentrations from 3.03 (µg/m$^3$) in 2017 to 3.16 (µg/m$^3$) in 2018 (4.3\% increase) was observed. The change in V$_{10}$ was considerably smaller than in CO, decreasing from 0.22 (µg/m$^3$) in 2017 to 0.18 (µg/m$^3$) in 2018 (18.2\% decrease). Similarly, SO$_{2}$ decreased from 1.99 (µg/m$^3$) in 2017 to 1.73 (µg/m$^3$) in 2018 (13.1\% decrease). The Mean Absolute Error is included to give a context of the model performance between pollutants with considerably different maximum values, as seen in Figure \ref{S-fig:RMSEInterpolationPowerScatterPlots}. The Mean Squared Error is also included to highlight the model's performance concerning the more extreme values seen within the dataset, which are experienced due to the varying geographical sizes of the districts being estimated. }\label{tab:Table 3 All highway types results length}
\end{table}

The second variation of the model splits the road network into the different types of highways (roads) detailed in the OpenStreetMaps database \citep{OpenStreetMap:2021:KeyHighway}, and created a length element per road type in the feature vector. In relation to Figure \ref{fig:SpatialModelVisualisation} the feature vector still only considers the blue road network but now creates a separate feature for each road type, such as the solid black lines denoting the residential roads within the blue road network. An example feature vector for the composition model can be seen in Table \ref{S-tab:Table 7 Curated Example Highway Composition Feature Vector}, with predictions made by the model shown in Table \ref{S-tab:Curated Highway Composition Model Results}. A multiple linear regression \citep{freedman:2009:statistical} implemented this model. The results from this model can be seen in Table \ref{tab:Curated Highway Composition Model Results}, detailing the mean absolute error, mean squared error and the R\textsuperscript{2} (coefficient of determination) \citep{draper:1998:applied}. When contrasting this with Table~\ref{tab:Table 3 All highway types results length} which uses the total length of roads, it becomes clear that this model in which lengths are separated per type of road performs better. 

\begin{table}
\resizebox{\linewidth}{!}{
\begin{filecontents*}{CSVFiles/AllHighwayTypesResultsCompositionRounded.csv}
Pollution Name,MAE Composition,MSE Composition,R2Composition
CO,37.885,7105.308,0.712
NO,2275.254,24531005.517,0.832
NO₂,2787.602,33815646.291,0.915
NOₓ,6244.618,176639841.556,0.881
NV₁₀,1358.646,7912333.282,0.944
NV₂₅,840.539,2965264.271,0.933
O₃,4429.423,90029016.993,0.948
PM₁₀,1672.108,12032403.918,0.942
PM₂₅,1104.896,5172685.475,0.932
SO₂,357.653,632527.654,0.707
V₁₀,313.205,423503.083,0.941
V₂₅,265.856,297593.321,0.943
\end{filecontents*}
\pgfplotstabletypeset[
    multicolumn names=l, 
    col sep=comma, 
    string type, 
    header = has colnames, 
    columns={Pollution Name, MAE Composition, MSE Composition, R2Composition},
    columns/Pollution Name/.style={column type=l, column name=Pollutant Name},
    columns/MAE Composition/.style={column type={S[round-precision=2, table-format=1.2e1,scientific-notation = true, table-number-alignment=center]}, column name=Mean Absolute Error (µg/m$^3$)},
    columns/MSE Composition/.style={column type={S[round-precision=2, table-format=1.2e1,scientific-notation = true, table-number-alignment=center]}, column name=Mean Squared Error (µg/m$^3$)},
    columns/R2Composition/.style={column type={S[round-precision=2, table-format=1.2, table-number-alignment=center]}, column name=$R^2$ Coefficient of Determination},
    every head row/.style={before row=\toprule, after row=\midrule},
    every last row/.style={after row=\bottomrule}
    ]{CSVFiles/AllHighwayTypesResultsCompositionRounded.csv}}
    \smallskip
    \caption{{\bfseries Results from the composition feature vector model using all road types.} The $R^2$ for all air pollutants has improved over the length model detailed in Table \ref{tab:Table 3 All highway types results length}, alongside reducing both the Mean Absolute Error and Mean Squared Error, indicating an improved model framework for estimating air pollutants, with the same relative performance between air pollutants. Note that this model is also based on road lengths, but the lengths here are not the total but based on road type.} \label{tab:Curated Highway Composition Model Results}
\end{table}

The third model created was a spatial variant of the second model, using the same multiple linear regression technique as the composition model.  We created the spatial model to capture whether the district was in the centre of an urbanised zone, on the outskirts or wholly removed, as the primary sources of a district's air pollution might be outside the district's boundaries. The feature vector for the spatial model comprised the same features as the composition model; however, the process conducted for the blue road network within Figure \ref{fig:SpatialModelVisualisation} was repeated for the yellow road network, providing a total length of the adjacent district's road networks, thereby doubling the number of features present with the feature vector over the composition model, as seen in Table \ref{S-tab:Table 10 Curated Example Spatial Feature Vector}. Predictions made by the spatial model are shown in Table \ref{S-tab:Table 11 Curated Spatial Predicition}. The use of an additional set of elements that give values for the sum of the road network length by road type in adjacent MSOAs to the estimated district has the effect of something akin to a network in which the neighbours influence the values of the current location, causing the values used in the composition model feature vector to be more contextualised. The results from this model can be seen in Table \ref{tab:Table 11 Curated Model Results Spatial}, detailing the mean absolute error, mean squared error and the R\textsuperscript{2} (coefficient of determination) \citep{draper:1998:applied}.

While the improvements in the $R^2$ score are minimal, there is no overhead cost in creating the additional features for the vector as data collection concerning the roads has already taken place. Similar computational costs between the models also arguably make the time investment worthwhile, as there are minimal barriers to implementing the spatial variant of the composition model. 

Overall the three different variations of the model help alleviate a specific situation that could be encountered and have varying benefits. The length model has the benefit of being conceptually simple, allowing multiple datasets to be integrated due to the universal property of road length between datasets. The composition model offers improvements over the length model but restricts the datasets that can be used where the classification schema is consistent across the study period. Finally, the spatial model offers some further performance improvements but requires considerably more data than the composition model, alongside adding further complexity to the model itself.

\begin{table}
\resizebox{\linewidth}{!}{
\begin{filecontents*}{CSVFiles/AllHighwayTypesResultsSpatialRounded.csv}
Pollutant Name,MAE Spatial,MSE Spatial,R2 Spatial
CO,38.35,6698.946,0.729
NO,2320.79,23905675.199,0.836
NO₂,2845.108,31823322.218,0.92
NOₓ,6363.865,169035020.271,0.886
NV₁₀,1398.695,7467525.593,0.947
NV₂₅,863.295,2779375.839,0.937
O₃,4582.988,86357497.527,0.95
PM₁₀,1721.591,11341938.593,0.946
PM₂₅,1133.152,4837131.307,0.936
SO₂,361.93,592487.886,0.725
V₁₀,323.345,399460.291,0.945
V₂₅,273.781,280215.091,0.946
\end{filecontents*}
\pgfplotstabletypeset[
    multicolumn names=l, 
    col sep=comma, 
    string type, 
    header = has colnames,
    columns={Pollutant Name, MAE Spatial, MSE Spatial, R2 Spatial},
    columns/Pollutant Name/.style={column type=l, column name=Pollutant Name},
    columns/MAE Spatial/.style={column type={S[round-precision=2, table-format=1.2e1,scientific-notation = true, table-number-alignment=center]}, column name=Mean Absolute Error (µg/m$^3$)},
    columns/MSE Spatial/.style={column type={S[round-precision=2, table-format=1.2e1,scientific-notation = true, table-number-alignment=center]}, column name=Mean Squared Error (µg/m$^3$)},
    columns/R2 Spatial/.style={column type={S[round-precision=2, table-format=1.2, table-number-alignment=center]}, column name=$R^2$ Coefficient of Determination},
    every head row/.style={before row=\toprule, after row=\midrule},
    every last row/.style={after row=\bottomrule}
    ]{CSVFiles/AllHighwayTypesResultsSpatialRounded.csv}}
    \smallskip
    \caption{{\bfseries Results from the spatial composition feature vector model using all road types.} The $R^2$ has improved for some of the pollutants included within the study over the composition model detailed in Table \ref{tab:Curated Highway Composition Model Results} other than NO$_\text{2}$, O$_\text{3}$, however not all the metrics have improved in the same manner as the transition from the length to composition model, for example with the Mean Absolute Error and Mean Squared Error increasing, potentially indicating a less robust model framework for estimating air pollutants that requires additional data over the other model's details in Table \ref{tab:Table 3 All highway types results length} and Table \ref{tab:Curated Highway Composition Model Results}.}\label{tab:Table 11 Curated Model Results Spatial}
\end{table}

\subsection{Feature Selection}
\label{sec:methodology:featureselection}

The OpenStreetMaps dataset is exceptionally comprehensive with regards to road information. The feature vector in the previously described models included all road types within the OpenStreetMaps dataset. There are 99 different road types but some are misclassified (shown in tables in Section \ref{S-sec:featureSelectionDetails}).
This section describes the work conducted to reduce the input dataset’s size to only what was needed to achieve the goal of estimating air pollution within a district. 

\begin{figure}
  \begin{subfigure}{0.48\textwidth}
    \captionsetup[figure]{font=small}
    \includegraphics[width=\linewidth]{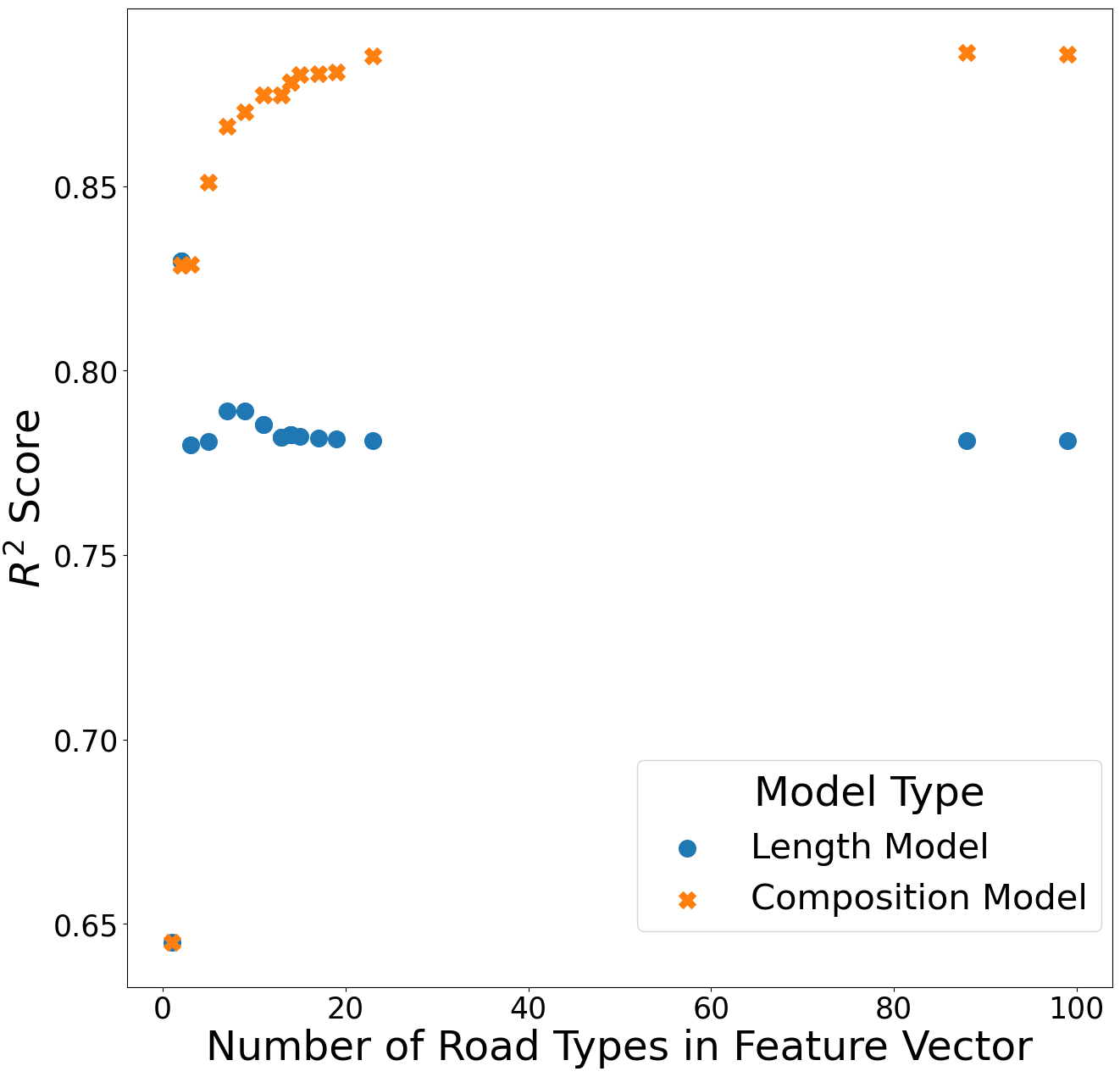}
    \caption{{\bfseries Starting with all 99 road types.} } \label{fig:ScatterPlotNumHighwaysR2ScoreAllHighways}
  \end{subfigure}%
  \hspace*{\fill}   
  \begin{subfigure}{0.48\textwidth}
    \captionsetup[figure]{font=small}
    \includegraphics[width=\linewidth]{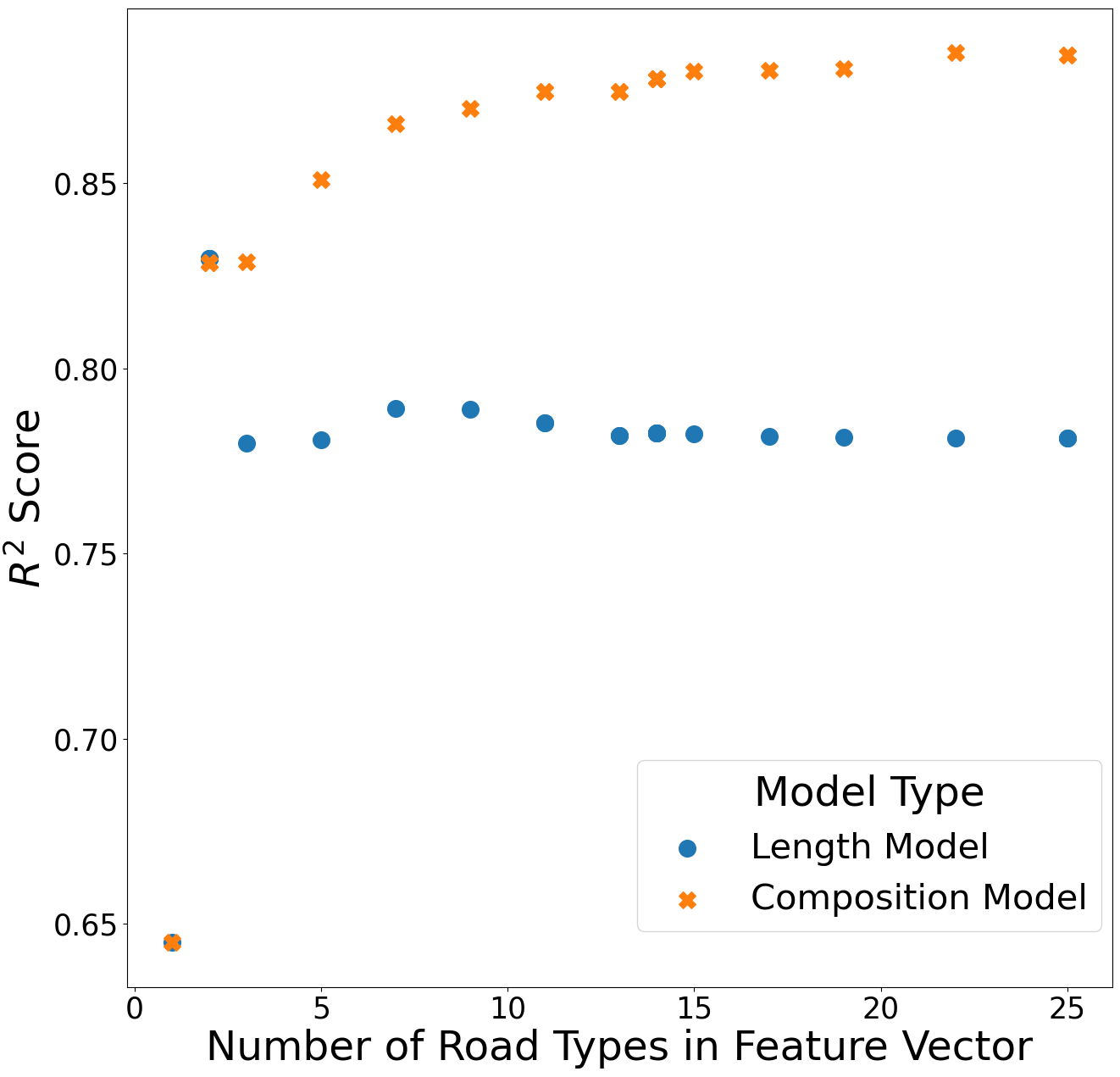}
    \caption{{\bfseries 25 road types relevant to all air pollutants.} } \label{fig:ScatterPlotNumHighwaysR2Score24HighwaysSubset}
  \end{subfigure}%

\caption{{\bfseries Model performance vs $R^2$ score, with road type inclusion based upon road type mutual information with air pollution.} Shown is the performance of the length and composition model when removing certain roads from being included as part of the feature vector. Inclusion in the feature vector depends on the sum of mutual information of the road type with pollutants considered in the study. A threshold is incremented to remove the road types with less mutual information than others. The requirement for a non-zero mutual information sum removes 11 road types with no relation to pollution, such as the OpenStreetMaps classification of turning circles, leaving 88 road types. Many road types have a minimal sum for mutual information, so incrementing the threshold to 0.1 removes 65 road types, with classifications such as living street removed. This process is then repeated until a single road type remains, in this case, the track road classification, which can be seen at which point the length and composition model reduces to the same model. Figure \ref{fig:ScatterPlotNumHighwaysR2Score24HighwaysSubset} details the same experiment detailed in Figure \ref{fig:ScatterPlotNumHighwaysR2ScoreAllHighways} however only road types that have a non zero value for mutual information with all pollutants were considered, leaving 25 road types, removing road types such as lane that only have a non zero mutual information values with SO$_\text{2}$} \label{fig:highwaySubsettingMIForPollutants}
\end{figure}

Given the number of variables (road types) that exist in the OpenStreetMaps dataset, we relied on information theory \cite{shannon1948mathematical} and in particular mutual information between the road types and the different air pollutants to determine which road types have the highest relation with each air pollutant. As mentioned above, the initial number of road types in the dataset was 99. Removing road types that had no relevance to the air pollutants being studied, represented by a mutual information summation value of 0, kept 88 road types. The types removed were mostly misclassifications by a user inputting a road type, such as "fence" or "residential;footway". 
Our next step was incrementing the threshold for removal for a road types summation value and comparing the resulting model’s performance by its $R^2$ score. Figure \ref{fig:ScatterPlotNumHighwaysR2ScoreAllHighways} shows a plot of the resulting $R^2$ score for both the length and composition models against differing numbers of road types subset by the value of the threshold for inclusion of the mutual information summation. Table \ref{S-tab:Every5thMutualInformationPerPollutant} shows mutual information values for the pollutants across a range of road types, and Table \ref{S-tab:R2ValuesAll99HighwaysFeatureSelection} shows the model's performance with reduced input datasets.

We then repeated the experiments; however, we included only road types relevant to all 12 pollutants in the test.  25 out of 99 pollutants had relevance to all 12 pollutants to be estimated; relevance is measured by ensuring that the road type had non-zero mutual information for all 12 pollutants.  The results of this experiment are shown Figure \ref{fig:ScatterPlotNumHighwaysR2Score24HighwaysSubset}. Table \ref{S-tab:Table 14 R2 Values Highway Types 22 relevant highway types} shows the results from the previous three models with road types with zero mutual information with at least one of the pollutants.

In Figure \ref{fig:highwaySubsettingMIForPollutants}, the best model uses two road types, track and unclassified, to predict the pollution within a district, seen distinctly with an increase in the $R^2$ score for the length model to 0.83, up from 0.78. We can see that some road types cause the length model to perform worse, but even when the result improves, it never performs as well as the composition model. Eventually, with only one road type remaining, the length and composition model are reduced to the same linear regression model, producing an $R^2$ score of 0.65.

We performed pairwise mutual information between the road types to further reduce the dataset. The goal was to determine the road network types that contained the most information about other road types within the dataset. The road types we performed the mutual information on were the 25 road types with some mutual information with every pollutant covered in the study. 

\begin{figure}[ht]
\begin{center}
\includegraphics[width=0.95\textwidth]{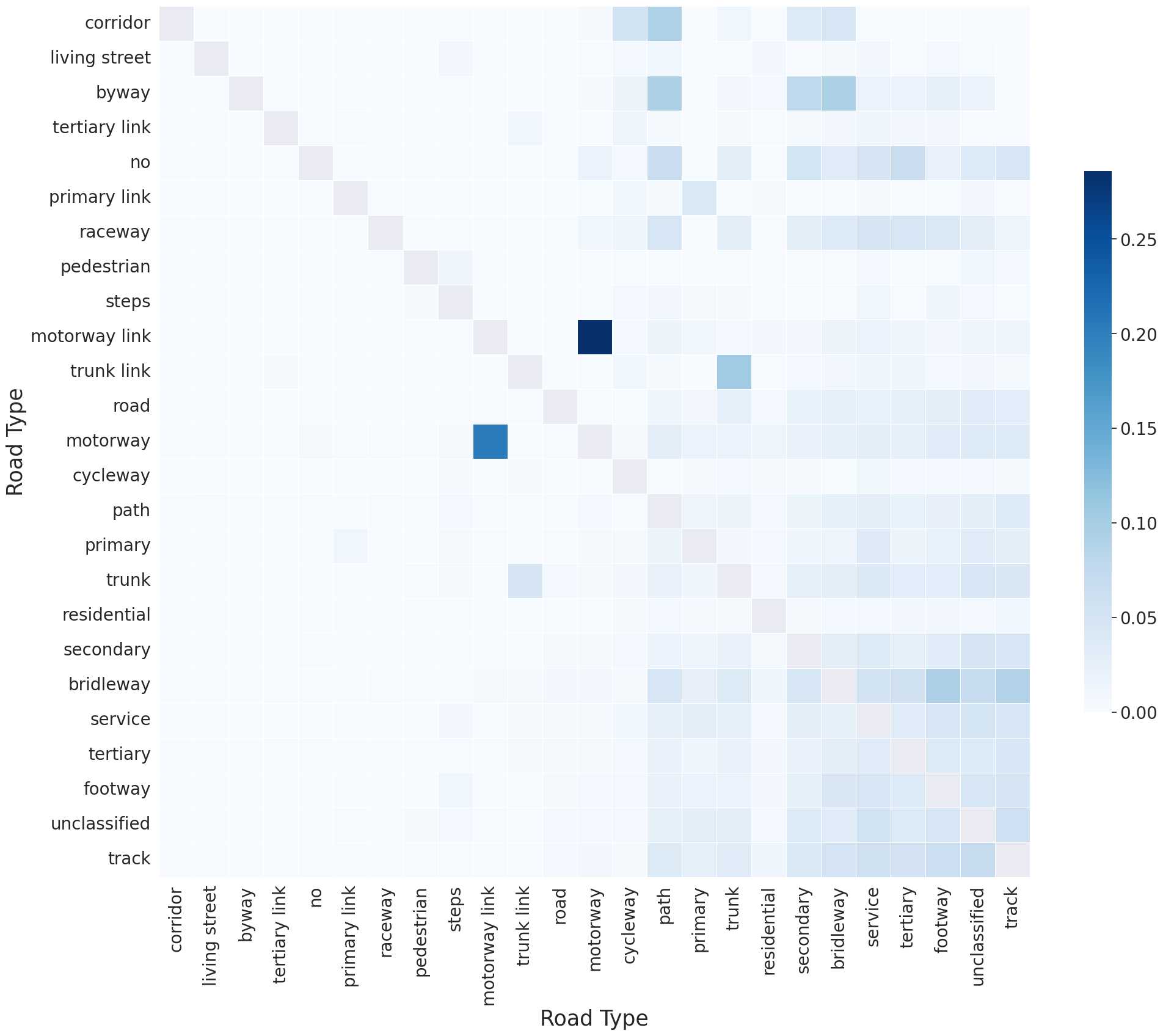}
\caption{{\bfseries Mutual information between road types.} The heatmap shows the mutual information between all 25 road types with a non-zero mutual information value with all the pollutants in the study. The heatmap values are normalised by row with the diagonal removed (which has value 1), allowing for an understanding of which road type has the highest mutual information to the other road types in the dataset. Motorway link has the highest mutual information with the motorway road type. While the motorway type has strong mutual information with the motorway link type, there are also strong mutual information scores with other road types, such as track and path. This difference in the strength of mutual information indicates that the motorway link dataset contains information mainly about the motorway road type, but the motorway dataset contains information about motorway link and other road types such as track, path etc., giving it a more desirable use in retaining information about all datasets (values are normalized by row and the diagonal removed)}
\label{fig:NormalizedHeatmap}
\end{center}
\end{figure}

Figure \ref{fig:NormalizedHeatmap} shows a heatmap of the mutual information values between the 25 road types. The values in the heatmap have been normalised row-wise. In the heatmap, there are 625 data points. 457 points have a value of mutual information greater than zero. Some data points have a value of 0 for mutual information; for example, living street and raceway have zero pairwise mutual information.

The heatmap was modelled as a network of the 25 road types to determine the road types that contain the most mutual information about other road types, as shown in Figure \ref{S-fig:NetorkModelMutualInformation}. Each node in the network represents a road type, and the edge weight between the two nodes is the mutual information between the two road types. The degree of each node is equal to the sum of the weights of the edges adjacent to the given node, giving a value for the amount of mutual information a given road type contains about other road types in the network. 

The node with the highest degree was chosen for inclusion in the model feature vector and then removed from the network as the mutual information of the edges no longer needed to be included in the dataset. This process was repeated, with the next highest node based on degree weight chosen, until no nodes remained. The results of this process on the network representation can be seen in Figure \ref{S-fig:NetworkModelRemoveTimesteps}.

\begin{figure}[ht]
\begin{center}
\includegraphics[width=0.5\textwidth]{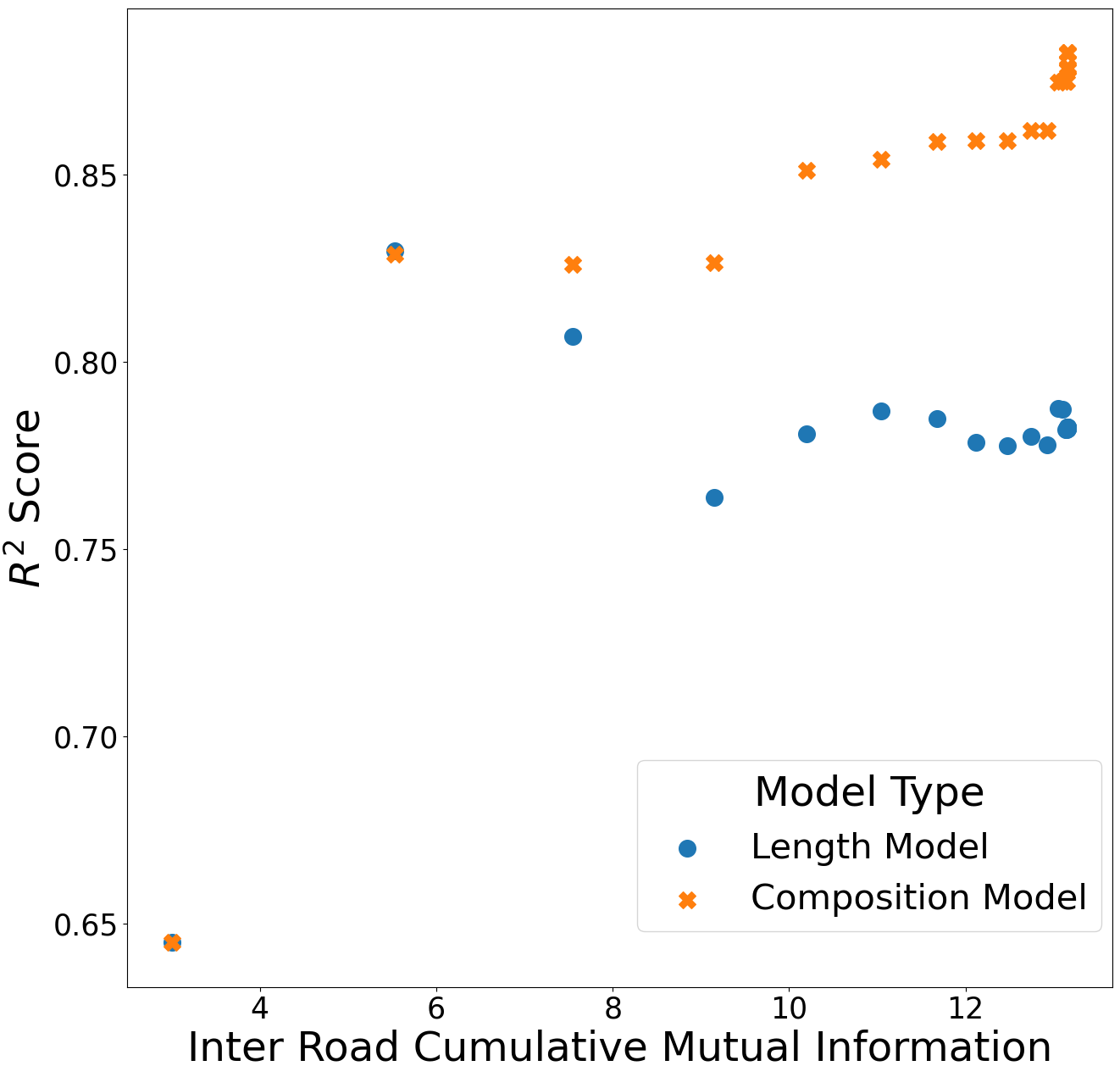}
\caption{{\bfseries Total inter road cumulative mutual information in input feature vector vs $R^2$ score.} Shown is the performance of the length and composition model when removing road types from being included as part of the feature vector based on the road type's mutual information with other roads, aiming to reduce the data required while maintaining performance by including road types containing information about other road types. A single road type was removed at each step, with the x-axis detailing the total remaining mutual information about other road types within the datasets used for the feature vector. Eventually, the model reduces into the same model when only a single road type remains. The performance improvement of the length model as some road types are removed indicates that some of the datasets included are noisy and affect model performance; this issue is not present with the composition model that can differentiate different road types, seen with a gradual reduction in model performance. The full experiment results can be seen in Table \ref{S-tab:Inter road mutual information tests}}
\label{fig:ScatterPlotNumHighwayTypesInterRoadSubset}
\end{center}
\end{figure}

The test results in Figure \ref{fig:ScatterPlotNumHighwayTypesInterRoadSubset} show that a model underpinned by the ``track'' and ``unclassified road'' road type produce a model with an $R^2$ score of 0.83 while only making use of 25.9\% of the total road network length in the study area. Indicating that these two types of roads are sufficient to provide a good estimation of air pollution.

\subsection{Missing Districts Tests}
\label{sec:methodology:singleyeartests}

We also conducted experiments to explore whether some districts within the study area could be used for training, having the remaining districts estimated from the model created. The idea was to reduce the input dataset further by removing similar districts from the training set, retaining only a subset of urban, suburban or rural districts. The population density was used as a proxy metric for how urbanised a district is, with example districts shown in Table \ref{S-tab:Table 17 Population Density}. The goal was to understand the minimal number of roads that need to be measured across different districts to predict pollution across all districts accurately. We did not include the spatial variant in these tests, as it would require additional data acquisition over the minimal amount of measuring needed for just the district itself. We used the data from 2018 for the tests. For a set of 16 random states, different MSOAs were randomly sampled from the 7201 total MSOA districts. When selecting the additional districts to be included in the test, we ensured that the districts used in the previous iterations were still present. For example, the districts used in one random state in the 100 training size set were also used in the 500 training size set, as visualised in Figure \ref{S-fig:InterYearExperimentTrainingSetsVisualisation}.

Table \ref{S-tab:Table 18 Inter year tests} shows the results for the mean $R^2$ score for both the length and the composition model for reducing training set size across 16 random state tests. We chose to explore test set sizes ranging from 100 MSOAs to 7200 MSOAs, with the training set being made up of the remaining MSOAs, for example, 100 MSOAs in the test set and 7101 MSOAs in the training set.

\begin{figure}
  \begin{subfigure}{0.48\textwidth}
    \includegraphics[width=\linewidth]{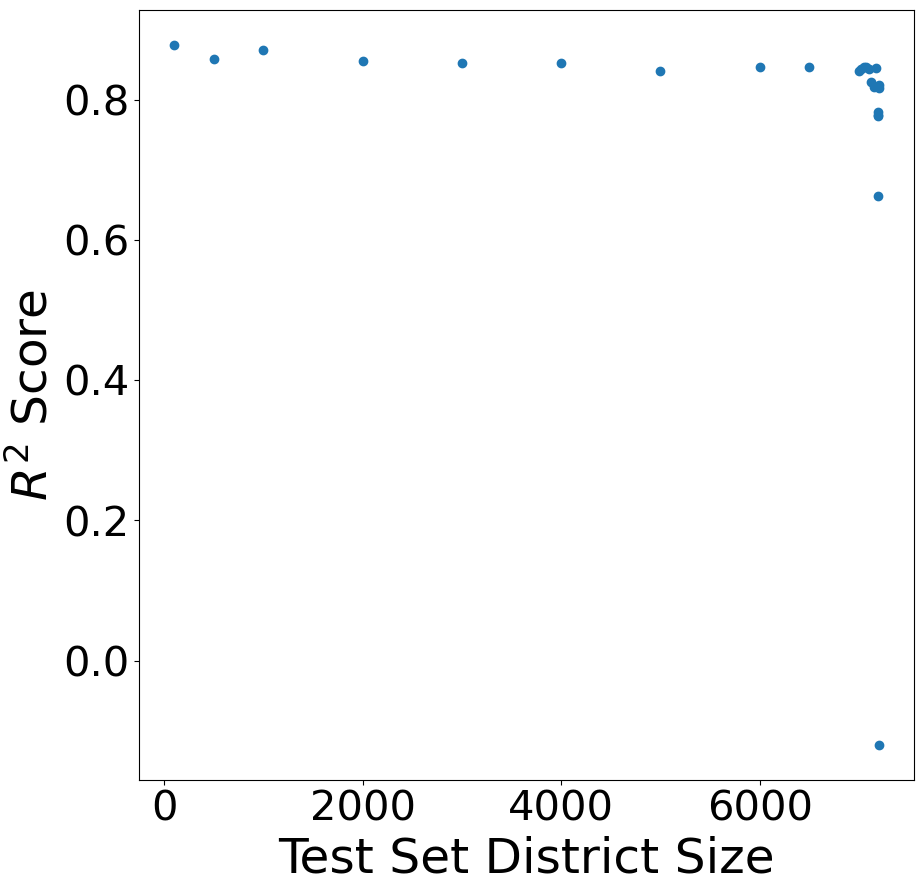}
    \caption{Length Model} \label{fig:TestSizeRegionsVsR2ScoreLength}
  \end{subfigure}%
  \hspace*{\fill}   
  \begin{subfigure}{0.48\textwidth}
    \includegraphics[width=\linewidth]{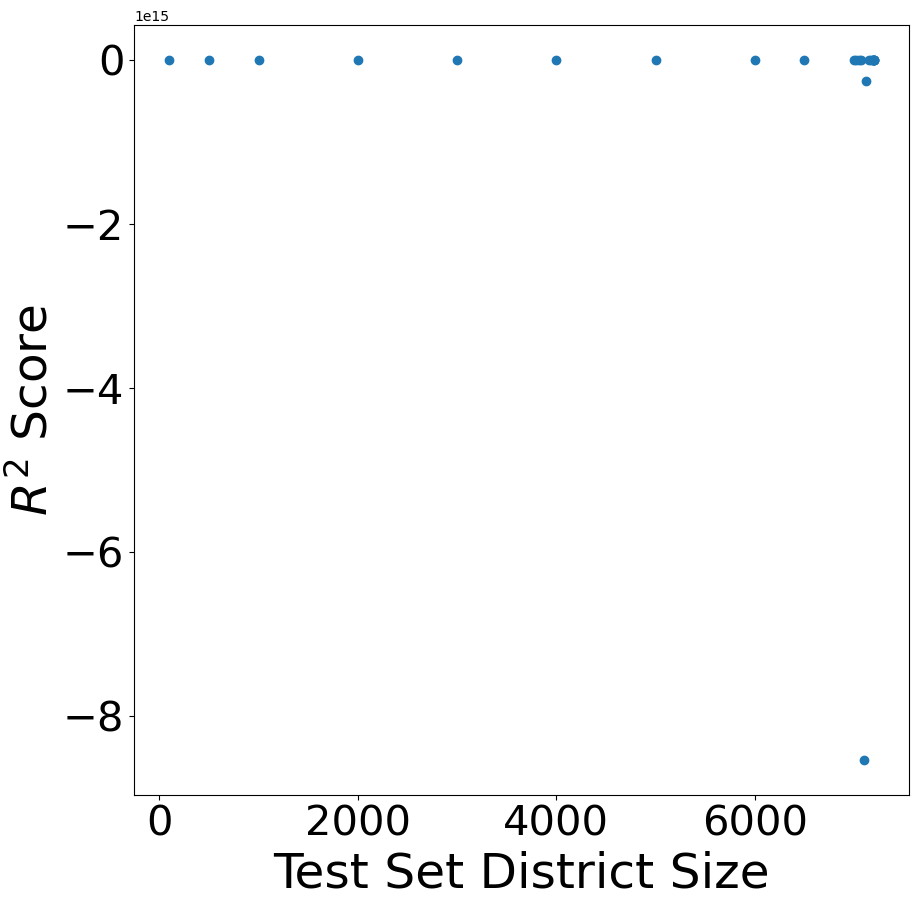}
    \caption{Composition Model} \label{fig:TestSizeRegionsVsR2ScoreComposition}
  \end{subfigure}%

\caption{{\bfseries Test set size districts vs $R^2$ score.} Shown is the performance of the length and composition model as the input districts for the training set are reduced from 7101 to 1. Of note is the difference in the scale of the y-axis between the figures. As the test set size increases (and so the training set size decreases), the performance of both models remains stable until the test set comprises 7,175 out of 7,201 (0.36\%) districts. At that point, both models begin to reduce in performance. The length model maintains its performance at around 0.8 $R^2$ score for longer, with the composition model degrading to an average $R^2$ score of 0.65. When the test set size is 7,200, meaning only a single district is used as the training set, both models break down worse than just estimating the average air pollution value; however, the composition performs significantly worse than the length model. Thereby indicating that while the composition model performs better than the length model initially, it does require more input data to perform well} \label{fig:TestSizeRegionsVsR2Score}
\end{figure}

Figure \ref{fig:TestSizeRegionsVsR2Score} shows the scatter plots for the data detailed in Table \ref{S-tab:Table 18 Inter year tests}. The random tests show that the length model is more robust than the composition model, indicating that the composition of the road network between different districts can vary greatly, resulting in poor performance in some random training sets. The length model maintains performance as the training size is reduced from the initial 7101 MSOAs to 26 MSOAs, with an $R^2$ score of 0.85. After this, the performance of the model begins to decrease. Across a range of different random states, around 26 MSOAs need to be used to train the model to successfully predict all MSOAs pollution, representing 0.36\% of the total number of MSOAs. 

\clearpage

\begin{figure}[ht]
\begin{center}
\includegraphics[width=1\textwidth]{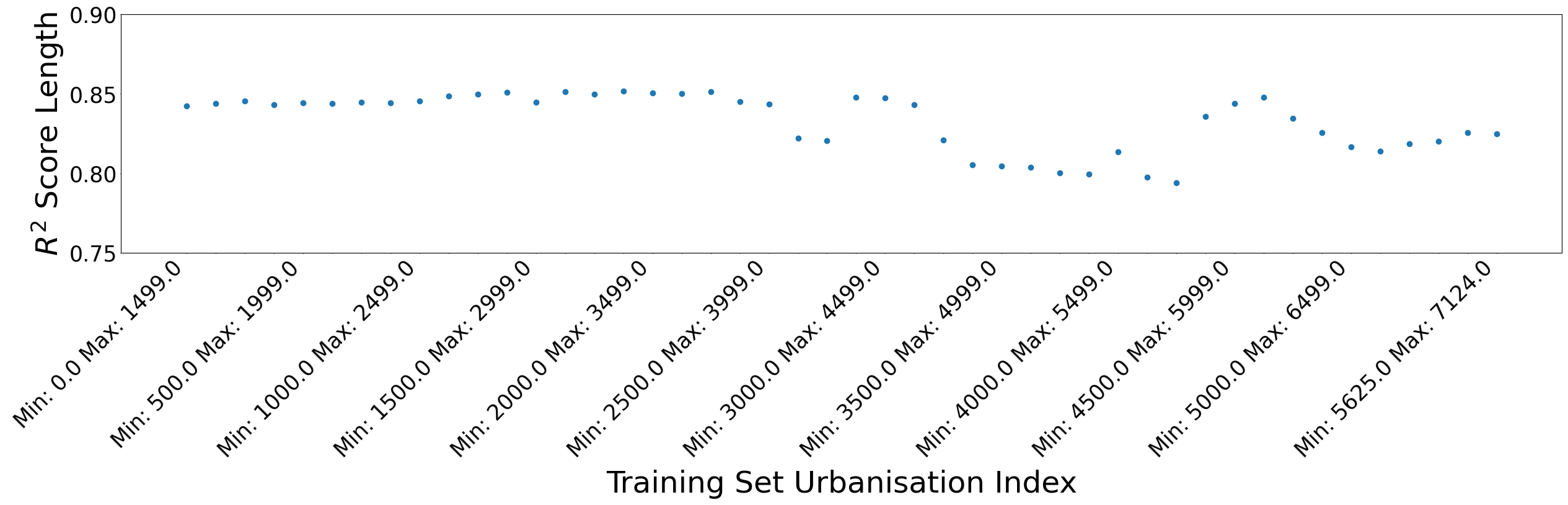}
\caption{{\bfseries Test sizes vs $R^2$ score for length model for changing population density within the training set.} Shown is the length model's performance across a range of training sets chosen based on the MSOAs population density. The population density for an MSOA is used as a proxy for how urbanised an MSOA is; the higher the population density, the more urbanised an MSOA is. 1,500 MSOAs were chosen to be part of each training set, with the variation being in the average population density of the training set MSOAs. The first point (Min 0 Max 1,499) is where the least urbanised MSOAs are the training set. The last point (Min 5,625 Max 7,124) is where the most urbanised MSOAs are the training set. The consistent performance of the length model as the average urbanisation of the training set MSOAs increases indicates how urbanised an MSOA does not affect the model performance. Table \ref{S-tab:trainingSetSizePopulationDensity} shows the full experiment results}
\label{fig:ScatterPlotPopulationDensity}
\end{center}
\end{figure}

As seen in Figure \ref{fig:TestSizeRegionsVsR2Score}, the $R^2$ score for the length model averaged 0.83, providing a  baseline performance for the next experiment. We trained a set of models with 1500 districts with similar population densities to explore the effect of changing population density distribution in the training set on model performance. Starting from the least urbanised MSOAs, to the most urbanised MSOAs, with increments of 125, a model was trained and the $R^2$ score computed. The results from this experiment are shown in Figure \ref{fig:ScatterPlotPopulationDensity}, where the average $R^2$ score across the 46 tests was 0.83. The $R^2$ score remained similar to the expected 0.83 from the random tests as the training set changed from least to most urbanised. Thus, pointing toward the idea that the $R^2$ score of the length model is not related to the population density distribution of the MSOAs included in the subset of training districts.

\clearpage
\section{Use Cases}
\label{sec:usecases}

From the tests conducted as part of this work, the critical road types to predict the pollution in an area are the ``minor roads''; the road types ``track'' and ``unclassified'' within the OpenStreetMaps classification schema. Not only do the road types ``track'' and ``unclassified'' have the most mutual information about the 12 pollutants analysed within this study, with a normalised summation value of 12 and 10.12 respectively, but they also contain the most mutual information about the other road types with a total value of 2.97 and 3.00 respectively. The ``track'' and ``unclassified'' road type input data produce a model with an $R^2$ score of 0.83 for both the Length and Composition model. The model then degrades to an $R^2$ score of 0.65 when only the ``track'' road type is included, where the length and composition model is the same. While the $R^2$ score of the complete composition model, with all 25 road types, improves to 0.89 compared to the length models $R^2$ score of 0.78, an argument is that the increased amount of data needed to attain the increase of 0.11 on the $R^2$ is not worthwhile. The entire length of the road network within the MSOA boundaries in 2018 is 1,140,835,548m, out of which 295,948,514m, represents the ``track'' and ``unclassified'' road network, representing a 74.1\% reduction in input data for minimal performance loss. However, it is also vital to consider the context in which the model is intended for use. The model was never designed to accurately predict the pollution in an area to a specific value, but rather to give an idea of where it is most polluted within an area to help identify areas for intervention. Within this context, the loss of 0.11 $R^2$ score is meaningless in the model’s practical use. 

The missing districts tests have clear implications on the approach of deploying monitoring stations to create a target vector. The experiments show that a distribution of urban, suburban and rural districts across the training set is unnecessary. In the UK case, only about 0.36\% of districts, or 26/7201 districts, were required to accurately predict pollution across all districts at the annual level. Using the model proposed would allow policymakers to have full spatial coverage of air pollution while only deploying monitoring stations in 0.36\% of districts, representing significant monetary savings over deploying a comprehensive air pollution monitoring network like the UK's AURN.

The findings of this study show that a linear regression model underpinned by a single structural property, length, of the track and unclassified road network within 0.36\% of districts within England and Wales was enough data to identify an ordering for which districts are the most polluted. Furthermore, the model discussed presents a low-cost method of achieving similar results to more expensive models such as the one currently used by DEFRA. The model has apparent practical uses for policymakers that want to pursue clean air initiatives but lack the capital to invest in comprehensive dedicated monitoring networks.

\section{Discussion}
\label{sec:discussion}

In this work, we presented a solution for the estimation of annual air pollution using a dataset related to the structure of roads, we have focused on a model based on length of roads and also on a composition model which consider other factors associated to road classifications. 

Our contributions can be split into 3 categories. {\em (1)} We have shown that one can get to good annual estimations on a very low budget. This contrasts with current approaches in the UK which costs millions of pounds (£). Making an approach that is cheaper and accessible to multiple stakeholders is more inclusive, allowing society to use the estimations in secondary applications (health exposure, traffic, city planning, etc.) {\em (2)} Our model is quite transparent, and the characteristics used as well as the estimator based on simple  regressions is quite explainable. This leads to a better reusability and generality of the approach. {\em (3)} In a more general sense, we believe this work has an impact on showing that in data science and in particular, environmental data science, lack of data about a particular phenomenon can be overcome by using the information embedded in other datasets which may not have been collected with that intent. This happens due to mutual information, in which aspects of one phenomenon may be present in other pieces of information. 

While the use of road length is inferior at estimating air pollution to approaches that use dynamic data, the method presented has the benefit of a nominal implementation cost due to the data's open source nature, minimal training computation, and accessible model design. The missing districts experiments give a framework for using a minimal amount of monitoring stations across an area, in the case of the UK, 0.36\% of MSOAs, and use the model presented to fill in missing districts, producing a full spatial map of air pollution at the annual temporal level. The full spatial map ensures that all districts, urban, suburban and rural alike, have an estimate for ambient air pollution, helping to address inequality related to a current focus on urban areas with monitoring station placement.

While the composition model surpassed the length model in performance, there is a trade-off in operationalising the composition model over the length model due to the data used. The length model has the benefit of using a universal structural property of roads, its length, which is present across datasets such as OpenStreetMaps used in our study or other datasets concerning road networks such as Ordnance Survey Opens Roads \footnote{OS Open Roads Dataset: https://www.ordnancesurvey.co.uk/business-government/products/open-map-roads}. However, when using the composition model, there is an immediate limitation of being tied to a single dataset and the classification schema used for road types. For example, in the OS Opens Roads, there are only 6 classifications of roads, but as seen in Section \ref{sec:methodology:featureselection}, there were 99 different classifications in OpenStreetMaps. The road classification used also exposes the model to potential system shocks. One of them is the potential for OpenStreetMaps to change their classification schema, which could make it challenging to have a complete dataset that spanned multiple years. This issue is simple to solve with the length model, with the ability to combine datasets depending on their availability to extend the temporal coverage of the data used.

While the UK and many other global north countries can invest in  comprehensive monitoring-station networks, this is not always the case for lower-income counties, particularly those in the global south. Data concerning annual air pollution globally would help to inform policymakers' decisions and help to combat some of the 4.2 million death/year attributed to ambient air pollution \citep{WHO:2018:Burden}. The framework presented, paired with the global availability of road structural properties' data through OpenStreetMaps, provides a basis for future work to design a global annual air pollution model from similar secondary datasets to those used in this study.

Future work could explore the possibility of using the same framework to estimate air pollution at a higher temporal level, such as the daily or hourly level. However, there are likely limits to the approach presented in this paper at a finer temporal level due to the static nature of road transport infrastructure. Future work would highlight the need to transition from data concerning the static elements of the road, such as its length, to more dynamic aspects of the road, such as the traffic counts of different types of vehicles. However, this data is more expensive to acquire and more invasive to privacy. Therefore, critical evaluations are needed whether the benefits derived from the model underpinned by more invasive dynamic data are worth the additional benefits of more accurate air pollution environmental intelligence.

\paragraph{Funding Statement}
This research was supported by an EPSRC studentship awarded to LJ. Berrisford as part of the UKRI Centre for Doctoral Training in Environmental Intelligence.

\paragraph{Competing Interests}
None.

\paragraph{Data Availability Statement}
The road network data can be accessed via [\url{https://download.geofabrik.de/europe/great-britain.html}]. The boundaries for the MSOA used can be accessed via [\url{https://statistics.ukdataservice.ac.uk/dataset/2011-census-geography-boundaries-middle-layer-super-output-areas-and-intermediate-zones}].

\paragraph{Ethical Standards}
The research meets all ethical guidelines, including adherence to the legal requirements of the study country.

\paragraph{Author Contributions}
Conceptualization: L.B; R.M. Methodology: L.B; R.M; E.R. Data curation: L.B. Data visualisation: L.B. Writing original draft: L.B. Writing - Review and Editing L.B; E.R; R.M. Software L.B. All authors approved the final submitted draft.

\bibliographystyle{apalike}
\bibliography{Sample-refs}
\end{document}


\maketitle

\section{Study Area Choice}
\label{sec:MAUP}

The first significant design decision we made during the study was the district boundaries under which the length of the road network and air pollution would be aggregated. This problem is known as the Modifiable Aerial Unit Problem (MAUP) \footnote{Wong, David WS (2004). The modifiable areal unit problem (MAUP), WorldMinds: Geographical perspectives on 100 problems. 571–575}. Three schemas were considered for aggregating the study area into districts, each motivated by specific design philosophy.

\begin{figure}[ht]
\begin{center}
\includegraphics[width=0.3\textwidth]{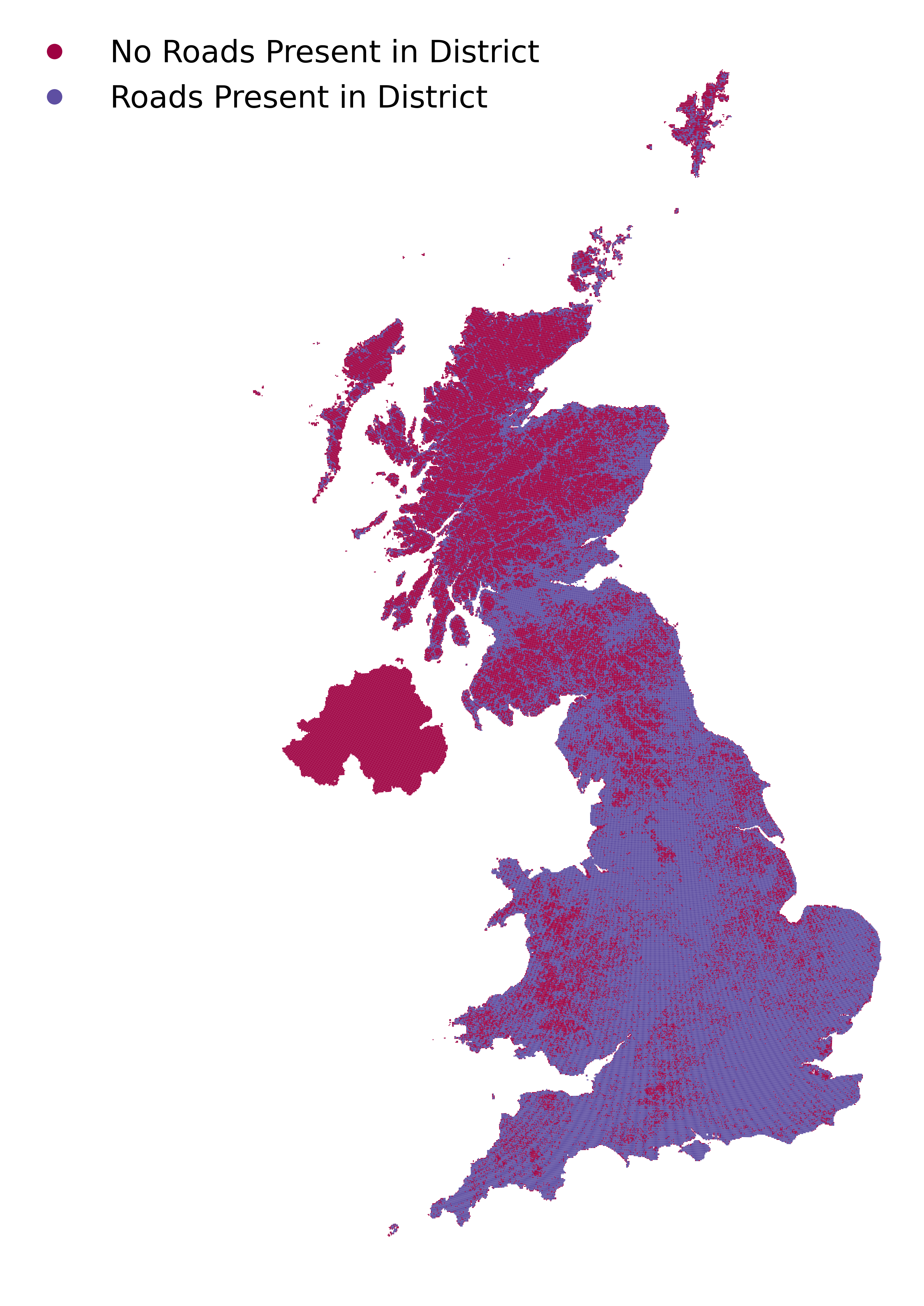}
\includegraphics[width=0.3\textwidth]{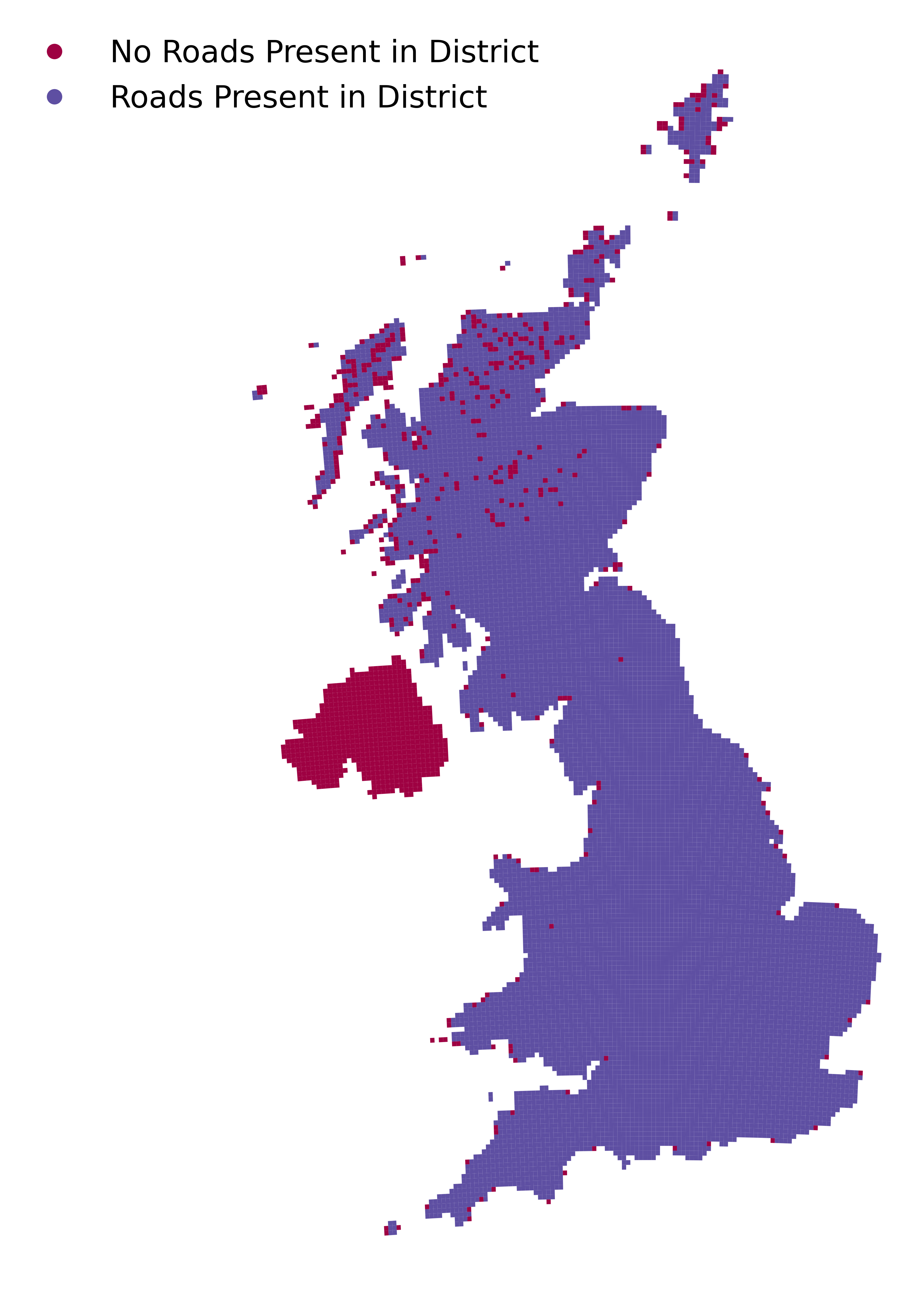}
\includegraphics[width=0.3\textwidth]{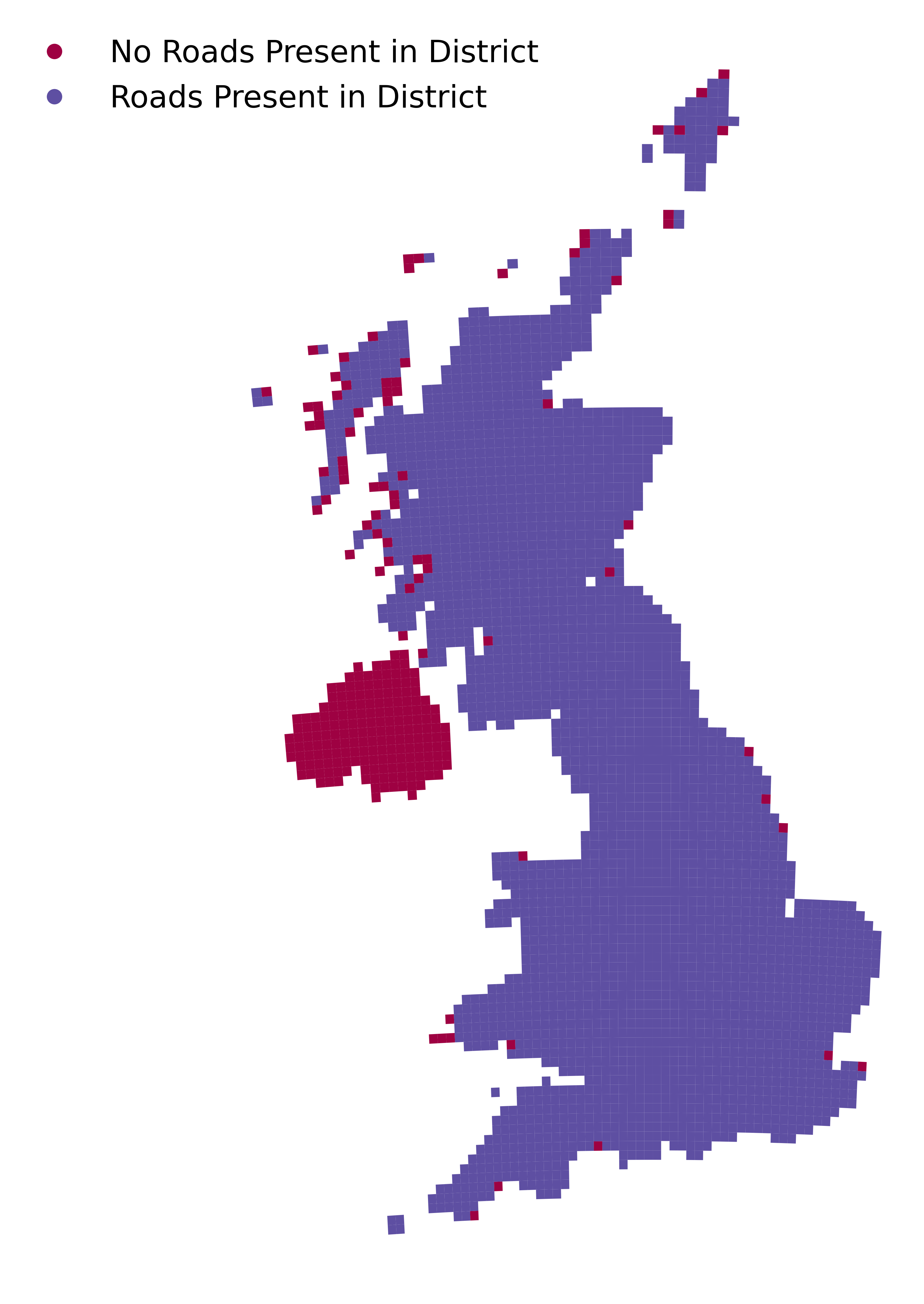}
\caption{{\bfseries Ordnance Survey grids (1km / 5km / 10km) modifiable aerial unit problem (MAUP) examples.} While the Ordnance Survey Grids gave a grid-based approach to aggregating the air pollution and road length to a high spatial resolution, it had occurrences where there were grids without any roads, shown with the crimson districts in the Figure. The grids with missing roads resulted in grids where we could not create a feature vector. Subsequently, air pollution could not be estimated, resulting in a non-complete spatial map of air pollution concentrations across the study area.}
\label{fig:MAUPExamplesOSGrid}
\end{center}
\end{figure}

\begin{figure}[ht]
\begin{center}
\includegraphics[width=0.3\textwidth]{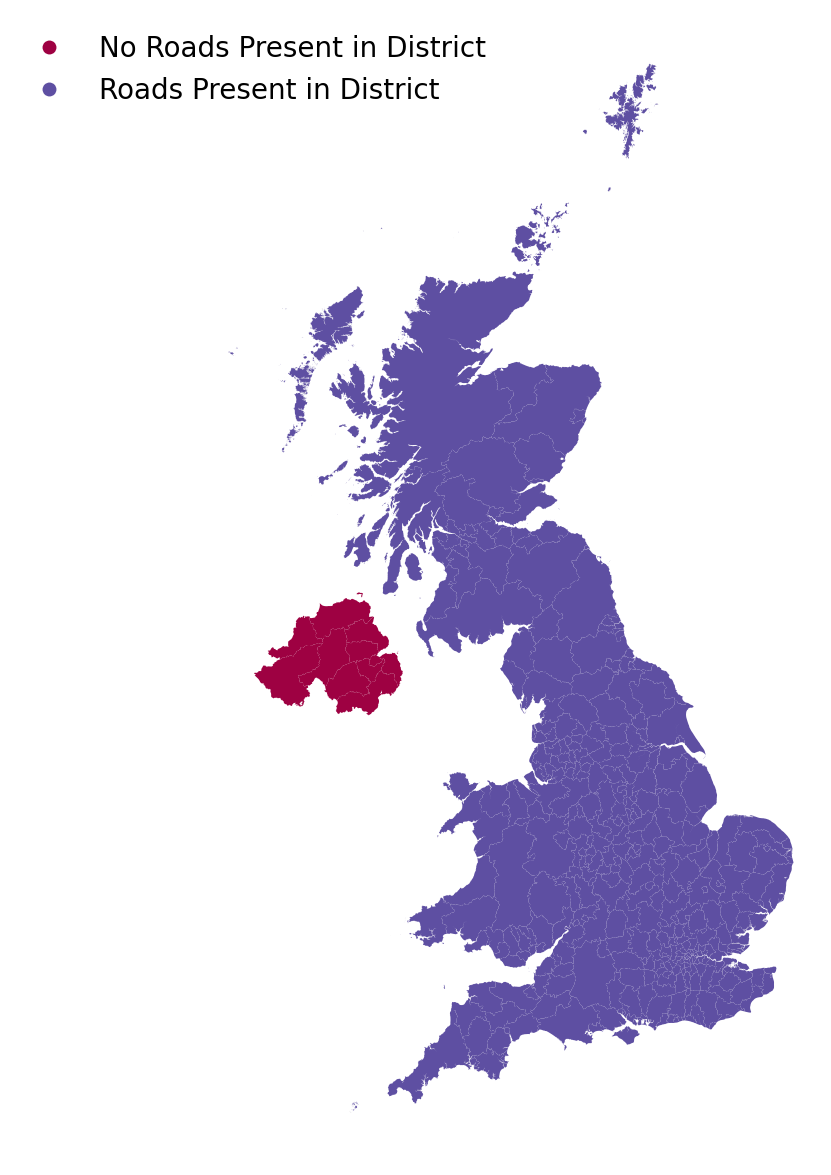}\\
\caption{ {\bfseries Local Authority District (LAD) modifiable aerial unit problem (MAUP) examples.} The LAD district-based framework for aggregating the air pollution and road length ensured that every district had a feature vector allowing for a full spatial map of air pollution to be estimated, unlike the grid-based approach detailed in \ref{fig:MAUPExamplesOSGrid}. However, it had the disadvantage of only comprising 382 (including Northern Ireland LADs), offering a coarse spatial resolution for the district-based air pollution estimations.}
\label{fig:MAUPExamplesLAD}
\end{center}
\end{figure}

\begin{figure}[ht]
\begin{center}
\includegraphics[width=0.3\textwidth]{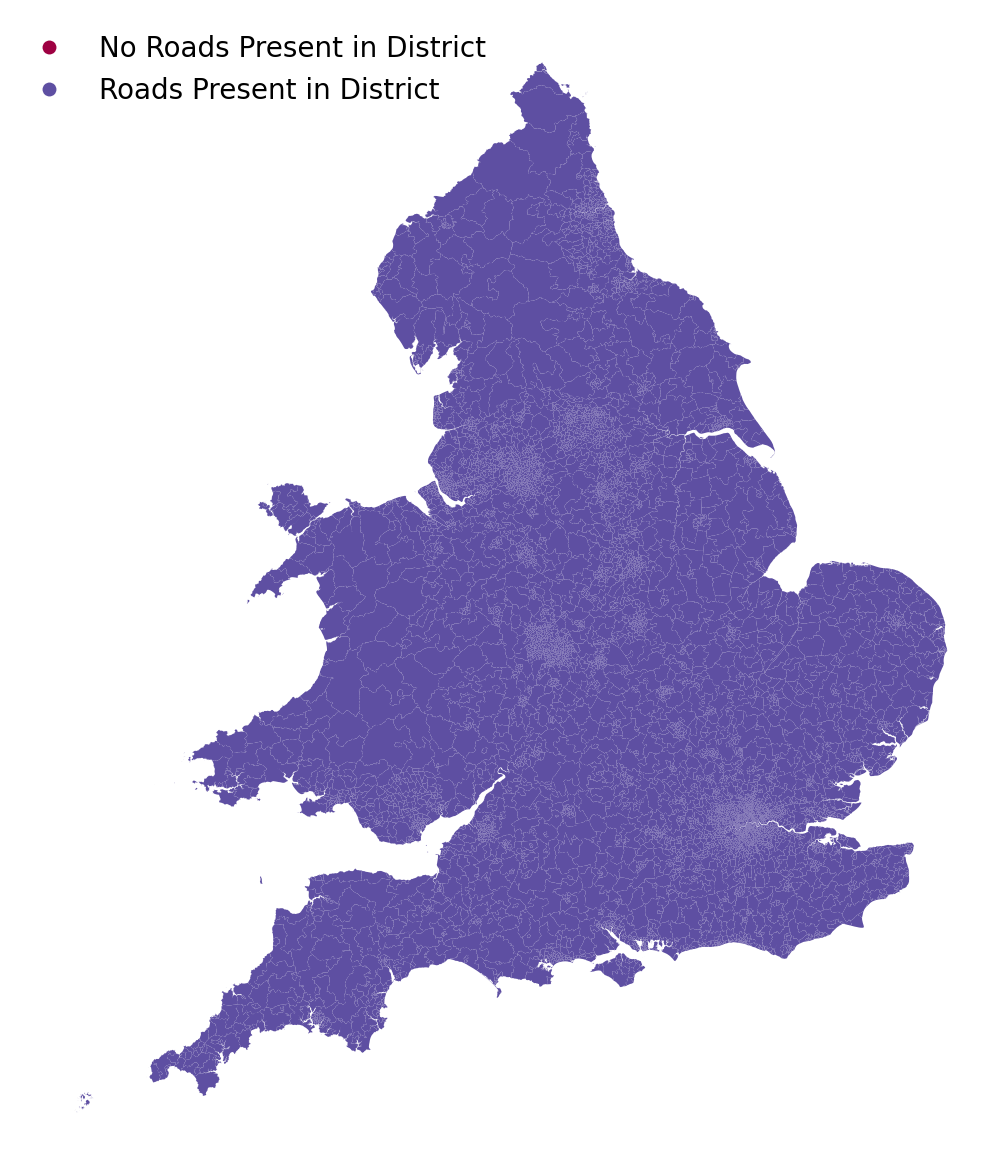}\\
\caption{{\bfseries Middle Layer Super Output Area (MSOA) modifiable aerial unit problem (MAUP) examples.} The MSOA district-based framework for aggregating the air pollution and road length has the benefit over the grid-based approach detailed in Figure \ref{fig:MAUPExamplesOSGrid} of having a feature vector for every district, allowing a complete spatial map of air pollution estimations, alongside providing a higher number of districts at 7201, a considerable spatial resolution improvement over Figure \ref{fig:MAUPExamplesLAD}. The use of MSOA also has the benefit of being used within the census reporting units and so provides easy matching to sociodemographic data such as population present within each district, leading to the decision to use the MSOA district-based framework for aggregation for the study.}
\label{fig:MAUPExamplesMSOA}
\end{center}
\end{figure}

\begin{itemize}
    \item Ordnance Survey (OS) National Grid (motivated by geographic consistency) Figure \ref{fig:MAUPExamplesOSGrid}– Considered to give a consistent size in each district with boundary areas considered including 1km/5km/10km district areas.
    \item Local Authority District (LAD) (motivated by administrative and policy capture) Figure \ref{fig:MAUPExamplesLAD}– The use of Local Authority Districts as the district boundary allowed the road network maintained by the same political entity to be aggregated. The reasoning is that local legislators decide the choice of road built within an area, and so if there were a disposition to build a particular type of road, this would be captured by the district.
    \item Middle Super Output Areas (MSOA) (motivated by population metrics) Figure \ref{fig:MAUPExamplesMSOA}– The districts within the MSOA are based upon the population in an area, with each district having a similar population with varying geographical size.
\end{itemize}

\clearpage

\section{Data Details}
\label{sec:datadetails}

\subsection{Air Pollution Ground Observation Data}
\label{sec:AirPollutionDataDetails}

\begin{figure}[ht]
\begin{center}
\includegraphics[width=0.8\textwidth]{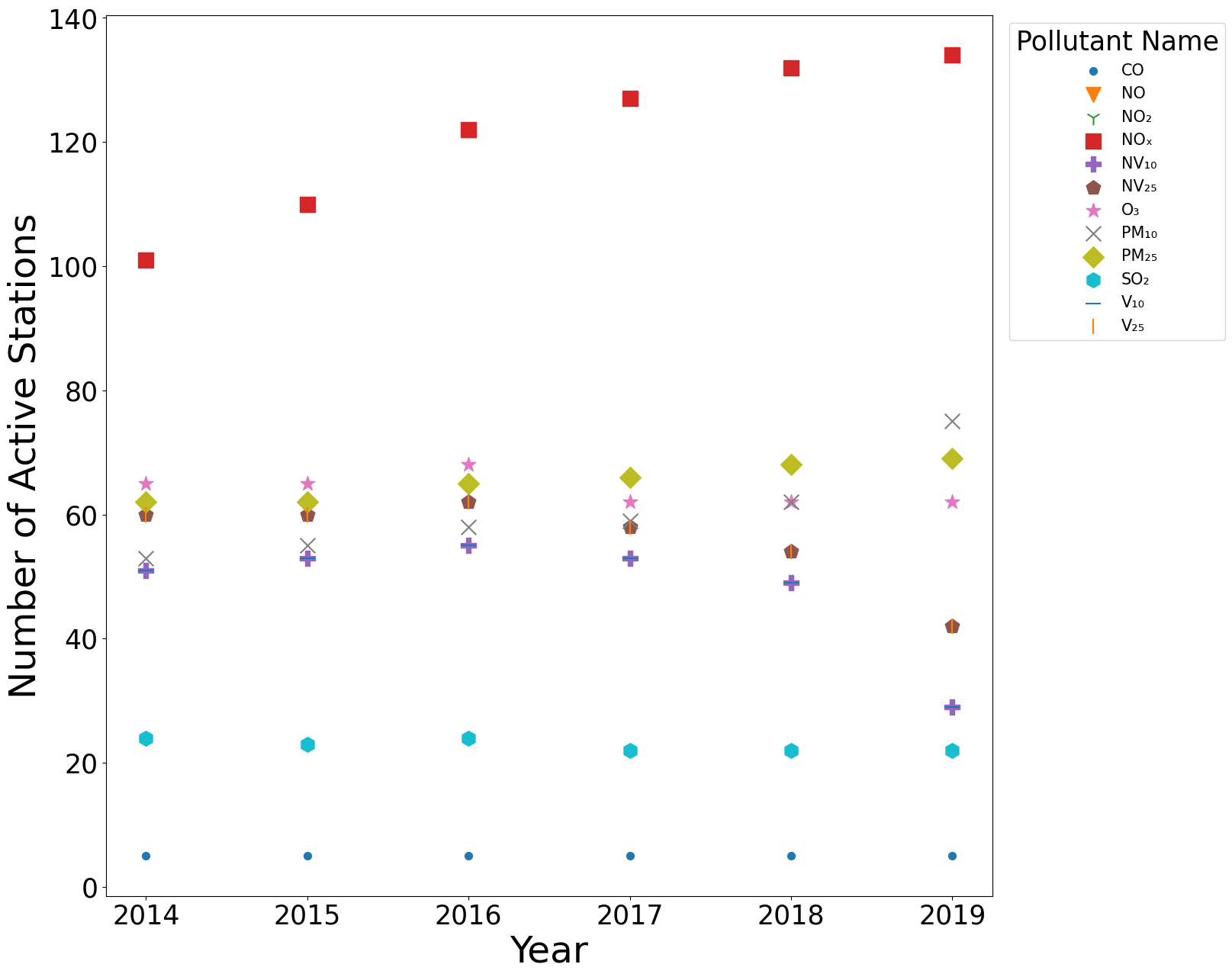}
\caption{{\bfseries Number of active AURN monitoring stations by year.} The number of AURN stations, according to UK Air (DEFRA), that are online in any given year. Some air pollution monitoring stations are decommissioned each year, and others are commissioned. Of note is that a single air pollution monitoring station can monitor multiple different air pollutants.}
\label{fig:ActiveAURNStations}
\end{center}
\end{figure}

In 2019 there were 173 functional monitoring stations across the UK in the Automatic Urban and Rural Network (AURN). Each monitoring station only measures a subset of the 12 air pollutants of interest in this study. Figure \ref{fig:ActiveAURNStations} shows the count for the number of stations online in each given year during the 2014-2019 period coloured by air pollutants. 

Another critical aspect of the AURN is the type of location in which the station is placed. There are six different station location types \footnote{Tony Clark et al. (2020). Automatic Urban and Rural Network: Site Operator’s Manual - 6.2.3 Site Classifications, Ricardo Energy and
Environment.}. The stations used in this study were also restricted to stations within the area covered by the MSOA, meaning only stations within England and Wales were included. Local Authority Environmental Health Offices are responsible for running the stations. While locations within the MSOA are closer to a monitoring station outside the MSOA boundary, it was essential to restrict the stations to the top-level political authority due to the local authority running the station. We made this choice in case of differences in standard operating procedures by health officers within different local authorities \footnote{UK AIR Information Resource (2021). Automatic Urban and Rural Network (AURN), Department for Environment, Food and Rural Affairs.
\url{https://uk-air.defra.gov.uk/networks/network-info?view=aurn} ”[Online; accessed August-2021]”}.

\clearpage 

\subsection{Air Pollution Data Interpolation}
\label{sec:AirPollutionDataInterpolation}
\begin{figure}[ht]\captionsetup[subfigure]{font=footnotesize}
\captionsetup[subfigure]{labelformat=empty}
\begin{center}
\begin{subfigure}{0.26\textwidth}
\includegraphics[width=\linewidth]{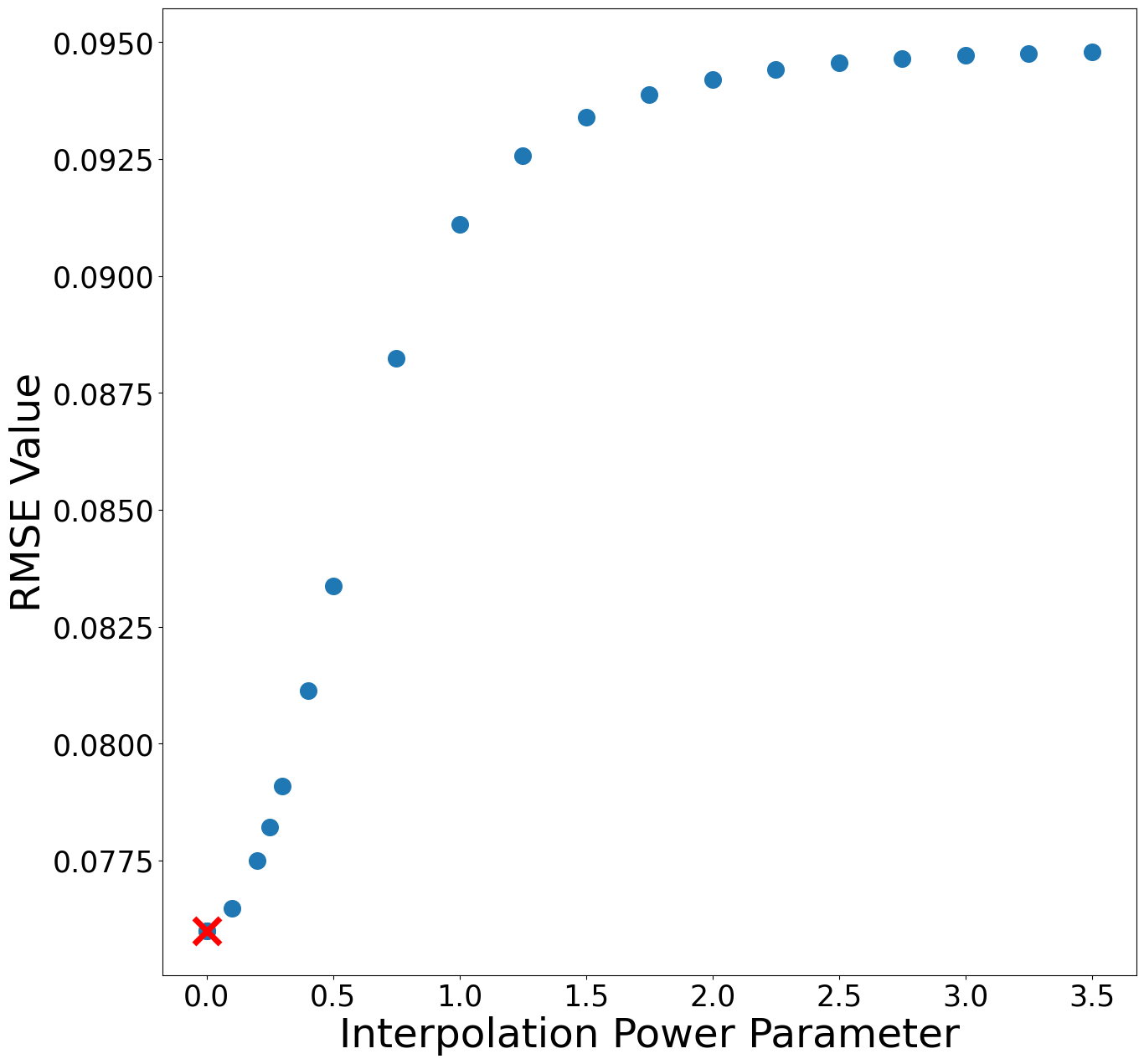}
\caption{Pollutant CO, Chosen: 0.0} \label{fig:RMSEPlotCO}
\end{subfigure}
\begin{subfigure}{0.26\textwidth}
\includegraphics[width=\linewidth]{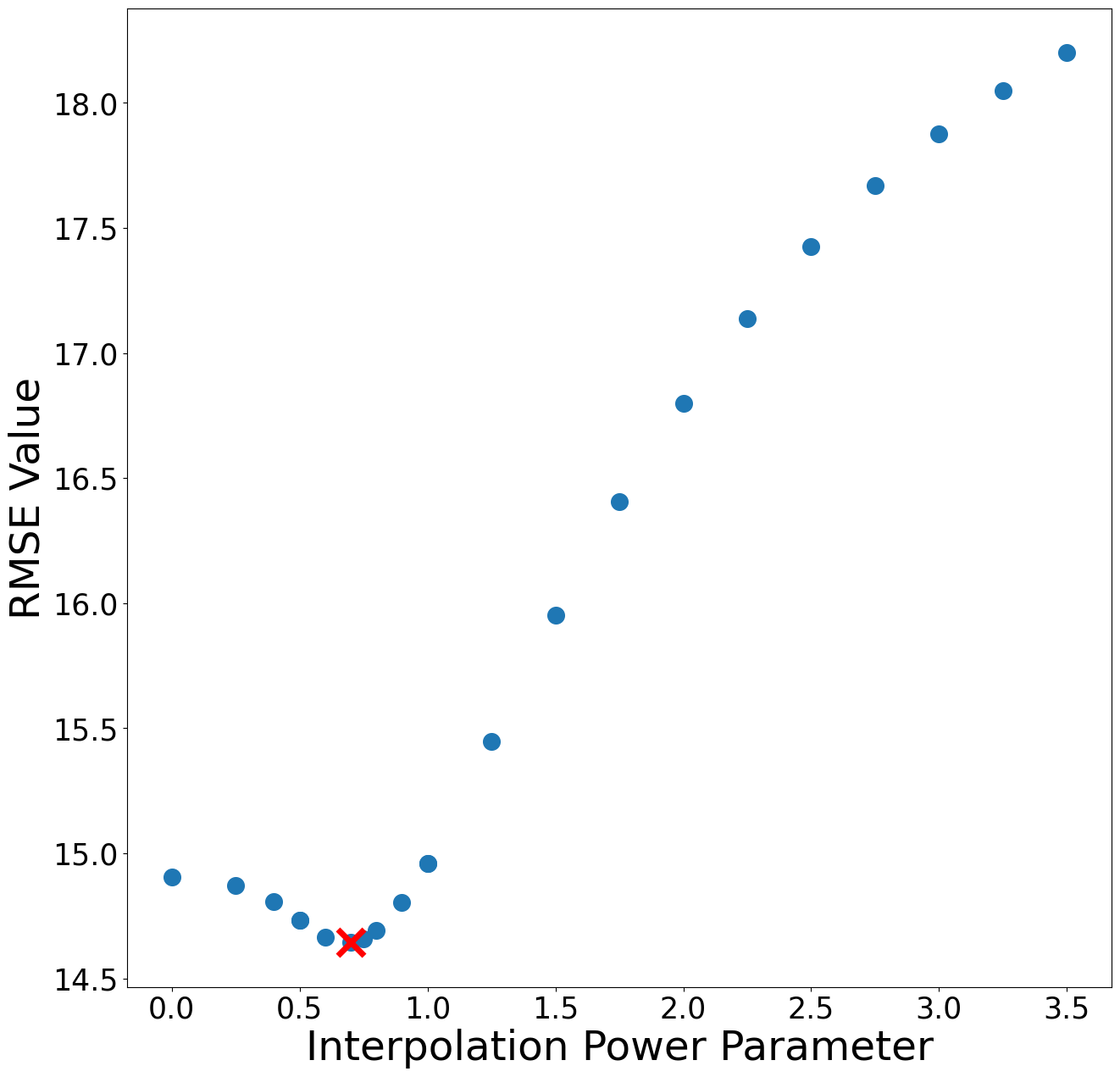}
\caption{Pollutant NO, Chosen: 0.7} \label{fig:RMSEPlotNO}
\end{subfigure}
\begin{subfigure}{0.26\textwidth}
\includegraphics[width=\linewidth]{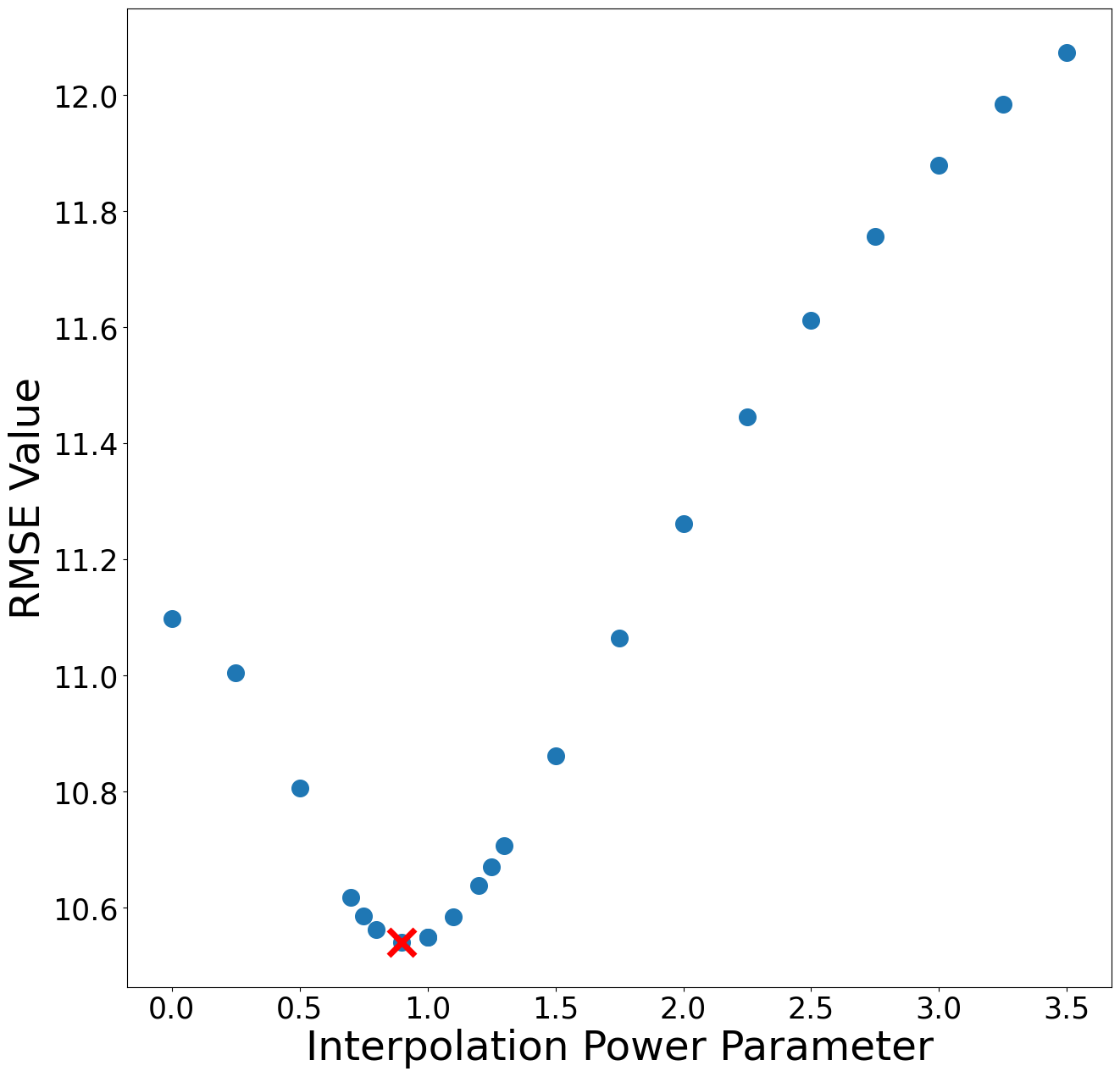}
\caption{Pollutant NO$_{2}$, Chosen: 0.9} \label{fig:RMSEPlotNO2}
\end{subfigure}\\
\begin{subfigure}{0.26\textwidth}
\includegraphics[width=\linewidth]{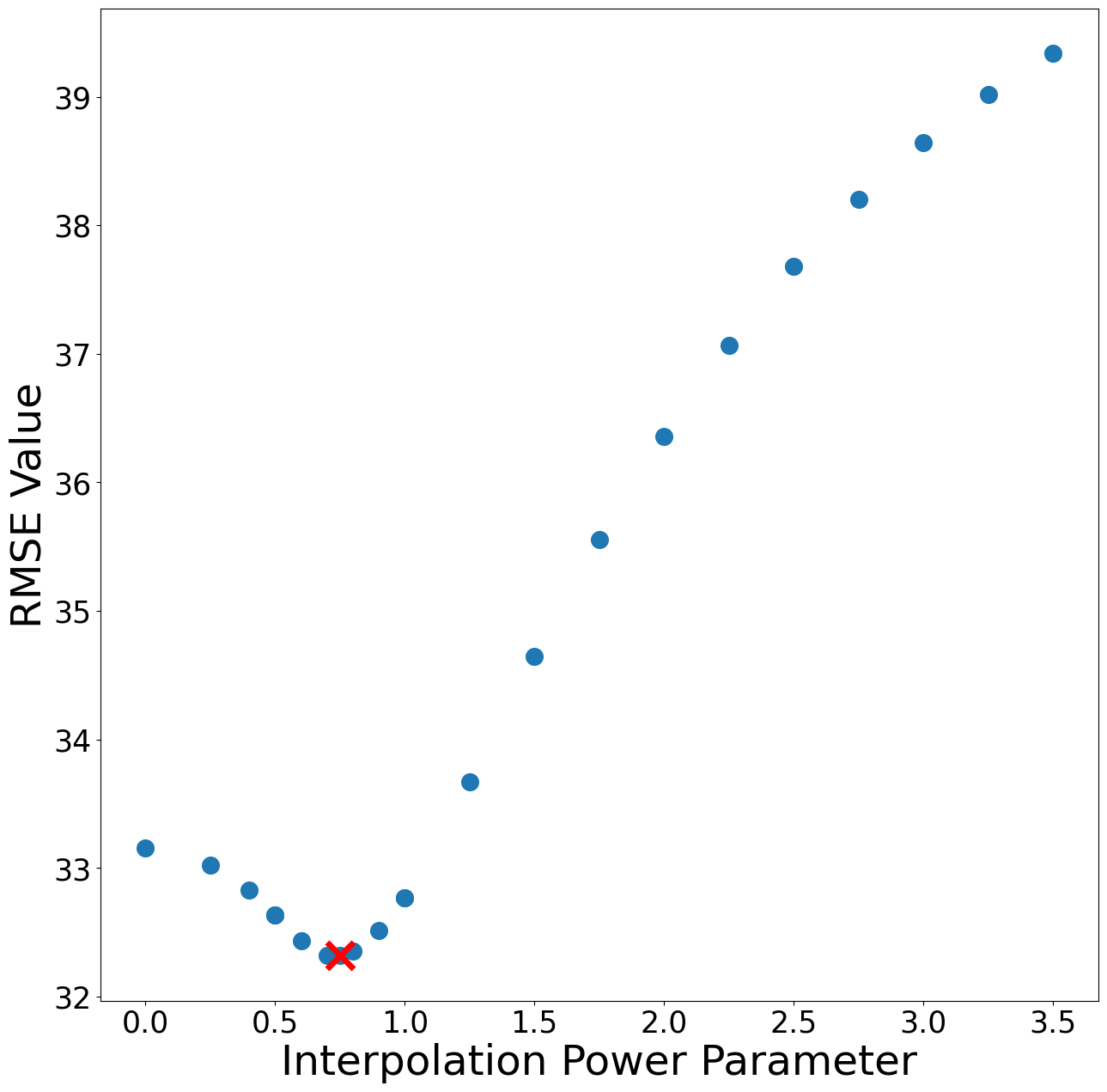}
\caption{Pollutant NO$_{x}$, Chosen: 0.7} \label{fig:RMSEPlotNOX}
\end{subfigure}
\begin{subfigure}{0.26\textwidth}
\includegraphics[width=\linewidth]{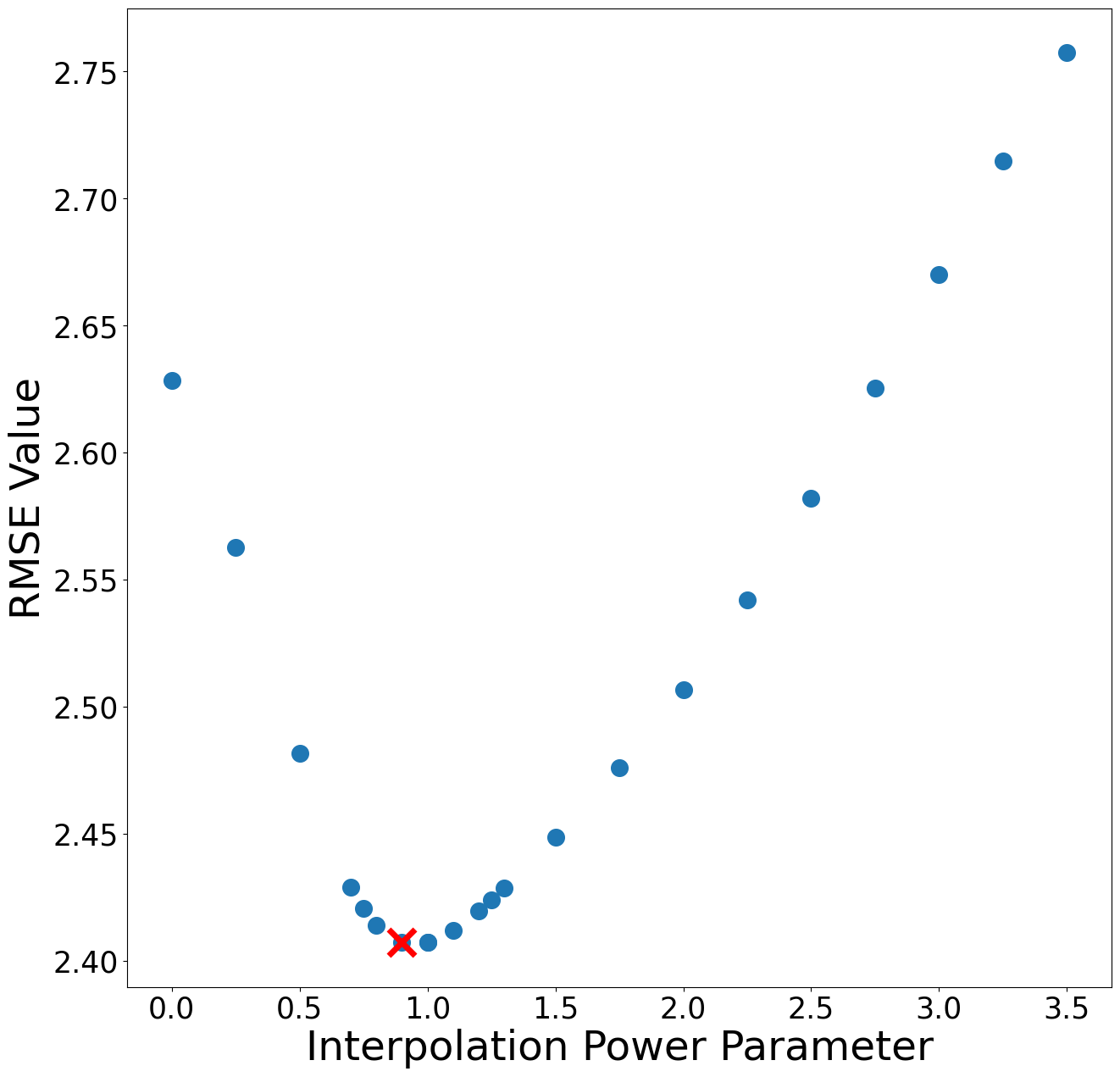}
\caption{Pollutant NV$_{10}$, Chosen: 0.9} \label{fig:RMSEPlotNV10}
\end{subfigure}
\begin{subfigure}{0.26\textwidth}
\includegraphics[width=\linewidth]{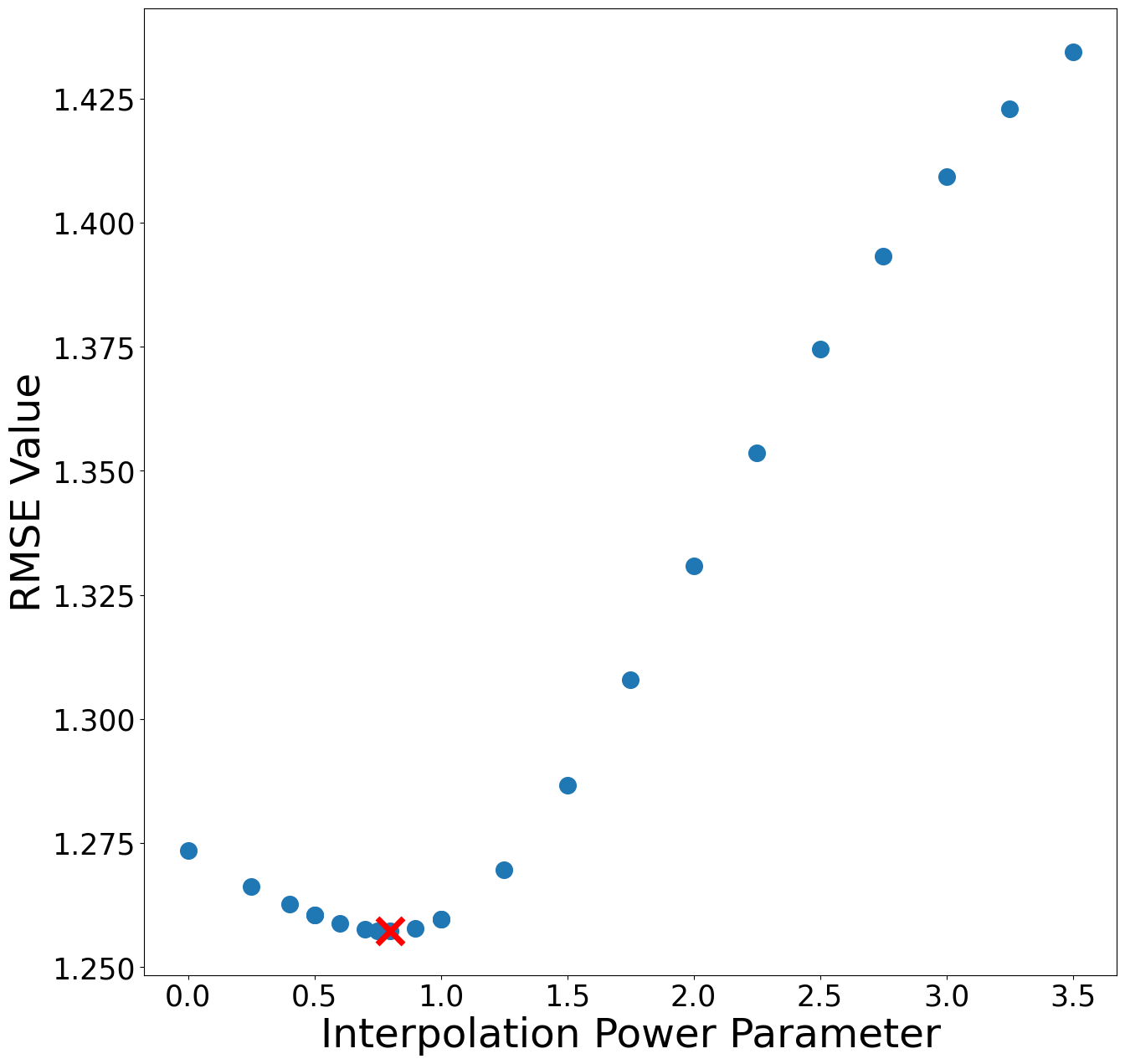}
\caption{Pollutant NV$_{25}$, Chosen: 0.8} \label{fig:RMSEPlotNV25}
\end{subfigure}\\
\begin{subfigure}{0.26\textwidth}
\includegraphics[width=\linewidth]{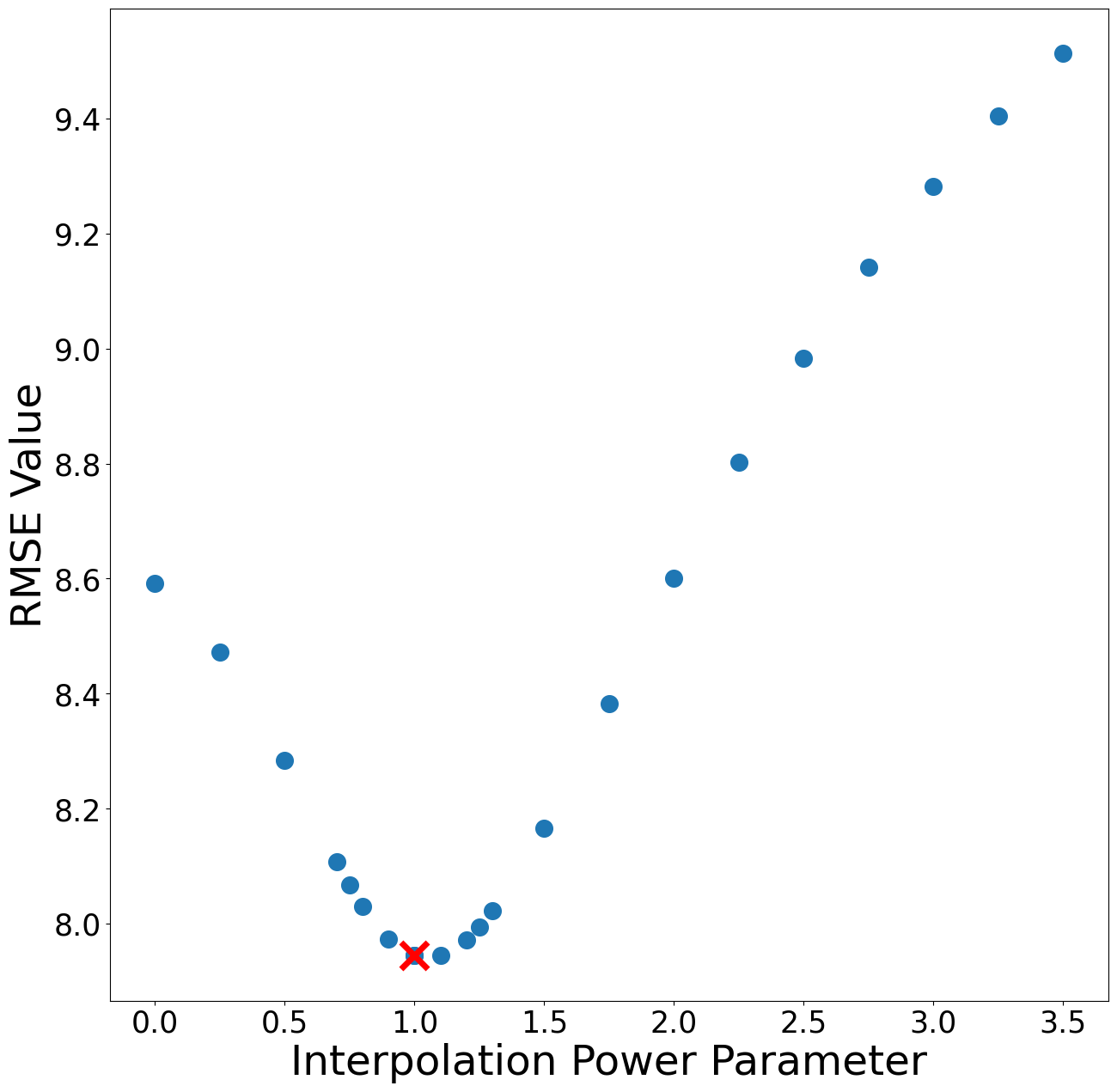}
\caption{Pollutant O$_{3}$, Chosen: 1.0} \label{fig:RMSEPlotO3}
\end{subfigure}
\begin{subfigure}{0.26\textwidth}
\includegraphics[width=\linewidth]{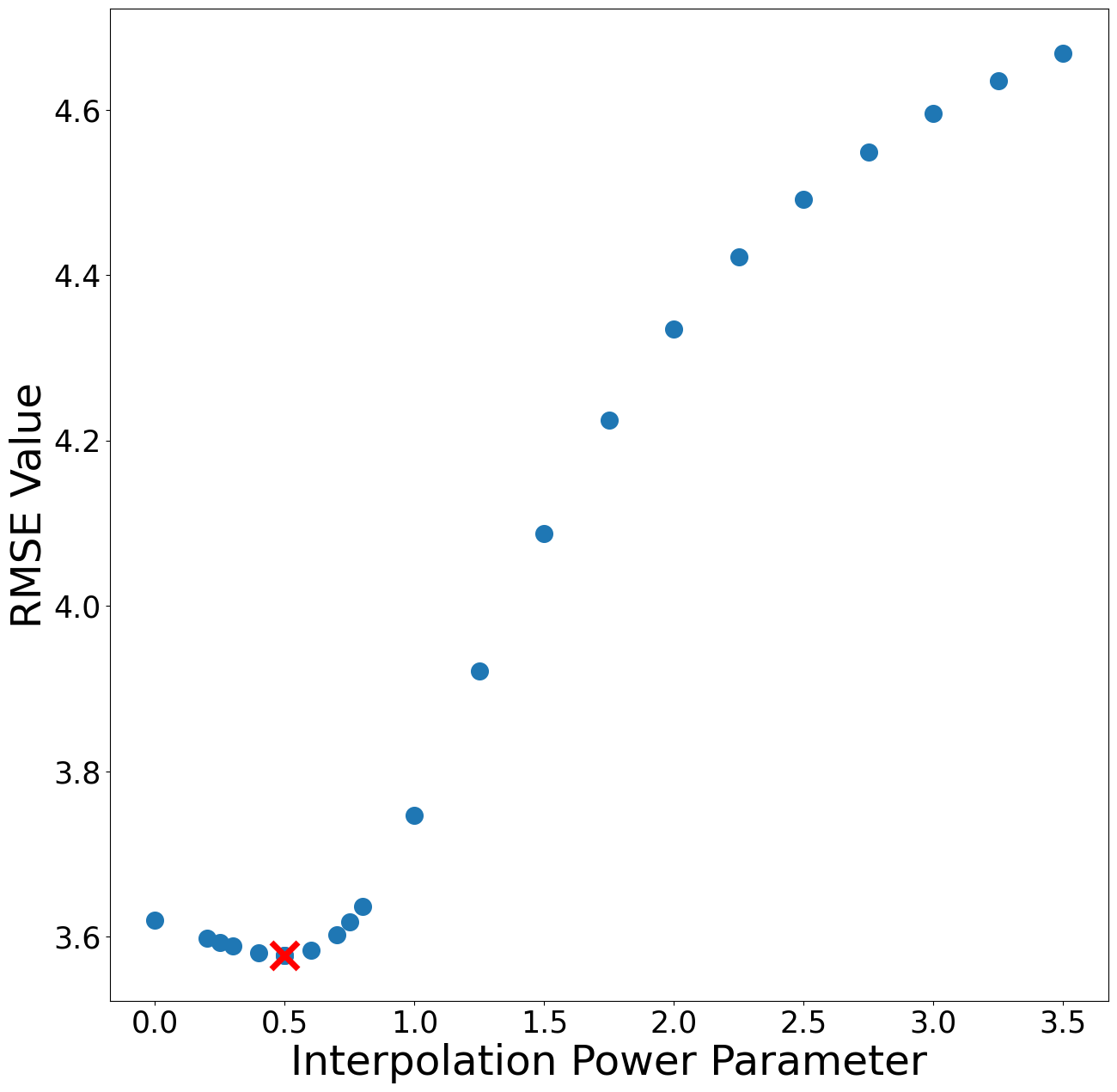}
\caption{Pollutant PM$_{10}$, Chosen: 0.5} \label{fig:RMSEPlotPM10}
\end{subfigure}
\begin{subfigure}{0.26\textwidth}
\includegraphics[width=\linewidth]{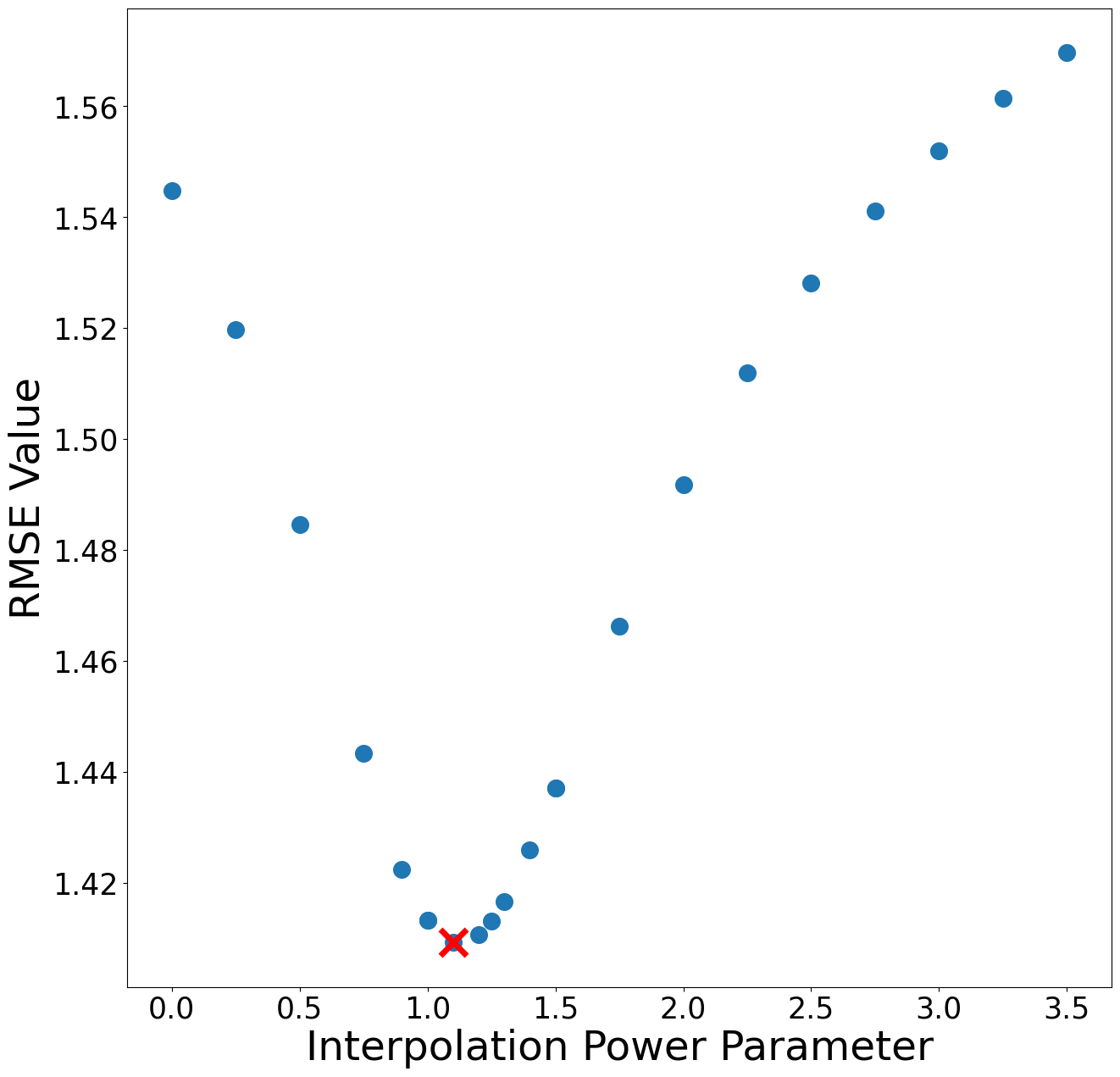}
\caption{Pollutant PM$_{25}$, Chosen: 1.1} \label{fig:RMSEPlotPM25}
\end{subfigure}\\
\begin{subfigure}{0.26\textwidth}
\includegraphics[width=\linewidth]{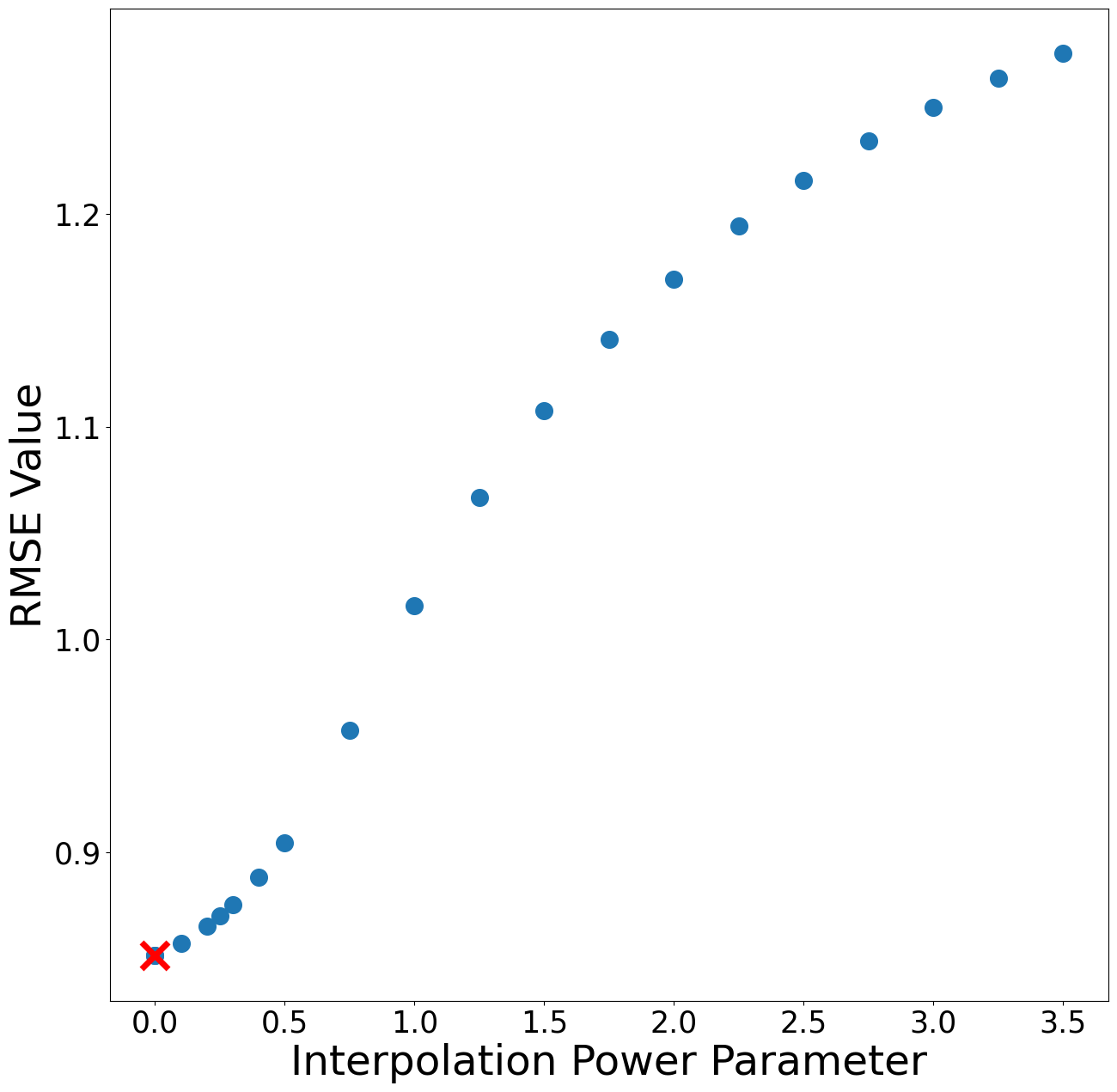}
\caption{Pollutant SO$_{2}$, Chosen: 0.0} \label{fig:RMSEPlotSO2}
\end{subfigure}
\begin{subfigure}{0.26\textwidth}
\includegraphics[width=\linewidth]{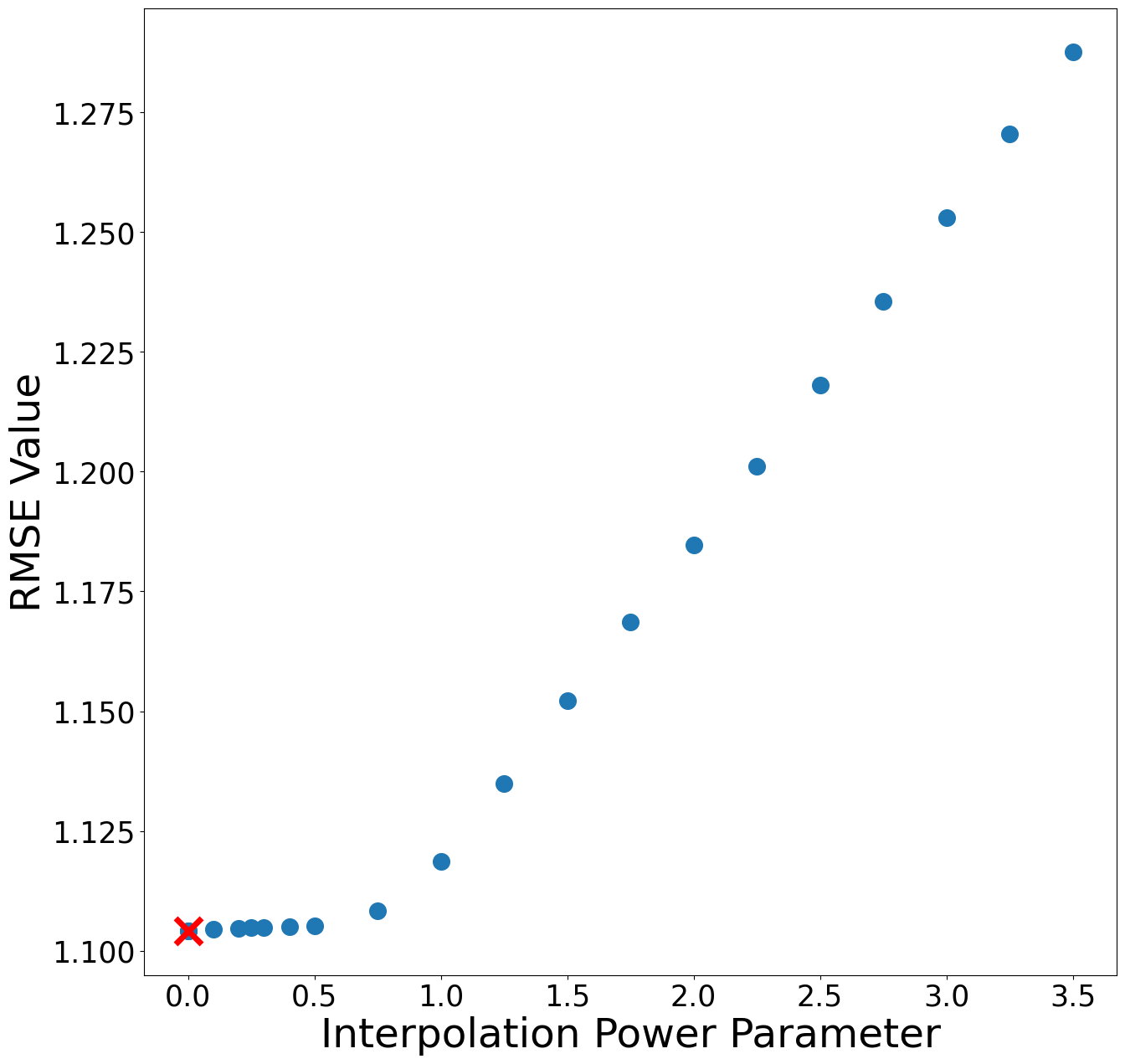}
\caption{Pollutant V$_{10}$, Chosen: 0.0} \label{fig:RMSEPlotV10}
\end{subfigure}
\begin{subfigure}{0.26\textwidth}
\includegraphics[width=\linewidth]{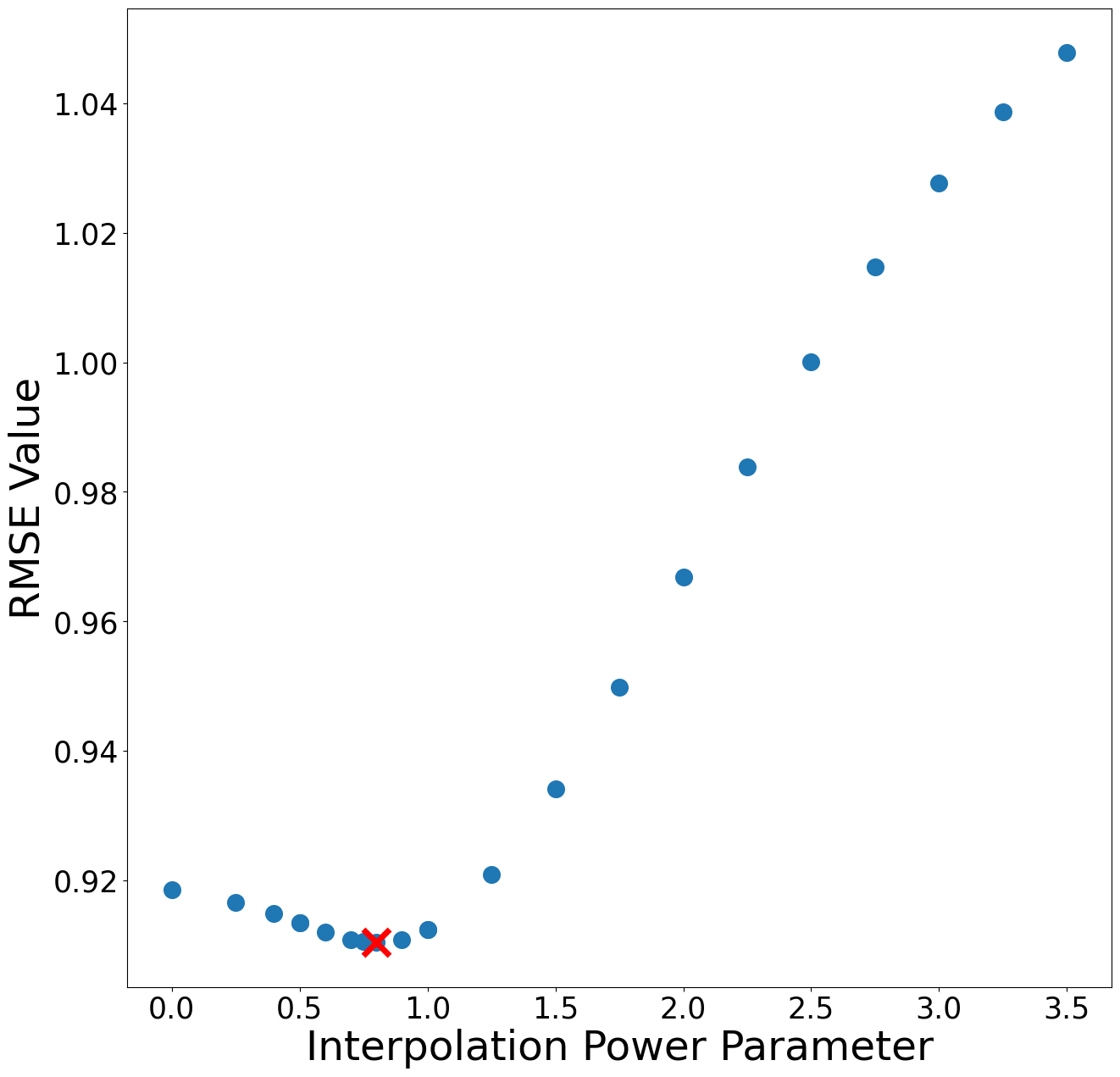}
\caption{Pollutant V$_{25}$, Chosen: 0.8} \label{fig:RMSEPlotV25}
\end{subfigure}
\caption{ {\bfseries Root mean square error (RMSE) vs interpolation power parameter during leave one out validation.} Each subfigure details the results of changing the power parameter for Inverse Distance Weighted (IDW) interpolation on the Root Mean Square Error during a leave-one-out validation process for the annual pollution level for each pollutant in the study in 2019. The overlayed red \textcolor{red}{X} denotes the power parameter chosen that minimises the RMSE. Each air pollution monitoring station was individually left out of creating an interpolated air pollution map, with the location of the monitoring station being used to sample the interpolated map and compared with the actual observation seen, with the error contributing to the overall RMSE for that power parameter value. The power parameter that minimised the error between the interpolated and actual observations across all stations was selected as the best performing and used to create the interpolated air pollution map for all years in the study for that pollutants. It can be seen that when there are a minimal number of stations, as is the case for CO (Figure \ref{fig:ActiveAURNStations}), a power parameter of 0 is chosen, indicating that estimating the average across all points performed best.}
\label{fig:RMSEInterpolationPowerScatterPlots}
\end{center}
\end{figure}

Figure \ref{fig:RMSEInterpolationPowerScatterPlots} shows the 12 different pollutants scatter plots of the values for the Root Mean Square Error (RMSE) against different values of the power parameter, used during the interpolation leave one out validations process. 

\begin{figure}[ht]\captionsetup[subfigure]{font=footnotesize}
\captionsetup[subfigure]{labelformat=empty}
\begin{center}
\begin{subfigure}{0.32\textwidth}
\includegraphics[width=\linewidth]{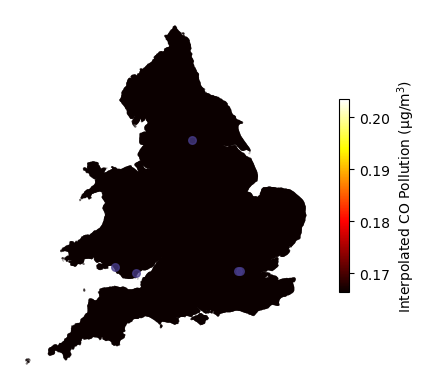}
\caption{Pollutant CO, Station Count: 5} \label{fig:InterpolationIDWSurfaceCO}
\end{subfigure}
\begin{subfigure}{0.32\textwidth}
\includegraphics[width=\linewidth]{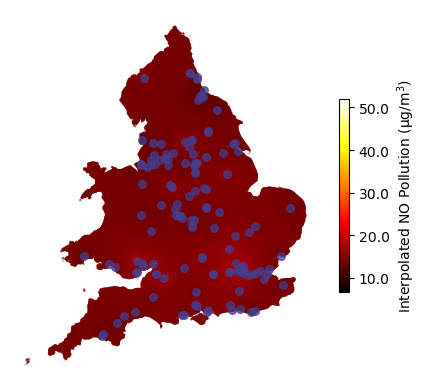}
\caption{Pollutant NO, Station Count: 132} \label{fig:InterpolationIDWSurfaceNO}
\end{subfigure}
\begin{subfigure}{0.32\textwidth}
\includegraphics[width=\linewidth]{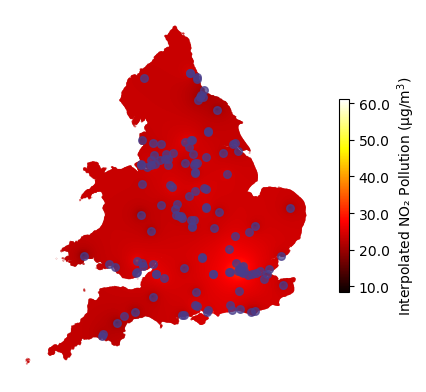}
\caption{Pollutant NO$_{2}$, Station Count: 132} \label{fig:InterpolationIDWSurfaceNO2}
\end{subfigure}\\
\begin{subfigure}{0.32\textwidth}
\includegraphics[width=\linewidth]{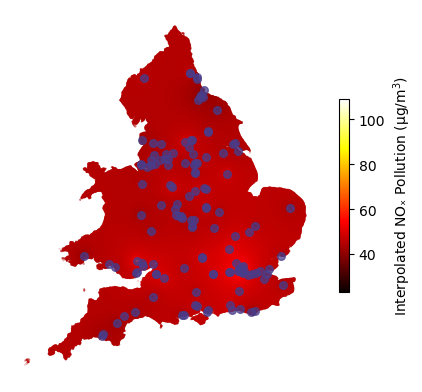}
\caption{Pollutant NO$_{X}$, Station Count: 132} \label{fig:InterpolationIDWSurfaceNOX}
\end{subfigure}
\begin{subfigure}{0.32\textwidth}
\includegraphics[width=\linewidth]{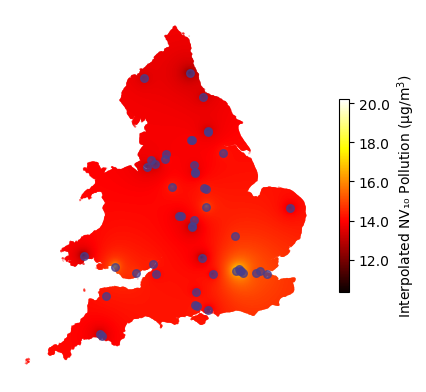}
\caption{Pollutant NV$_{10}$, Station Count: 49} \label{fig:InterpolationIDWSurfaceNV10}
\end{subfigure}
\begin{subfigure}{0.32\textwidth}
\includegraphics[width=\linewidth]{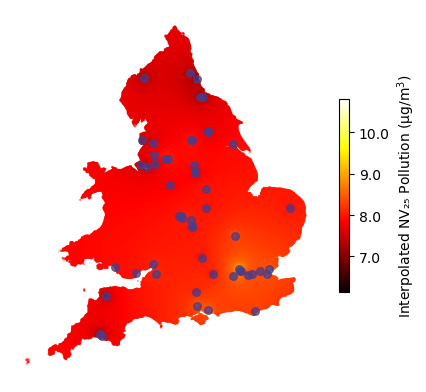}
\caption{Pollutant NV$_{25}$, Station Count: 54} \label{fig:InterpolationIDWSurfaceNV25}
\end{subfigure}\\
\begin{subfigure}{0.32\textwidth}
\includegraphics[width=\linewidth]{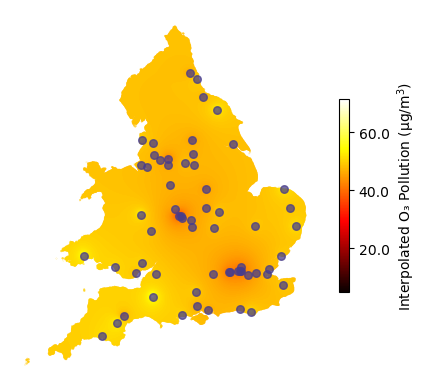}
\caption{Pollutant O$_{3}$, Station Count: 62} \label{fig:InterpolationIDWSurfaceO3}
\end{subfigure}
\begin{subfigure}{0.32\textwidth}
\includegraphics[width=\linewidth]{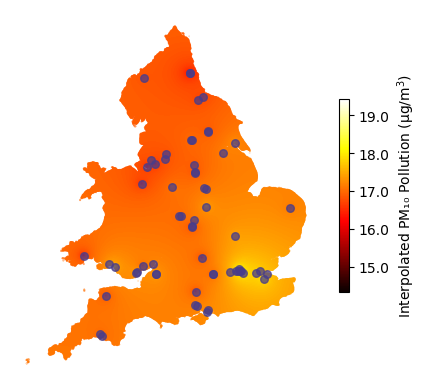}
\caption{Pollutant PM$_{10}$, Station Count: 62} \label{fig:InterpolationIDWSurfacePM10}
\end{subfigure}
\begin{subfigure}{0.32\textwidth}
\includegraphics[width=\linewidth]{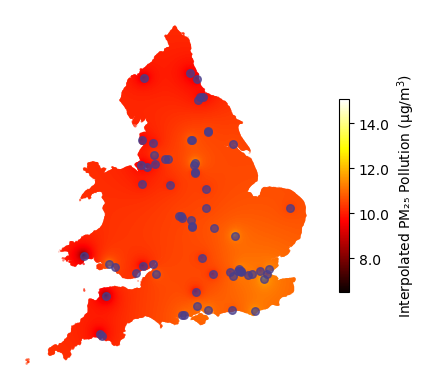}
\caption{Pollutant PM$_{25}$, Station Count: 68} \label{fig:InterpolationIDWSurfacePM25}
\end{subfigure}\\
\begin{subfigure}{0.32\textwidth}
\includegraphics[width=\linewidth]{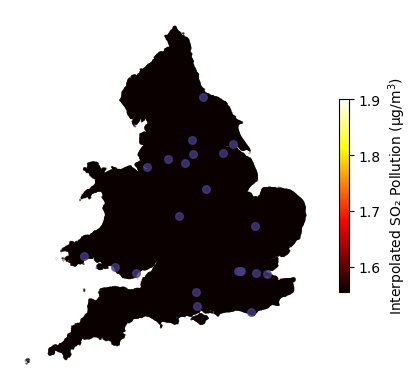}
\caption{Pollutant SO$_{2}$, Station Count: 22} \label{fig:InterpolationIDWSurfaceSO2}
\end{subfigure}
\begin{subfigure}{0.32\textwidth}
\includegraphics[width=\linewidth]{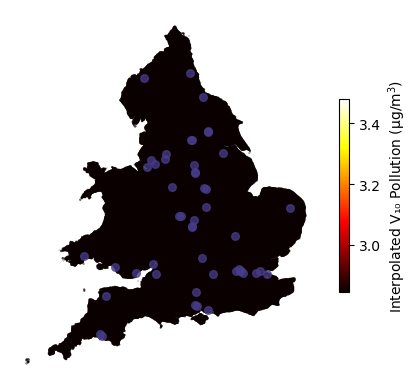}
\caption{Pollutant V$_{10}$, Station Count: 49} \label{fig:InterpolationIDWSurfaceV10}
\end{subfigure}
\begin{subfigure}{0.32\textwidth}
\includegraphics[width=\linewidth]{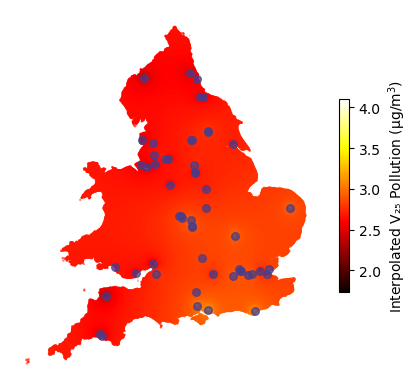}
\caption{Pollutant V$_{25}$, Station Count: 52} \label{fig:InterpolationIDWSurfaceV25}
\end{subfigure}
\caption{ { \bfseries Interpolation surface for air pollutants at the annual level in 2018. Purple points represent air pollution ground observation monitoring stations location.}  Each subfigure represents the interpolated raster that was created at the annual pollution for each air pollutant in the study during 2018 with IDW interpolation using the power parameter determined during the experiments detailed in Figure \ref{fig:RMSEInterpolationPowerScatterPlots}.}
\label{fig:InterpolationIDWSurfaces}
\end{center}
\end{figure}

Figure \ref{fig:InterpolationIDWSurfaces} shows the resulting interpolated raster from the monitoring station ground observations, with the given pollutants power parameter that minimised the RMSE. 

Of note is that some air pollutants, such as Carbon Monoxide (CO), had minimal ground observation monitoring stations. Therefore, the interpolation estimated the mean of the stations across the study area, highlighted by a power parameter of 0.

\subsection{Target Vector Data}

Table \ref{tab:Table 2 Target Vector Example} details each MSOAs aggregate pollution value calculated from the 2019 interpolated raster using the uniform point grid with a count for the number of points sampled.

\begin{table}
\resizebox{\linewidth}{!}{
\begin{filecontents*}{CSVFiles/curatedExampleTargetVector.csv}
msoa11nm,Raster Value Pollution,Count
Brighton and Hove 029,43.98337,1
Waltham Forest 013,843.5677,15
Solihull 003,898.82,20
Walsall 013,1103.7821,25
Sheffield 061,1338.5426,29
Welwyn Hatfield 009,1758.347,34
Peterborough 021,1863.5809,40
Kirklees 042,2253.5342,48
Redcar and Cleveland 021,2413.2217,60
Sutton 011,4464.2163,83
Calderdale 018,5871.66,122
Sefton 013,9449.627,214
Cheshire West and Chester 004,20918.29,474
Taunton Deane 005,46874.105,1032
Selby 002,108746.2,2385
\end{filecontents*}
\pgfplotstabletypeset[
    multicolumn names=l, 
    col sep=comma, 
    string type, 
    header = has colnames,
    columns={msoa11nm, Raster Value Pollution, Count},
    columns/msoa11nm/.style={column type=l, column name=MSOA District Name},
    columns/Raster Value Pollution/.style={column type={S[round-precision=2, , table-format=1.2e1,scientific-notation = true, table-number-alignment=center]}, column name=Aggregate Pollution Sum (µg/m$^3$)},
    columns/Count/.style={column type={S[round-precision=0, table-format=4.0,  table-number-alignment=center]}, column name=Point Sample Count},
    every head row/.style={before row=\toprule, after row=\midrule},
    every last row/.style={after row=\bottomrule}
    ]{CSVFiles/curatedExampleTargetVector.csv}}
    \smallskip
\caption{ { \bfseries Example target vector for NO$_{x}$ in 2018 at the annual temporal level.} Sample (every 500th MSOA by sample point count) aggregate pollution sum for MSOA district boundaries with associated number of point samples from the raster based on the uniform point grid shown in Figure \ref{fig:ToyExampleOfRasterSampleWithUniformPointGrid}, that the model aims to estimate.} \label{tab:Table 2 Target Vector Example}
\end{table}

\begin{figure}[ht]
\begin{center}
\includegraphics[width=0.8\textwidth]{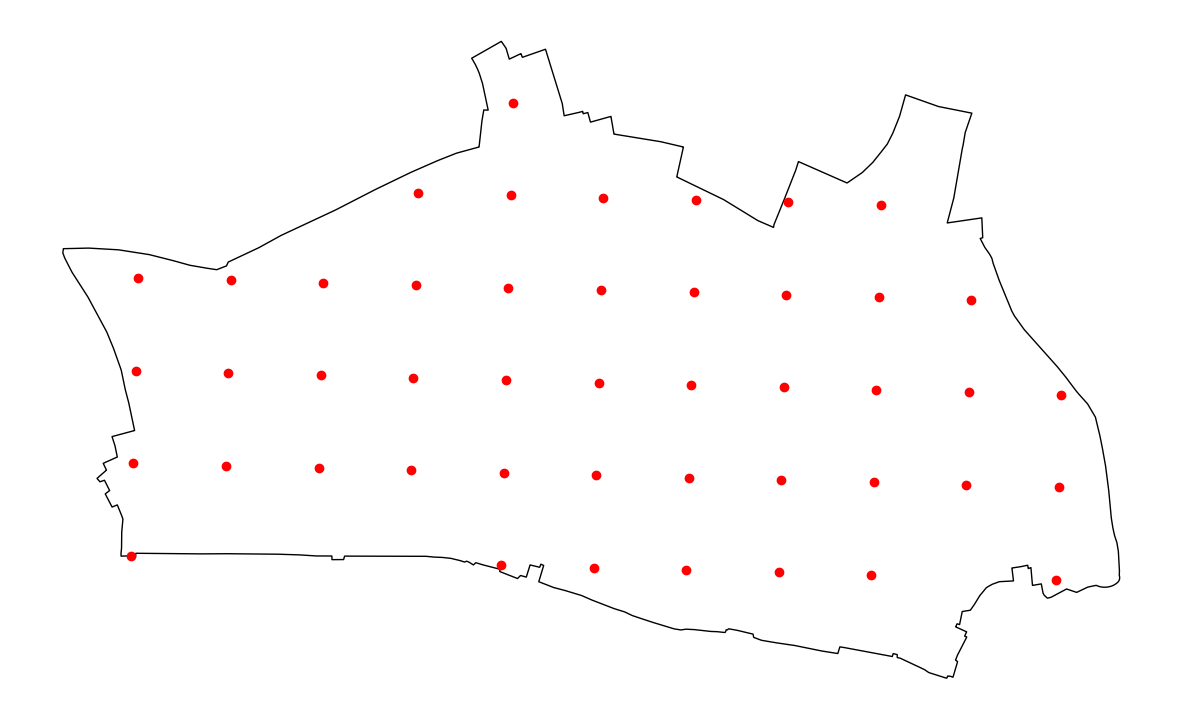}
\caption{{\bfseries Uniform point grid raster sample location for MSOA City of London 001. } The red points represent the sample locations at which measurements from the interpolated raster are taken to give a total estimate for air pollution of a given pollutant in a district. The number of sample locations for MSOA City of London 001 is 47.}
\label{fig:ToyExampleOfRasterSampleWithUniformPointGrid}
\end{center}
\end{figure}

\clearpage

\section{Model Variant Details}
\label{sec:modelDetails}

\subsection{Model Variant Feature Vector}
\label{sec:modelDetailsFeatureVector}

Table \ref{tab:Table 4 Curated Example Length Feature Vector} shows an example feature vector for the length model. Table \ref{tab:Table 7 Curated Example Highway Composition Feature Vector} shows an example feature vector for the composition model. Table \ref{tab:Table 10 Curated Example Spatial Feature Vector} shows an example feature vector for the spatial variant of the composition model. 

\begin{table}
\resizebox{\linewidth}{!}{
\begin{filecontents*}{CSVFiles/CuratedExampleLengthFeatureVector2018.csv}
msoa11nm,Length
Hammersmith and Fulham 007,10243
Islington 002,31022
Lewisham 023,39613
Hounslow 014,46625
South Gloucestershire 026,53441
Swansea 022,60536
Rhondda Cynon Taf 002,68278
Cornwall 044,77191
Bridgend 017,87979
Spelthorne 012,103251
Merthyr Tydfil 002,124519
Blaby 010,159499
Babergh 002,243968
Teignbridge 004,408278
East Devon 001,760696
\end{filecontents*}
\pgfplotstabletypeset[
    multicolumn names=l, 
    col sep=comma, 
    string type, 
    header = has colnames,
    columns={msoa11nm, Length},
    columns/msoa11nm/.style={column type=l, column name=MSOA District Name},
    columns/Length/.style={column type={S[round-precision=2, table-format=1.2e1,scientific-notation = true,  table-number-alignment=center]}, column name=Total Road Length (m)},
    every head row/.style={before row=\toprule, after row=\midrule},
    every last row/.style={after row=\bottomrule}
    ]{CSVFiles/CuratedExampleLengthFeatureVector2018.csv}}
    \smallskip
    \caption{{\bfseries Example feature vector for the length model for 2018.}  The feature vector is sorted in ascending order by total road length (m), with every 500th MSOA shown.} \label{tab:Table 4 Curated Example Length Feature Vector}
\end{table}

\begin{table}
\resizebox{\linewidth}{!}{
\begin{filecontents*}{CSVFiles/CuratedExampleCompositionFeatureVector2018.csv}
Unnamed: 0,msoa11nm,abandoned,access,bridleway,bus_guideway,bus_stop,byway,construction,corridor,crossing,cycleway,demolished,dismantled,disused,elevator,escalator,escape,footway,gallop,junction,layby,living_street,motorway,motorway_link,no,none,ohm:military:Trench,path,pedestrian,planned,platform,primary,primary_link,proposed,raceway,residential,rest_area,road,secondary,secondary_link,ser,service,services,stepping_stones,steps,tertiary,tertiary_link,track,traffic_island,traffic_signals,trail,trunk,trunk_link,unclassified,unsurfaced,virtual,yes
2677,Hammersmith and Fulham 007,0.0,0.0,0.0,0.0,0.0,0.0,0.0,0.0,0.0,540.8564960250585,0.0,0.0,0.0,0.0,0.0,0.0,241.4451543193805,0.0,0.0,0.0,0.0,0.0,0.0,0.0,0.0,0.0,45.64298683774842,0.0,0.0,0.0,1049.8715012469413,0.0,0.0,0.0,7365.932296679964,0.0,0.0,0.0,0.0,0.0,568.0635842052201,0.0,0.0,0.0,0.0,0.0,0.0,0.0,0.0,0.0,72.46803913225965,0.0,359.3094555289591,0.0,0.0,0.0
3098,Islington 002,0.0,0.0,0.0,0.0,0.0,0.0,99.08486623185226,0.0,0.0,3161.233011922556,0.0,0.0,0.0,0.0,0.0,0.0,7568.169725328244,0.0,0.0,0.0,0.0,0.0,0.0,0.0,0.0,0.0,381.0428813568794,727.5085278245436,0.0,0.0,77.38582156684178,0.0,0.0,0.0,13445.725125343091,0.0,0.0,818.5174476292066,12.148684855378274,0.0,1102.8306062818572,0.0,0.0,365.4716059664461,1454.5820427555902,0.0,0.0,0.0,0.0,0.0,1528.4448610122324,37.003006368254894,243.5829613351177,0.0,0.0,0.0
3533,Lewisham 023,0.0,0.0,0.0,0.0,0.0,0.0,0.0,0.0,0.0,324.1520152926347,0.0,0.0,0.0,0.0,0.0,0.0,5514.0866380499165,0.0,0.0,0.0,0.0,0.0,0.0,0.0,0.0,0.0,92.8515168775288,0.0,0.0,0.0,2765.3713204839246,0.0,0.0,0.0,18237.530299161524,0.0,0.0,245.8956079591211,0.0,0.0,6294.740026379831,0.0,0.0,58.39894470278828,3726.761813599632,0.0,510.23511062509886,0.0,0.0,0.0,1843.850004830102,0.0,0.0,0.0,0.0,0.0
3006,Hounslow 014,0.0,0.0,0.0,0.0,0.0,0.0,0.0,0.0,0.0,0.0,0.0,0.0,0.0,0.0,0.0,0.0,5724.133609112252,0.0,0.0,0.0,0.0,0.0,0.0,0.0,0.0,0.0,92.62764511457404,0.0,0.0,0.0,4696.545774947096,68.79165030127928,0.0,0.0,20658.302999763346,0.0,0.0,1741.733152710274,88.29185460546461,0.0,7136.669275244287,0.0,0.0,59.678543059047385,2635.4625377506586,0.0,1770.454961189777,0.0,0.0,0.0,0.0,0.0,1953.143219643481,0.0,0.0,0.0
5391,South Gloucestershire 026,0.0,0.0,0.0,0.0,0.0,0.0,0.0,0.0,0.0,1825.067288940192,0.0,0.0,0.0,0.0,0.0,0.0,4925.642482752844,0.0,0.0,0.0,0.0,0.0,0.0,0.0,0.0,0.0,89.1496375351624,239.48995969581603,0.0,0.0,4568.67443586104,118.5659081845362,0.0,0.0,27298.0157568447,0.0,0.0,225.9086971684364,0.0,0.0,5580.645733759378,0.0,0.0,37.45450314861704,4650.429767911206,0.0,2253.1430378076257,0.0,0.0,0.0,0.0,0.0,1628.8490508801283,0.0,0.0,0.0
6023,Swansea 022,0.0,0.0,316.83879721784183,0.0,0.0,0.0,0.0,0.0,0.0,63.143894576833006,0.0,0.0,0.0,0.0,0.0,0.0,5720.002820597366,0.0,0.0,0.0,0.0,0.0,0.0,0.0,0.0,0.0,12.28198525872194,209.42662154700923,0.0,0.0,2852.199465419336,0.0,0.0,0.0,30315.41723938116,0.0,0.0,0.0,0.0,0.0,13130.905301841469,0.0,0.0,290.721004483397,5233.137697673509,0.0,1104.4478676631943,0.0,0.0,0.0,0.0,0.0,1288.2313055353843,0.0,0.0,0.0
4787,Rhondda Cynon Taf 002,0.0,0.0,649.2180697116559,0.0,0.0,0.0,437.1911398631426,0.0,0.0,3990.1516051090775,0.0,0.0,0.0,0.0,0.0,0.0,1016.71614843776,0.0,0.0,0.0,0.0,0.0,0.0,0.0,0.0,0.0,196.6218371443008,360.03014294216763,0.0,0.0,6125.451541690992,0.0,0.0,0.0,26236.02527133404,0.0,0.0,11295.820425000376,0.0,0.0,4338.805678230934,0.0,0.0,52.42125170739238,0.0,0.0,0.0,0.0,0.0,0.0,1050.5661470284904,0.0,12529.258165869718,0.0,0.0,0.0
1613,Cornwall 044,0.0,0.0,0.0,0.0,0.0,0.0,0.0,0.0,0.0,3513.389499271934,0.0,0.0,0.0,0.0,0.0,0.0,7475.413413877134,0.0,0.0,0.0,0.0,0.0,0.0,0.0,0.0,0.0,1129.094540279484,1108.2264916719596,0.0,0.0,0.0,0.0,0.0,0.0,21116.50974542136,0.0,0.0,1390.1382099245625,0.0,0.0,13016.60353272635,0.0,0.0,69.04529245733804,5239.662086048289,0.0,0.0,0.0,0.0,0.0,9073.919043157865,138.85622876572504,13920.651379424397,0.0,0.0,0.0
792,Bridgend 017,0.0,0.0,0.0,0.0,0.0,0.0,0.0,0.0,0.0,3321.4437904059623,0.0,0.0,0.0,0.0,0.0,0.0,3462.872939352933,0.0,0.0,0.0,0.0,0.0,0.0,0.0,0.0,0.0,8169.126025271354,0.0,0.0,0.0,3461.186010626301,0.0,0.0,0.0,44039.18455279839,0.0,0.0,2496.252818821947,0.0,0.0,12707.249413340372,0.0,0.0,28.28084889106364,9173.583769846642,0.0,440.35594759388295,0.0,0.0,0.0,0.0,0.0,680.2059826017176,0.0,0.0,0.0
5668,Spelthorne 012,0.0,0.0,37.99837446782932,0.0,0.0,0.0,0.0,0.0,0.0,2039.403602676888,0.0,0.0,0.0,0.0,0.0,0.0,12187.874929796166,0.0,0.0,0.0,0.0,7744.926203275638,0.0,0.0,0.0,0.0,61.91579345683044,0.0,0.0,0.0,49.36085854283825,0.0,0.0,0.0,22042.74728801298,0.0,0.0,8057.441453596571,0.0,0.0,16271.681090931892,0.0,0.0,0.0,11732.548522488149,0.0,18259.09165802697,0.0,0.0,0.0,2763.2659430741824,0.0,2003.6346984562808,0.0,0.0,0.0
3819,Merthyr Tydfil 002,0.0,0.0,0.0,0.0,0.0,0.0,0.0,0.0,0.0,9063.646718995094,0.0,0.0,0.0,0.0,0.0,0.0,4362.647348096966,0.0,0.0,0.0,0.0,0.0,0.0,0.0,0.0,0.0,0.0,0.0,0.0,0.0,526.824679800677,0.0,6184.359530153647,0.0,48798.99875466455,0.0,1773.044029838005,0.0,0.0,0.0,20556.607820268207,0.0,0.0,13.377940522129046,10522.46824816736,0.0,0.0,0.0,0.0,0.0,13662.379546409858,0.0,9055.135867760317,0.0,0.0,0.0
496,Blaby 010,0.0,0.0,6574.251936586073,0.0,0.0,0.0,289.28926871423283,0.0,0.0,1680.9723148986154,0.0,0.0,0.0,0.0,0.0,0.0,55527.36538271078,0.0,0.0,0.0,0.0,7223.598771750249,0.0,0.0,0.0,0.0,1696.021314881427,0.0,0.0,0.0,2950.0549226537487,0.0,0.0,0.0,30796.4865065287,0.0,76.21460469293191,9946.186712351784,0.0,0.0,4979.390959641119,0.0,0.0,0.0,19374.16634047333,0.0,3409.847949702585,0.0,0.0,0.0,0.0,0.0,14975.916717914904,0.0,0.0,0.0
110,Babergh 002,0.0,0.0,1034.2662337592917,0.0,0.0,0.0,0.0,0.0,0.0,0.0,0.0,0.0,0.0,0.0,0.0,0.0,50535.58811280195,0.0,0.0,0.0,0.0,0.0,0.0,0.0,0.0,0.0,0.0,0.0,0.0,0.0,3950.918985744597,0.0,0.0,0.0,15056.455525989752,0.0,0.0,19096.45231055484,0.0,0.0,32029.85729577793,0.0,0.0,0.0,34352.05480933697,0.0,26081.448595953712,0.0,0.0,0.0,3642.322130746501,0.0,58188.956808543335,0.0,0.0,0.0
6128,Teignbridge 004,0.0,0.0,3127.106993794103,0.0,0.0,0.0,0.0,0.0,0.0,2469.0724768185764,0.0,0.0,0.0,0.0,0.0,0.0,38784.86165520995,0.0,0.0,0.0,410.4690009492059,0.0,0.0,0.0,0.0,0.0,19617.798625322048,304.1777953133644,0.0,0.0,1170.3604975403878,0.0,0.0,0.0,35506.403680734875,0.0,0.0,21829.7077987166,0.0,0.0,56365.25589686238,0.0,0.0,94.87516592659571,3254.88712249014,0.0,69857.5926724325,0.0,0.0,0.0,42368.994018768826,4721.462747448846,108395.02562266968,0.0,0.0,0.0
2088,East Devon 001,0.0,0.0,19171.751242758823,0.0,0.0,0.0,0.0,0.0,0.0,0.0,0.0,0.0,0.0,0.0,0.0,0.0,99313.13495011828,0.0,0.0,0.0,0.0,0.0,0.0,0.0,0.0,0.0,2825.712958763004,0.0,0.0,0.0,9280.31329936466,34.3952728047232,17231.425749766997,1899.1713452462363,12804.862809038925,0.0,0.0,4334.950484015188,125.5111910435022,0.0,49698.44526589625,0.0,0.0,76.77635154566711,29974.562426986613,0.0,66289.01929610869,0.0,0.0,0.0,21871.56122907312,1093.1816224397876,424671.5552064724,0.0,0.0,0.0
\end{filecontents*}
\pgfplotstabletypeset[
    multicolumn names=l, 
    col sep=comma, 
    string type, 
    header = has colnames, 
    columns={msoa11nm,  motorway, tertiary, trunk},
    columns/msoa11nm/.style={column type=l, column name=MSOA District Name},
    columns/motorway/.style={column type={S[round-precision=2, table-format=1.2e1,scientific-notation = true,  table-number-alignment=center]}, column name=Total Road Motorway Length (m)},
    columns/tertiary/.style={column type={S[round-precision=2, table-format=1.2e1,scientific-notation = true,  table-number-alignment=center]}, column name=Total Road Tertiary Length (m)},
    columns/trunk/.style={column type={S[round-precision=2, table-format=1.2e1,scientific-notation = true,  table-number-alignment=center]}, column name=Total Road Trunk Length (m)},
    every head row/.style={before row=\toprule, after row=\midrule},
    every last row/.style={after row=\bottomrule}
    ]{CSVFiles/CuratedExampleCompositionFeatureVector2018.csv}}
    \smallskip
    \caption{{\bfseries Example feature vector for the composition model for 2018.}  The feature vector is sorted in ascending order by total road length (m), with a selection of different road types for every 500th MSOA.} \label{tab:Table 7 Curated Example Highway Composition Feature Vector}
\end{table}

\begin{table}
\resizebox{\linewidth}{!}{
\begin{filecontents*}{CSVFiles/CuratedExampleSpatialFeatureVector2018.csv}
Unnamed: 0,msoa11nm,abandoned,access,bridleway,bus_guideway,bus_stop,byway,construction,corridor,crossing,cycleway,demolished,dismantled,disused,elevator,escalator,escape,footway,gallop,junction,layby,living_street,motorway,motorway_link,no,none,ohm:military:Trench,path,pedestrian,planned,platform,primary,primary_link,proposed,raceway,residential,rest_area,road,secondary,secondary_link,ser,service,services,stepping_stones,steps,tertiary,tertiary_link,track,traffic_island,traffic_signals,trail,trunk,trunk_link,unclassified,unsurfaced,virtual,yes,N_abandoned,N_access,N_bridleway,N_bus_guideway,N_bus_stop,N_byway,N_construction,N_corridor,N_crossing,N_cycleway,N_demolished,N_dismantled,N_disused,N_elevator,N_escalator,N_escape,N_footway,N_gallop,N_junction,N_layby,N_living_street,N_motorway,N_motorway_link,N_no,N_none,N_ohm:military:Trench,N_path,N_pedestrian,N_planned,N_platform,N_primary,N_primary_link,N_proposed,N_raceway,N_residential,N_rest_area,N_road,N_secondary,N_secondary_link,N_ser,N_service,N_services,N_stepping_stones,N_steps,N_tertiary,N_tertiary_link,N_track,N_traffic_island,N_traffic_signals,N_trail,N_trunk,N_trunk_link,N_unclassified,N_unsurfaced,N_virtual,N_yes
2677,Hammersmith and Fulham 007,0.0,0.0,0.0,0.0,0.0,0.0,0.0,0.0,0.0,540.8564960250585,0.0,0.0,0.0,0.0,0.0,0.0,241.4451543193805,0.0,0.0,0.0,0.0,0.0,0.0,0.0,0.0,0.0,45.64298683774842,0.0,0.0,0.0,1049.8715012469413,0.0,0.0,0.0,7365.932296679964,0.0,0.0,0.0,0.0,0.0,568.0635842052201,0.0,0.0,0.0,0.0,0.0,0.0,0.0,0.0,0.0,72.46803913225965,0.0,359.3094555289591,0.0,0.0,0.0,0.0,0.0,0.0,0.0,0.0,0.0,0.0,0.0,0.0,1481.570983258418,0.0,0.0,0.0,0.0,0.0,0.0,12297.451124091973,0.0,0.0,0.0,0.0,0.0,0.0,0.0,0.0,0.0,1293.7055245271192,862.8905702336003,0.0,0.0,4248.373526510904,7.8565374207180385,0.0,0.0,56362.99601056763,0.0,0.0,2271.85779621992,0.0,0.0,4668.632673678359,0.0,0.0,132.47380709999874,1151.2070287243614,0.0,546.3824967386576,0.0,0.0,0.0,3596.716800067221,64.38257393561693,1086.7013404950876,0.0,0.0,0.0
3098,Islington 002,0.0,0.0,0.0,0.0,0.0,0.0,99.08486623185226,0.0,0.0,3161.233011922556,0.0,0.0,0.0,0.0,0.0,0.0,7568.169725328244,0.0,0.0,0.0,0.0,0.0,0.0,0.0,0.0,0.0,381.0428813568794,727.5085278245436,0.0,0.0,77.38582156684178,0.0,0.0,0.0,13445.725125343091,0.0,0.0,818.5174476292066,12.148684855378274,0.0,1102.8306062818572,0.0,0.0,365.4716059664461,1454.5820427555902,0.0,0.0,0.0,0.0,0.0,1528.4448610122324,37.003006368254894,243.5829613351177,0.0,0.0,0.0,0.0,0.0,0.0,0.0,0.0,0.0,0.0,0.0,0.0,1547.2741538711848,0.0,0.0,0.0,0.0,0.0,0.0,9381.340937363662,0.0,0.0,0.0,0.0,0.0,0.0,0.0,0.0,0.0,9.004958736845143,380.6367543630324,0.0,0.0,2707.697045080158,0.0,0.0,0.0,21635.190947653035,0.0,0.0,1066.230092662085,204.62055813103825,0.0,7068.940197985252,0.0,0.0,151.29447152957366,4570.786075787704,1.1947447056790366,45.74106765392913,0.0,0.0,0.0,1979.4640321831696,0.0,1545.829862795228,0.0,0.0,0.0
3533,Lewisham 023,0.0,0.0,0.0,0.0,0.0,0.0,0.0,0.0,0.0,324.1520152926347,0.0,0.0,0.0,0.0,0.0,0.0,5514.0866380499165,0.0,0.0,0.0,0.0,0.0,0.0,0.0,0.0,0.0,92.8515168775288,0.0,0.0,0.0,2765.3713204839246,0.0,0.0,0.0,18237.530299161524,0.0,0.0,245.8956079591211,0.0,0.0,6294.740026379831,0.0,0.0,58.39894470278828,3726.761813599632,0.0,510.23511062509886,0.0,0.0,0.0,1843.850004830102,0.0,0.0,0.0,0.0,0.0,0.0,0.0,0.0,0.0,0.0,0.0,0.0,0.0,0.0,310.78293155703545,0.0,0.0,0.0,0.0,0.0,0.0,9343.072856922368,0.0,0.0,0.0,0.0,0.0,0.0,0.0,0.0,0.0,60.77458463166695,0.0,0.0,0.0,842.5707177631035,0.0,0.0,0.0,47816.74528020234,0.0,0.0,1010.689581224257,0.0,0.0,15693.09301747892,0.0,0.0,40.0828752233078,9605.66957901429,0.0,1883.6512692575268,0.0,0.0,0.0,112.2597234887442,0.0,542.1580947505473,0.0,0.0,0.0
3006,Hounslow 014,0.0,0.0,0.0,0.0,0.0,0.0,0.0,0.0,0.0,0.0,0.0,0.0,0.0,0.0,0.0,0.0,5724.133609112252,0.0,0.0,0.0,0.0,0.0,0.0,0.0,0.0,0.0,92.62764511457404,0.0,0.0,0.0,4696.545774947096,68.79165030127928,0.0,0.0,20658.302999763346,0.0,0.0,1741.733152710274,88.29185460546461,0.0,7136.669275244287,0.0,0.0,59.678543059047385,2635.4625377506586,0.0,1770.454961189777,0.0,0.0,0.0,0.0,0.0,1953.143219643481,0.0,0.0,0.0,0.0,0.0,0.0,0.0,0.0,0.0,0.0,0.0,0.0,931.466785902226,0.0,0.0,0.0,0.0,0.0,0.0,13897.862346961518,0.0,0.0,0.0,0.0,0.0,0.0,0.0,0.0,0.0,0.0,103.71343684664087,0.0,0.0,4746.4668225766745,0.0,0.0,0.0,51294.70051332466,0.0,0.0,1182.5272059567346,0.0,0.0,15166.50433379455,0.0,0.0,30.290455521311667,11360.500658291436,0.0,2139.859581319188,0.0,0.0,0.0,0.0,0.0,1843.0005389593289,0.0,0.0,0.0
5391,South Gloucestershire 026,0.0,0.0,0.0,0.0,0.0,0.0,0.0,0.0,0.0,1825.067288940192,0.0,0.0,0.0,0.0,0.0,0.0,4925.642482752844,0.0,0.0,0.0,0.0,0.0,0.0,0.0,0.0,0.0,89.1496375351624,239.48995969581603,0.0,0.0,4568.67443586104,118.5659081845362,0.0,0.0,27298.0157568447,0.0,0.0,225.9086971684364,0.0,0.0,5580.645733759378,0.0,0.0,37.45450314861704,4650.429767911206,0.0,2253.1430378076257,0.0,0.0,0.0,0.0,0.0,1628.8490508801283,0.0,0.0,0.0,0.0,0.0,0.0,0.0,0.0,0.0,0.0,0.0,0.0,1811.6143424032816,0.0,0.0,0.0,0.0,0.0,0.0,21003.042689005924,0.0,0.0,0.0,0.0,0.0,0.0,0.0,0.0,0.0,0.0,0.6035719349316306,0.0,0.0,3462.1890147216727,0.0,0.0,0.0,56093.95720654423,0.0,0.0,2511.30157655262,0.0,0.0,12776.09928313336,0.0,0.0,313.47139227102184,3777.128923653967,0.0,85.57323904449653,0.0,0.0,0.0,1362.7612157017486,0.0,763.1133466189815,0.0,0.0,0.0
6023,Swansea 022,0.0,0.0,316.83879721784183,0.0,0.0,0.0,0.0,0.0,0.0,63.143894576833006,0.0,0.0,0.0,0.0,0.0,0.0,5720.002820597366,0.0,0.0,0.0,0.0,0.0,0.0,0.0,0.0,0.0,12.28198525872194,209.42662154700923,0.0,0.0,2852.199465419336,0.0,0.0,0.0,30315.41723938116,0.0,0.0,0.0,0.0,0.0,13130.905301841469,0.0,0.0,290.721004483397,5233.137697673509,0.0,1104.4478676631943,0.0,0.0,0.0,0.0,0.0,1288.2313055353843,0.0,0.0,0.0,0.0,0.0,6625.06421292031,0.0,0.0,0.0,0.0,0.0,0.0,11368.38698305902,0.0,0.0,0.0,0.0,0.0,0.0,38365.50425148836,0.0,0.0,0.0,0.0,0.0,0.0,0.0,0.0,0.0,4222.574756165115,30.429909039811083,0.0,0.0,15806.398967483456,246.0892193715626,0.0,0.0,74784.58512147388,0.0,93.47863272033658,4110.405446780564,0.0,0.0,38145.10115022088,0.0,0.0,791.3955771505948,11990.48312543874,0.0,7419.371215473147,0.0,0.0,0.0,0.0,0.0,7633.515547772167,0.0,0.0,0.0
4787,Rhondda Cynon Taf 002,0.0,0.0,649.2180697116559,0.0,0.0,0.0,437.1911398631426,0.0,0.0,3990.1516051090775,0.0,0.0,0.0,0.0,0.0,0.0,1016.71614843776,0.0,0.0,0.0,0.0,0.0,0.0,0.0,0.0,0.0,196.6218371443008,360.03014294216763,0.0,0.0,6125.451541690992,0.0,0.0,0.0,26236.02527133404,0.0,0.0,11295.820425000376,0.0,0.0,4338.805678230934,0.0,0.0,52.42125170739238,0.0,0.0,0.0,0.0,0.0,0.0,1050.5661470284904,0.0,12529.258165869718,0.0,0.0,0.0,0.0,0.0,195.2481250520248,0.0,0.0,0.0,2036.8605103327995,0.0,0.0,15775.77494445389,0.0,0.0,0.0,0.0,0.0,0.0,8409.91268196761,0.0,0.0,0.0,0.0,0.0,0.0,0.0,0.0,0.0,4126.982608564758,305.51477952297154,0.0,0.0,21908.04149455496,215.97051991412584,6344.5301288911205,0.0,120408.62236199554,0.0,1592.2779656199982,1703.5524917480325,0.0,0.0,22844.193541684424,0.0,0.0,473.9551310049673,2859.384072965413,0.0,9461.338026642758,0.0,0.0,0.0,14424.976739990108,122.53761790713924,32042.793078162813,0.0,0.0,0.0
1613,Cornwall 044,0.0,0.0,0.0,0.0,0.0,0.0,0.0,0.0,0.0,3513.389499271934,0.0,0.0,0.0,0.0,0.0,0.0,7475.413413877134,0.0,0.0,0.0,0.0,0.0,0.0,0.0,0.0,0.0,1129.094540279484,1108.2264916719596,0.0,0.0,0.0,0.0,0.0,0.0,21116.50974542136,0.0,0.0,1390.1382099245625,0.0,0.0,13016.60353272635,0.0,0.0,69.04529245733804,5239.662086048289,0.0,0.0,0.0,0.0,0.0,9073.919043157865,138.85622876572504,13920.651379424397,0.0,0.0,0.0,0.0,0.0,44283.76128004717,0.0,0.0,0.0,0.0,0.0,0.0,1427.7724088441212,0.0,0.0,0.0,0.0,0.0,0.0,25921.41598324071,0.0,0.0,0.0,0.0,0.0,0.0,0.0,0.0,0.0,47242.323254585135,0.0,0.0,0.0,188.2118396957631,9.584989365039014,0.0,0.0,23649.318200171536,0.0,0.0,14739.776155438974,39.21629754359755,0.0,74429.16439499888,0.0,0.0,83.57180272726177,39388.7003819722,0.0,66979.0217138564,0.0,0.0,0.0,32153.137492068337,1008.3875228660904,161099.96338972345,0.0,0.0,0.0
792,Bridgend 017,0.0,0.0,0.0,0.0,0.0,0.0,0.0,0.0,0.0,3321.4437904059623,0.0,0.0,0.0,0.0,0.0,0.0,3462.872939352933,0.0,0.0,0.0,0.0,0.0,0.0,0.0,0.0,0.0,8169.126025271354,0.0,0.0,0.0,3461.186010626301,0.0,0.0,0.0,44039.18455279839,0.0,0.0,2496.252818821947,0.0,0.0,12707.249413340372,0.0,0.0,28.28084889106364,9173.583769846642,0.0,440.35594759388295,0.0,0.0,0.0,0.0,0.0,680.2059826017176,0.0,0.0,0.0,0.0,0.0,0.0,0.0,0.0,0.0,0.0,0.0,0.0,0.0,0.0,0.0,0.0,0.0,0.0,0.0,0.0,0.0,0.0,0.0,0.0,0.0,0.0,0.0,0.0,0.0,0.0,0.0,0.0,0.0,0.0,0.0,0.0,0.0,0.0,0.0,0.0,0.0,0.0,0.0,0.0,0.0,0.0,0.0,0.0,0.0,0.0,0.0,0.0,0.0,0.0,0.0,0.0,0.0,0.0,0.0
5668,Spelthorne 012,0.0,0.0,37.99837446782932,0.0,0.0,0.0,0.0,0.0,0.0,2039.403602676888,0.0,0.0,0.0,0.0,0.0,0.0,12187.874929796166,0.0,0.0,0.0,0.0,7744.926203275638,0.0,0.0,0.0,0.0,61.91579345683044,0.0,0.0,0.0,49.36085854283825,0.0,0.0,0.0,22042.74728801298,0.0,0.0,8057.441453596571,0.0,0.0,16271.681090931892,0.0,0.0,0.0,11732.548522488149,0.0,18259.09165802697,0.0,0.0,0.0,2763.2659430741824,0.0,2003.6346984562808,0.0,0.0,0.0,0.0,0.0,146.79975988073744,0.0,0.0,0.0,0.0,0.0,0.0,7132.626931246814,0.0,0.0,0.0,0.0,0.0,0.0,14351.891843394988,0.0,0.0,0.0,0.0,5147.087305797667,0.0,0.0,0.0,0.0,1930.77912942362,19.58671192741788,0.0,0.0,3317.5174348788246,0.0,0.0,0.0,27945.09499152625,0.0,0.0,10996.19307697393,0.0,0.0,16054.944688424572,0.0,0.0,21.83642417055816,7373.904497264204,0.0,5731.988341432127,0.0,0.0,0.0,0.0,0.0,4833.065749734095,0.0,0.0,0.0
3819,Merthyr Tydfil 002,0.0,0.0,0.0,0.0,0.0,0.0,0.0,0.0,0.0,9063.646718995094,0.0,0.0,0.0,0.0,0.0,0.0,4362.647348096966,0.0,0.0,0.0,0.0,0.0,0.0,0.0,0.0,0.0,0.0,0.0,0.0,0.0,526.824679800677,0.0,6184.359530153647,0.0,48798.99875466455,0.0,1773.044029838005,0.0,0.0,0.0,20556.607820268207,0.0,0.0,13.377940522129046,10522.46824816736,0.0,0.0,0.0,0.0,0.0,13662.379546409858,0.0,9055.135867760317,0.0,0.0,0.0,0.0,0.0,72354.23926596256,0.0,0.0,0.0,242.3868254217444,0.0,0.0,35993.34776154505,0.0,0.0,0.0,0.0,0.0,0.0,336275.07534621714,0.0,0.0,0.0,0.0,0.0,0.0,0.0,0.0,0.0,132975.75682580238,0.0,0.0,0.0,69159.57607779975,893.2631357663422,12138.364033004153,0.0,173026.38188576253,288.1523508781758,14666.230784347556,66329.34930745518,0.0,0.0,209742.6279268851,0.0,0.0,600.5862465498581,29715.482258391334,0.0,594482.5170168788,0.0,0.0,0.0,159791.43353876722,3224.876894118133,791646.0315848478,0.0,0.0,0.0
496,Blaby 010,0.0,0.0,6574.251936586073,0.0,0.0,0.0,289.28926871423283,0.0,0.0,1680.9723148986154,0.0,0.0,0.0,0.0,0.0,0.0,55527.36538271078,0.0,0.0,0.0,0.0,7223.598771750249,0.0,0.0,0.0,0.0,1696.021314881427,0.0,0.0,0.0,2950.0549226537487,0.0,0.0,0.0,30796.4865065287,0.0,76.21460469293191,9946.186712351784,0.0,0.0,4979.390959641119,0.0,0.0,0.0,19374.16634047333,0.0,3409.847949702585,0.0,0.0,0.0,0.0,0.0,14975.916717914904,0.0,0.0,0.0,0.0,0.0,87142.1627103796,0.0,0.0,0.0,2968.4571595073803,0.0,0.0,19949.11133302301,0.0,0.0,275.1230910592294,0.0,0.0,0.0,449044.0983008504,0.0,0.0,0.0,577.796134642493,62331.14453939922,13151.375573834908,0.0,0.0,0.0,11040.510931011637,0.0,0.0,0.0,46257.316267501774,0.0,0.0,166.42159303223863,240069.83705823743,0.0,4872.564916070701,48751.71475013815,0.0,0.0,75420.37870096415,0.0,0.0,135.66905319219356,160833.5915401418,0.0,88955.76689449974,0.0,0.0,0.0,22815.23726528031,252.6128126392221,194056.8414422492,0.0,0.0,0.0
110,Babergh 002,0.0,0.0,1034.2662337592917,0.0,0.0,0.0,0.0,0.0,0.0,0.0,0.0,0.0,0.0,0.0,0.0,0.0,50535.58811280195,0.0,0.0,0.0,0.0,0.0,0.0,0.0,0.0,0.0,0.0,0.0,0.0,0.0,3950.918985744597,0.0,0.0,0.0,15056.455525989752,0.0,0.0,19096.45231055484,0.0,0.0,32029.85729577793,0.0,0.0,0.0,34352.05480933697,0.0,26081.448595953712,0.0,0.0,0.0,3642.322130746501,0.0,58188.956808543335,0.0,0.0,0.0,0.0,0.0,45764.88716662127,0.0,0.0,0.0,1114.714822418085,0.0,0.0,13093.663840505102,0.0,0.0,0.0,0.0,0.0,0.0,370651.3683225765,0.0,0.0,0.0,0.0,0.0,0.0,6788.877330362815,0.0,0.0,45189.19403976693,124.9476468675258,0.0,0.0,57328.72527907717,0.0,0.0,0.0,101832.60158724838,0.0,0.0,78278.9875309463,0.0,0.0,200898.1328233097,0.0,0.0,112.58966528687247,257886.44678120548,24.95371766099045,346693.11193791835,0.0,0.0,0.0,95385.0483261284,9018.883037100912,238673.6838468009,0.0,0.0,0.0
6128,Teignbridge 004,0.0,0.0,3127.106993794103,0.0,0.0,0.0,0.0,0.0,0.0,2469.0724768185764,0.0,0.0,0.0,0.0,0.0,0.0,38784.86165520995,0.0,0.0,0.0,410.4690009492059,0.0,0.0,0.0,0.0,0.0,19617.798625322048,304.1777953133644,0.0,0.0,1170.3604975403878,0.0,0.0,0.0,35506.403680734875,0.0,0.0,21829.7077987166,0.0,0.0,56365.25589686238,0.0,0.0,94.87516592659571,3254.88712249014,0.0,69857.5926724325,0.0,0.0,0.0,42368.994018768826,4721.462747448846,108395.02562266968,0.0,0.0,0.0,0.0,0.0,4239.185219121817,0.0,0.0,0.0,210.7896297054328,0.0,0.0,19184.86033035564,0.0,0.0,0.0,0.0,0.0,0.0,131530.54237801034,0.0,0.0,0.0,72.87497015012836,0.0,0.0,0.0,0.0,0.0,97335.6610064028,1145.5814981750011,0.0,0.0,35482.32566098864,0.0,0.0,2493.237168739452,83968.79792247864,0.0,31.16628472341717,23300.74658505124,0.0,0.0,133444.75806241663,0.0,0.0,826.7944954140304,22097.80581790613,135.14497948166735,139417.94957263692,0.0,0.0,0.0,39805.903187507945,5436.398642051862,302414.3514795913,0.0,0.0,0.0
2088,East Devon 001,0.0,0.0,19171.751242758823,0.0,0.0,0.0,0.0,0.0,0.0,0.0,0.0,0.0,0.0,0.0,0.0,0.0,99313.13495011828,0.0,0.0,0.0,0.0,0.0,0.0,0.0,0.0,0.0,2825.712958763004,0.0,0.0,0.0,9280.31329936466,34.3952728047232,17231.425749766997,1899.1713452462363,12804.862809038925,0.0,0.0,4334.950484015188,125.5111910435022,0.0,49698.44526589625,0.0,0.0,76.77635154566711,29974.562426986613,0.0,66289.01929610869,0.0,0.0,0.0,21871.56122907312,1093.1816224397876,424671.5552064724,0.0,0.0,0.0,0.0,0.0,160364.94142307853,0.0,0.0,0.0,4075.531478073335,0.0,0.0,11742.26669172819,0.0,0.0,0.0,0.0,0.0,0.0,989500.5506789336,0.0,0.0,0.0,0.0,43549.90207781824,1368.8671393705552,0.0,0.0,0.0,50314.815605315,743.1618114520559,0.0,0.0,141655.9230388037,94.2604053712452,300.38510044987225,1983.5043330013373,209463.5663310304,0.0,0.0,124640.61050119495,145.65669501102485,0.0,416931.1142016889,0.0,0.0,2840.676651030115,222263.448461543,243.65999163607933,479627.7547820426,0.0,0.0,0.0,98071.15008462484,6965.094322286908,2012417.3045014304,0.0,0.0,0.0
\end{filecontents*}
\pgfplotstabletypeset[
    multicolumn names=l, 
    col sep=comma, 
    string type, 
    header = has colnames, 
    columns={msoa11nm, motorway, tertiary, trunk, N_motorway, N_tertiary, N_trunk},
    columns/msoa11nm/.style={column type=l, column name=MSOA Name},
    columns/motorway/.style={column type={S[round-precision=2, table-format=1.2e1,scientific-notation = true,  table-number-alignment=center]}, column name=Total Road (TR) Motorway  Length (m)},
    columns/tertiary/.style={column type={S[round-precision=2, table-format=1.2e1,scientific-notation = true, table-number-alignment=center]}, column name=TR Tertiary Length (m)},
    columns/trunk/.style={column type={S[round-precision=2, table-format=1.2e1,scientific-notation = true, table-number-alignment=center]}, column name=TR Trunk Length (m)},
    columns/N_motorway/.style={column type={S[round-precision=2, table-format=1.2e1,scientific-notation = true, table-number-alignment=center]}, column name=TR Neighbour Motorway Length (m)},
    columns/N_tertiary/.style={column type={S[round-precision=2, table-format=1.2e1,scientific-notation = true, table-number-alignment=center]}, column name=TR Neighbour Tertiary Link Length (m)},
    columns/N_trunk/.style={column type={S[round-precision=2, table-format=1.2e1,scientific-notation = true, table-number-alignment=center]}, column name=TR Neighbour Trunk Length (m)},
    every head row/.style={before row=\toprule, after row=\midrule},
    every last row/.style={after row=\bottomrule}
    ]{CSVFiles/CuratedExampleSpatialFeatureVector2018.csv}}
    \smallskip
    \caption{{\bfseries Example feature vector for the spatial model for 2018.}  The feature vector is sorted in ascending order by total road length (m). The same feature vector values as in Table \ref{tab:Table 7 Curated Example Highway Composition Feature Vector} are included alongside their neighbouring MSOA road-type counterparts.} \label{tab:Table 10 Curated Example Spatial Feature Vector}
\end{table}

\clearpage

\subsection{Model Variant Example Output Estimations}
\label{sec:modelDetailsExampleOutput}

 Table \ref{tab:Lenght Model Example Outputs} shows example pollution predictions for the length model for different MSOAs. Table \ref{tab:Curated Highway Composition Model Results} shows example pollution predictions for the composition model for different MSOAs. Table \ref{tab:Table 11 Curated Spatial Predicition} shows example pollution predictions for the spatial model for different MSOAs.

\begin{table}
\resizebox{\linewidth}{!}{
\begin{filecontents*}{CSVFiles/CuratedmodelResultsLength.csv}
msoa11nm,Raster Value Pollution,Predicted 2018 Length,Length Difference
Leicester 026,1360.41,1358.07,2.35
Bradford 035,1619.04,888.49,730.55
Rotherham 003,2686.83,1281.24,1405.59
Southampton 012,2042.33,-75.63,2117.96
Knowsley 001,2100.21,4924.65,2824.44
Canterbury 002,5719.47,9280.29,3560.81
Manchester 004,3459.04,7818.62,4359.57
Oldham 026,2447.61,-2782.35,5229.96
Enfield 019,1382.42,-4738.22,6120.65
South Tyneside 008,702.8,-6492.63,7195.44
County Durham 052,14512.89,22991.17,8478.27
Warrington 001,15285.28,25993.94,10708.66
Sevenoaks 005,27622.77,43588.16,15965.39
Cotswold 003,115112.52,142113.86,27001.34
North East Derbyshire 010,35608.18,86885.82,51277.65
\end{filecontents*}
\pgfplotstabletypeset[
    multicolumn names=l, 
    col sep=comma, 
    string type, 
    header = has colnames, 
    columns={msoa11nm, Raster Value Pollution, Predicted 2018 Length, Length Difference},
    columns/msoa11nm/.style={column type=l, column name=MSOA District Name},
    columns/Raster Value Pollution/.style={column type={S[round-precision=2, table-format=1.2e1,scientific-notation = true, table-number-alignment=center]}, column name=Actual Pollution (µg/m$^3$)},
    columns/Predicted 2018 Length/.style={column type={S[round-precision=2, table-format=1.2e1,scientific-notation = true, table-number-alignment=center]}, column name=Predicted Pollution (µg/m$^3$)},
    columns/Length Difference/.style={column type={S[round-precision=2, table-format=1.2e1,scientific-notation = true, table-number-alignment=center]}, column name=Difference (µg/m$^3$)},
    every head row/.style={before row=\toprule, after row=\midrule},
    every last row/.style={after row=\bottomrule}
    ]{CSVFiles/CuratedmodelResultsLength.csv}}
    \smallskip
        \caption{{ \bfseries Length model air pollution concentration predictions for NO$_{x}$ in 2018.} Results from the length model with every 500th MSOA shown and sorted by the difference between predicted and actual pollution.} \label{tab:Lenght Model Example Outputs}
\end{table}

\begin{table}
\resizebox{\linewidth}{!}{
\begin{filecontents*}{CSVFiles/CuratedmodelResultsComposition.csv}
msoa11nm,Raster Value Pollution,Predicted 2018 Composition,Composition Difference
Bexley 001,2528.73,2528.63,0.1
Cheshire West and Chester 010,844.28,467.79,376.49
Tendring 018,20694.7,21378.01,683.31
St. Helens 004,4193.88,5203.98,1010.1
Blackburn with Darwen 009,1488.59,140.09,1348.5
Wellingborough 002,2888.54,4535.89,1647.35
Warwick 007,1539.73,3504.95,1965.22
Cardiff 018,1513.65,-775.26,2288.91
Lincoln 011,2875.08,179.68,2695.39
Bromley 010,3683.69,458.24,3225.44
Cardiff 049,1299.02,5215.58,3916.56
Wycombe 016,3394.1,-1933.82,5327.92
Scarborough 006,1182.78,-7271.65,8454.43
Teignbridge 018,13015.47,29841.54,16826.07
Derbyshire Dales 008,126067.2,164697.61,38630.42
\end{filecontents*}
\pgfplotstabletypeset[
    multicolumn names=l, 
    col sep=comma, 
    string type, 
    header = has colnames, 
    columns={msoa11nm, Raster Value Pollution, Predicted 2018 Composition, Composition Difference},
    columns/msoa11nm/.style={column type=l, column name=MSOA District Name},
    columns/Raster Value Pollution/.style={column type={S[round-precision=2, table-format=1.2e1,scientific-notation = true, table-number-alignment=center]}, column name=Actual Pollution (µg/m$^3$)},
    columns/Predicted 2018 Composition/.style={column type={S[round-precision=2, table-format=1.2e1,scientific-notation = true, table-number-alignment=center]}, column name=Predicted Pollution (µg/m$^3$)},
    columns/Composition Difference/.style={column type={S[round-precision=2, table-format=1.2e1,scientific-notation = true, table-number-alignment=center]}, column name=Difference (µg/m$^3$)},
    every head row/.style={before row=\toprule, after row=\midrule},
    every last row/.style={after row=\bottomrule}
    ]{CSVFiles/CuratedmodelResultsComposition.csv}}
    \smallskip
    \caption{{ \bfseries Composition model air pollution concentration predictions for NO$_{x}$ in 2018.} Results from the length model with every 500th MSOA shown and sorted by the difference between predicted and actual pollution.} \label{tab:Curated Highway Composition Model Results}
\end{table}

\begin{table}
\resizebox{\linewidth}{!}{
\begin{filecontents*}{CSVFiles/CuratedmodelResultsSpatial.csv}
msoa11nm,Raster Value Pollution,Predicted 2018 Spatial,Spatial Difference
Mid Sussex 008,3948.33,3948.73,0.41
Southwark 008,789.67,1155.08,365.41
Birmingham 138,735.85,-3.32,739.17
Warwick 007,1539.73,2607.07,1067.34
Harlow 010,1223.94,-207.44,1431.38
Croydon 015,811.05,-968.07,1779.11
Teignbridge 012,1685.67,-461.83,2147.5
Rotherham 012,6540.73,3974.93,2565.81
Cardiff 001,6850.06,3791.62,3058.44
Wiltshire 010,2027.79,-1663.64,3691.43
Isle of Wight 016,4197.77,-381.19,4578.95
Stockport 040,1958.26,-4070.91,6029.17
Shropshire 027,33122.29,42030.53,8908.24
West Somerset 003,29098.67,45527.85,16429.19
Derbyshire Dales 003,93641.53,131578.34,37936.81
\end{filecontents*}
\pgfplotstabletypeset[
    multicolumn names=l, 
    col sep=comma, 
    string type, 
    header = has colnames,  
    columns={msoa11nm, Raster Value Pollution, Predicted 2018 Spatial, Spatial Difference},
    columns/msoa11nm/.style={column type=l, column name=MSOA Name},
    columns/Raster Value Pollution/.style={column type={S[round-precision=2, table-format=1.2e1,scientific-notation = true, table-number-alignment=center]}, column name=Actual Pollution (µg/m$^3$)},
    columns/Predicted 2018 Spatial/.style={column type={S[round-precision=2, table-format=1.2e1,scientific-notation = true, table-number-alignment=center]}, column name=Predicted Pollution (µg/m$^3$)},
    columns/Spatial Difference/.style={column type={S[round-precision=2, table-format=1.2e1,scientific-notation = true, table-number-alignment=center]}, column name=Difference (µg/m$^3$)},
    every head row/.style={before row=\toprule, after row=\midrule},
    every last row/.style={after row=\bottomrule}
    ]{CSVFiles/CuratedmodelResultsSpatial.csv}}
    \smallskip
    \caption{{ \bfseries Spatial model air pollution concentration predictions for NO$_{x}$ in 2018.} Results from the length model with every 500th MSOA shown and sorted by the difference between predicted and actual pollution..} \label{tab:Table 11 Curated Spatial Predicition}
\end{table}

\clearpage
\subsection{Model Feature Selection}
\label{sec:featureSelectionDetails}

Table \ref{tab:Every5thMutualInformationPerPollutant} shows the mutual information between every 5th road type and all 12 pollutants, with the final column, summation, providing the total mutual information for that road type across each of the pollutants. The mutual information is also normalised between 0 and 1 and have been ordered from least to most total mutual information.

\begin{table}
\resizebox{\linewidth}{!}{
\begin{filecontents*}{CSVFiles/MutualInformationPerPollutant.csv}
Highway Type,MI PollutionCO,MI PollutionNO,MI PollutionNO2,MI PollutionNOX,MI PollutionNV10,MI PollutionNV25,MI PollutionO3,MI PollutionPM10,MI PollutionPM25,MI PollutionSO2,MI PollutionV10,MI PollutionV25,Summation
unclassified;residential,0.0,0.0,0.0,0.0,0.0,0.0,0.0,0.0,0.0,0.0,0.0,0.0,0.0
consultation,0.0,0.0,0.0,0.0,0.0,0.0,0.0,0.0,0.0,0.0,0.0,0.0,0.0
area,0.0,0.0,0.0,0.0,0.0,0.0,0.0,0.0,0.0,0.0,0.0,0.0,0.0
residential;footway,0.0,0.0,0.0,0.0,0.0,0.0,0.0,0.0,0.0,0.0,0.0,0.0,0.0
track;footway,0.0,0.0,0.0,0.0,0.0,0.0,0.0,0.0,0.0,0.0,0.0,0.0,0.0
motorway junction,0.0,0.0,0.0,0.0,0.0,0.0,0.0,0.0,0.0,0.0,0.0,0.0,0.0
driveway,0.0,0.0,0.0,0.0,0.0,0.0,0.0,0.0,0.0,0.0,0.0,0.0,0.0
depot,0.0,0.0,0.0,0.0,0.0,0.0,0.0,0.0,0.0,0.0,0.0,0.0,0.0
fence,0.0,0.0,0.0,0.0,0.0,0.0,0.0,0.0,0.0,0.0,0.0,0.0,0.0
path;service,0.0,0.0,0.0,0.0,0.0,0.0,0.0,0.0,0.0,0.0,0.0,0.0,0.0
conveyor,0.0,0.0,0.0,0.0,0.0,0.0,0.0,0.0,0.0,0.0,0.0,0.0,0.0
gallop,0.0,0.0,0.0,0.0,0.0,0.0,0.00012,0.0,0.0,0.0,0.0,0.0,0.00012
ser,0.0,0.0,0.0,0.0,0.0,6e-05,0.0,0.0,0.0,0.0,0.0,6e-05,0.00012
platform,0.0,6e-05,0.0,0.0,0.0,0.00017,0.0,0.0,0.0,0.0,0.0,0.0,0.00023
yes,0.0,6e-05,0.0,0.0,6e-05,0.0,6e-05,0.00012,0.0,0.0,0.0,6e-05,0.00035
traffic island,0.0,0.0,0.00012,7e-05,0.0,0.00024,1e-05,0.0,0.0,0.0,0.0,1e-05,0.00044
demolished,0.0,0.0,0.00023,0.00024,0.0,0.00018,6e-05,0.0,0.0,0.0,0.0,0.0,0.00072
layby,0.0,0.00029,0.0,0.00012,0.0,0.00029,0.0,0.0,0.0,0.0,0.0,0.00023,0.00094
junction,0.0,0.0,0.00011,6e-05,0.0,0.0,0.0,0.0,0.0,0.00141,0.0,0.0,0.00158
tank track,0.00159,0.0,0.0,0.0,0.0,0.0,0.0,0.0,0.0,0.0,0.0,0.0,0.00159
turning circle,0.0,0.0,0.0,0.0,0.0,0.0,0.0,0.0,0.0,0.0,0.00177,0.0,0.00177
bridalway,0.00077,0.0,0.0,0.0,0.0,0.0,0.0,0.0,0.0,0.0,0.00104,0.0,0.00181
escape,0.0,0.0,0.0,0.00013,0.00056,6e-05,0.00027,0.00049,0.00024,0.0,0.0,0.00026,0.00201
traffic signals,0.0,0.00017,0.00018,0.0003,6e-05,6e-05,0.0,0.00018,0.0,0.00163,0.0,0.0,0.00258
lay-by,0.0,0.0,0.0,0.0,0.0,0.0,0.0,0.0,0.0,0.00284,0.0,0.0,0.00284
trail,0.0,6e-05,0.0,6e-05,0.00023,0.00012,0.0,0.0,0.0,0.00262,0.0,0.0,0.00308
virtual,0.0,5e-05,0.00027,0.0,0.0,0.0,0.00023,2e-05,0.0,0.0,0.00317,8e-05,0.00383
show ring,0.0,0.0,0.0,0.0,0.0,0.0,0.0,0.0,0.0,0.00337,0.00051,0.0,0.00389
services,0.00261,0.0,0.00011,0.0,0.00012,0.0,6e-05,0.0,0.0,0.0,0.00146,0.00012,0.00447
access,0.0024,7e-05,0.00034,0.0,0.0,0.0004,5e-05,5e-05,0.0,0.0,0.00104,0.00023,0.00458
name only,0.00298,0.0,0.0,0.0,0.0,0.0,0.0,0.0,0.0,0.00162,0.0,0.0,0.00461
footway;steps,0.00491,0.0,0.0,0.0,0.0,0.0,0.0,0.0,0.0,0.0,0.0,0.0,0.00491
pre-existing before redevelopment,0.0,0.0,0.0,0.0,0.0,0.0,0.0,0.0,0.0,0.00542,0.0,0.0,0.00542
escalator,0.00385,6e-05,0.0,6e-05,0.00018,0.0,0.0,0.0,0.00023,0.00106,0.0,0.0,0.00543
ford,0.00357,0.0,0.0,0.0,0.0,0.0,0.0,0.0,0.0,0.0,0.00307,0.0,0.00663
Byway,0.0,0.0,0.0,0.0,0.0,0.0,0.0,0.0,0.0,0.00668,0.0,0.0,0.00668
residential;unclassified,0.00204,0.0,0.0,0.0,0.0,0.0,0.0,0.0,0.0,0.00232,0.00248,0.0,0.00684
path;track,0.00395,0.0,0.0,0.0,0.0,0.0,0.0,0.0,0.0,6e-05,0.0033,0.0,0.00731
manoeuvring forecourt,0.0,0.0,0.0,0.0,0.0,0.0,0.0,0.0,0.0,0.00558,0.0022,0.0,0.00778
stepping stones,0.00212,0.0,0.0,6e-05,0.0,0.00017,0.00029,0.00017,0.0,0.00307,0.00287,0.0,0.00876
dismantled,0.00127,0.0,0.0001,0.0,0.0003,0.0,0.0,0.0,0.0,0.00538,0.00218,0.0,0.00924
elevator,0.0,0.0,0.00064,0.00034,0.00018,0.00016,2e-05,0.0005,0.00045,0.00754,0.0,0.0,0.00983
rest area,0.00353,0.00015,0.0,0.0001,0.0001,5e-05,0.0004,0.00045,0.00016,0.0,0.00517,0.00039,0.01052
abandoned,0.00181,0.0,5e-05,0.0,0.0,0.0,0.0,0.0,0.0,0.0,0.00874,0.0,0.01059
passing place,0.00132,0.0,0.0,0.0,0.0,0.0,0.0,0.0,0.0,0.00594,0.00542,0.0,0.01268
bus guideway,0.0,0.00086,0.00099,0.0016,0.00095,0.00156,0.00181,0.00094,0.00138,0.00291,0.0,0.0006,0.01361
bus stop,0.00729,0.00011,0.0,0.00023,0.0,0.0,0.0,0.0,0.0,0.00186,0.00557,0.0,0.01508
lane,0.00661,0.0,0.0,0.0,0.0,0.0,0.0,0.0,0.0,0.00885,0.0,0.0,0.01546
disused,0.00455,0.00046,0.00021,0.00029,1e-05,0.00076,0.0,0.00046,0.00017,0.00703,0.00216,0.00011,0.0162
secondary link,0.00845,0.0,0.0,0.0,0.0,0.00023,0.0,0.0,0.0,0.00388,0.00342,0.00138,0.01737
corridor,0.00272,0.00045,0.00011,0.00064,0.00065,0.00111,0.00066,0.00069,0.00076,0.00246,0.00676,0.00096,0.01798
give way,0.00624,0.0,0.0,0.0,0.0,0.0,0.0,0.0,0.0,0.0,0.01219,0.0,0.01843
kerb,0.00355,0.0,0.0,0.0,0.0,0.0,0.0,0.0,0.0,0.00467,0.01108,0.0,0.01929
unsurfaced,0.00352,0.00025,0.00063,0.0001,0.0011,0.0,0.00032,0.0,0.0,0.00291,0.01082,0.00035,0.02001
planned,0.00702,0.00012,6e-05,0.00018,0.00018,6e-05,6e-05,6e-05,0.0,0.00558,0.00752,0.0,0.02083
construction route,0.01039,0.0,0.0,0.0,0.0,0.0,0.0,0.0,0.0,0.00453,0.00767,0.0,0.0226
centre line,0.00912,0.0,0.0,0.0,0.0,0.0,0.0,0.0,0.0,0.0073,0.00845,0.0,0.02487
none,0.00949,0.0,6e-05,0.0,0.0,0.0,0.0,0.0,0.0,0.00941,0.00638,0.0,0.02535
minor,0.00759,0.0,0.0,0.0,0.0,0.0,0.0,0.0,0.0,0.01212,0.007,0.0,0.02671
redeveloped,0.01094,0.0,0.0,0.0,0.0,0.0,0.0,0.0,0.0,0.01081,0.00505,0.0,0.02679
unclassified;track,0.00445,0.0,0.0,0.0,0.0,0.0,0.0,0.0,0.0,0.00949,0.01463,0.0,0.02857
former,0.00927,0.0,0.0,0.0,0.0,0.0,0.0,0.0,0.0,0.00743,0.01497,0.0,0.03167
byway open to all traffic,0.0042,0.0,0.0,0.0,0.0,0.0,0.0,0.0,0.0,0.0104,0.01723,0.0,0.03183
currently closed,0.00446,0.0,0.0,0.0,0.0,0.0,0.0,0.0,0.0,0.01017,0.01817,0.0,0.0328
TR,0.00731,0.0,0.0,0.0,0.0,0.0,0.0,0.0,0.0,0.00899,0.0172,0.0,0.0335
track; footway,0.0152,0.0,0.0,0.0,0.0,0.0,0.0,0.0,0.0,0.01523,0.00722,0.0,0.03765
culvet,0.01473,0.0,0.0,0.0,0.0,0.0,0.0,0.0,0.0,0.00571,0.01797,0.0,0.0384
closed footway,0.01034,0.0,0.0,0.0,0.0,0.0,0.0,0.0,0.0,0.01824,0.01032,0.0,0.0389
crossing,0.00929,0.00031,3e-05,8e-05,0.0,0.00013,0.00019,0.00031,8e-05,0.01794,0.01216,0.00031,0.04082
dismantaled,0.01489,0.0,0.0,0.0,0.0,0.0,0.0,0.0,0.0,0.01594,0.01207,0.0,0.0429
FIXME,0.01495,0.0,0.0,0.0,0.0,0.0,0.0,0.0,0.0,0.01576,0.01558,0.0,0.04629
steps;footway,0.01031,0.0,0.0,0.0,0.0,0.0,0.0,0.0,0.0,0.01765,0.01981,0.0,0.04777
proposed,0.0,0.0078,0.00187,0.00517,0.0055,0.00612,0.00564,0.00681,0.00561,0.0,0.0,0.00369,0.04822
ohm:military:Trench,0.0162,0.0,0.0,6e-05,0.0,0.0,0.0,0.0,0.00012,0.02207,0.01783,0.0,0.05628
living street,0.00555,0.00448,0.00752,0.00484,0.00266,0.00691,0.00455,0.00327,0.00275,0.00103,0.00887,0.00501,0.05744
byway,0.00276,0.0046,0.00514,0.00364,0.00393,0.00264,0.00303,0.0025,0.00364,0.00974,0.01133,0.00454,0.05748
construction,0.0,0.0139,0.01295,0.01131,0.01621,0.01246,0.01571,0.0163,0.01631,0.0,0.0,0.01735,0.13251
tertiary link,0.01867,0.00869,0.00907,0.00695,0.01261,0.01414,0.01101,0.01161,0.01408,0.01995,0.0213,0.01096,0.15904
no,0.01748,0.01391,0.0121,0.01197,0.01263,0.01149,0.01293,0.013,0.01179,0.01777,0.01293,0.01247,0.16047
primary link,0.01422,0.01137,0.01123,0.01987,0.01153,0.02192,0.01218,0.0197,0.01991,0.02096,0.01655,0.01587,0.19532
raceway,0.02374,0.01704,0.01718,0.01822,0.01772,0.01897,0.01953,0.01879,0.0175,0.01688,0.02174,0.01864,0.22595
pedestrian,0.02104,0.0132,0.02519,0.02325,0.02953,0.02395,0.02391,0.02578,0.02152,0.01946,0.01675,0.0215,0.26506
steps,0.03263,0.02841,0.02801,0.02261,0.01958,0.0203,0.02837,0.03116,0.03712,0.03179,0.02794,0.0498,0.35772
motorway link,0.0511,0.02536,0.02602,0.02604,0.03007,0.03002,0.03495,0.0294,0.02774,0.03627,0.04571,0.02865,0.39134
trunk link,0.02892,0.03797,0.04916,0.0464,0.04567,0.04509,0.04634,0.04184,0.04415,0.03837,0.03076,0.04255,0.4972
road,0.06754,0.06742,0.06692,0.07398,0.06959,0.06877,0.07616,0.07418,0.06978,0.06423,0.0649,0.07204,0.83552
motorway,0.09286,0.09656,0.09361,0.09837,0.09862,0.1002,0.09898,0.09945,0.09767,0.09896,0.09762,0.0985,1.17141
cycleway,0.10411,0.08835,0.08733,0.08756,0.10017,0.10611,0.11183,0.0975,0.1026,0.1027,0.09831,0.10061,1.18717
path,0.30154,0.29973,0.30263,0.30557,0.31519,0.30876,0.30727,0.31532,0.31351,0.30463,0.29624,0.30412,3.67451
primary,0.29605,0.30509,0.31736,0.31119,0.30551,0.29602,0.32367,0.31139,0.31808,0.30228,0.29332,0.31306,3.69301
trunk,0.354,0.35226,0.35278,0.36166,0.35993,0.36145,0.36828,0.3652,0.36826,0.36589,0.35411,0.36173,4.32554
residential,0.37724,0.33774,0.32994,0.35126,0.35336,0.36154,0.40945,0.36995,0.35763,0.37524,0.37174,0.35425,4.34935
secondary,0.44456,0.45361,0.45047,0.44861,0.45682,0.46513,0.46604,0.44938,0.44526,0.44512,0.44039,0.4578,5.4232
bridleway,0.45271,0.46221,0.46336,0.47635,0.46788,0.46431,0.47031,0.47071,0.481,0.46526,0.45277,0.46554,5.5924
service,0.53424,0.53913,0.53956,0.54487,0.53055,0.53727,0.54035,0.55129,0.54116,0.5292,0.52246,0.53549,6.44556
tertiary,0.56375,0.5597,0.54731,0.55709,0.55283,0.55361,0.58464,0.55669,0.56499,0.55459,0.54902,0.55939,6.70361
footway,0.60646,0.6207,0.62528,0.63974,0.61032,0.61055,0.60166,0.59391,0.61053,0.60834,0.59294,0.60718,7.32761
unclassified,0.83325,0.83205,0.84881,0.87426,0.86696,0.84782,0.85228,0.84493,0.84295,0.84057,0.83183,0.84958,10.16531
track,1.0,1.0,1.0,1.0,1.0,1.0,1.0,1.0,1.0,1.0,1.0,1.0,12.0
\end{filecontents*}
\pgfplotstabletypeset[
   multicolumn names=l, 
    col sep=comma, 
    string type, 
    header = has colnames, 
    columns={Highway Type, MI PollutionCO, MI PollutionNO, MI PollutionNO2, MI PollutionNOX, MI PollutionNV10, MI PollutionNV25, MI PollutionO3, MI PollutionPM10, MI PollutionPM25, MI PollutionSO2, MI PollutionV10, MI PollutionV25, Summation},
    columns/Highway Type/.style={column type=l, column name=Road Type},
    columns/MI PollutionCO/.style={column type={S[round-precision=2, round-integer-to-decimal, table-format=1.2e-1, table-number-alignment=center]}, column name=Mutual Information (MI) CO},
    columns/MI PollutionNO/.style={column type={S[round-precision=2, round-integer-to-decimal, table-format=1.2e-1, table-number-alignment=center]}, column name= MI NO},
    columns/MI PollutionNO2/.style={column type={S[round-precision=2, round-integer-to-decimal, table-format=1.2e-1, table-number-alignment=center]}, column name=MI NO2},
    columns/MI PollutionNOX/.style={column type={S[round-precision=2, round-integer-to-decimal, table-format=1.2e-1, table-number-alignment=center]}, column name=MI NOX},
    columns/MI PollutionNV10/.style={column type={S[round-precision=2, round-integer-to-decimal, table-format=1.2e-1, table-number-alignment=center]}, column name=MI NV10},
    columns/MI PollutionNV25/.style={column type={S[round-precision=2, round-integer-to-decimal, table-format=1.2e-1, table-number-alignment=center]}, column name=MI NV25},
    columns/MI PollutionO3/.style={column type={S[round-precision=2, round-integer-to-decimal, table-format=1.2e-1, table-number-alignment=center]}, column name=MI O3},
    columns/MI PollutionPM10/.style={column type={S[round-precision=2, round-integer-to-decimal, table-format=1.2e-1, table-number-alignment=center]}, column name=MI PM10},
    columns/MI PollutionPM25/.style={column type={S[round-precision=2, round-integer-to-decimal, table-format=1.2e-1, table-number-alignment=center]}, column name=MI PM25},
    columns/MI PollutionSO2/.style={column type={S[round-precision=2, round-integer-to-decimal, table-format=1.2e-1, table-number-alignment=center]}, column name=MI SO2},
    columns/MI PollutionV10/.style={column type={S[round-precision=2, round-integer-to-decimal, table-format=1.2e-1, table-number-alignment=center]}, column name=MI V10},
    columns/MI PollutionV25/.style={column type={S[round-precision=2, round-integer-to-decimal, table-format=1.2e-1, table-number-alignment=center]}, column name=MI V25},
    columns/Summation/.style={column type={S[round-precision=2, round-integer-to-decimal, table-format=2.2e-1, table-number-alignment=center]}, column name=Summation Mutual Information},
    every head row/.style={before row=\toprule, after row=\midrule},
    every last row/.style={after row=\bottomrule}
    ]{CSVFiles/MutualInformationPerPollutant.csv}}
    \smallskip
    \caption{{\bfseries Mutual information between road type and air pollutants.} Mutual information from every 5th road type concerning every air pollutant ordered by summation mutual information; the total mutual information a road type contains with every air pollutant.} 
    \label{tab:Every5thMutualInformationPerPollutant}
\end{table}

Table \ref{tab:R2ValuesAll99HighwaysFeatureSelection} shows results from repeating the experiments for all three models while reducing the input data set through the removal of specific road types reducing the data used to create the feature vector. The road types that were removed were chosen by incrementing a threshold value and removing road types that had a summation mutual information value below the threshold.    

\begin{table}
\resizebox{\linewidth}{!}{
\begin{filecontents*}{CSVFiles/CuratedR2ValuesHighwayTypesandAll99highways-Figure12CorresponidngTable.csv}
R2 Length,R2Composition,R2 Spatial,Threshold Value,Highway Type Length
0.781,0.886,0.886,-1.0,99
0.781,0.886,0.886,0.0,88
0.781,0.885,0.885,0.1,23
0.781,0.881,0.881,0.2,19
0.782,0.88,0.88,0.3,17
0.782,0.88,0.88,0.4,15
0.783,0.878,0.878,0.5,14
0.783,0.878,0.878,0.6,14
0.783,0.878,0.878,0.7,14
0.783,0.878,0.878,0.8,14
0.782,0.875,0.875,0.9,13
0.782,0.875,0.875,1.0,13
0.785,0.875,0.875,2.0,11
0.785,0.875,0.875,3.0,11
0.789,0.87,0.87,4.0,9
0.789,0.866,0.866,5.0,7
0.781,0.851,0.851,6.0,5
0.78,0.829,0.829,7.0,3
0.83,0.829,0.829,8.0,2
0.83,0.829,0.829,9.0,2
0.83,0.829,0.829,10.0,2
0.645,0.645,0.645,11.0,1
\end{filecontents*}
\pgfplotstabletypeset[
    multicolumn names=l, 
    col sep=comma, 
    string type, 
    header = has colnames,  
    columns={R2 Length, R2Composition, R2 Spatial, Threshold Value, Highway Type Length},
    columns/R2 Length/.style={column type={S[round-precision=2, table-format=1.2, table-number-alignment=center]}, column name=$R^2$ Score Length Model},
    columns/R2Composition/.style={column type={S[round-precision=2, table-format=1.2, table-number-alignment=center]}, column name=$R^2$ Score Composition Model},
    columns/R2 Spatial/.style={column type={S[round-precision=2, table-format=1.2, table-number-alignment=center]}, column name=$R^2$ Score Spatial Model},
    columns/Threshold Value/.style={column type={S[round-precision=1, round-integer-to-decimal, table-format=-2.1, table-number-alignment=center]}, column name=Threshold Value},
    columns/Highway Type Length/.style={column type={S[round-precision=0, table-format=2.0, table-number-alignment=center]}, column name=No. of Road Types},
    every head row/.style={before row=\toprule, after row=\midrule},
    every last row/.style={after row=\bottomrule}
    ]{CSVFiles/CuratedR2ValuesHighwayTypesandAll99highways-Figure12CorresponidngTable.csv}}
    \smallskip
    \caption{{\bfseries Experiment results for model performance on reducing feature vector input road types.} Results from reducing the included road types for the feature vector based on the total summation mutual information a road type has with all air pollutants, subsetted depending on the denoted threshold value.} 
    \label{tab:R2ValuesAll99HighwaysFeatureSelection}
\end{table}

Table \ref{tab:Table 14 R2 Values Highway Types 22 relevant highway types} shows the results from the same tests as Table \ref{tab:R2ValuesAll99HighwaysFeatureSelection}, however with the starting set of road types being restricted to road types that have a mutual information value with every pollutant. The initial set of road types numbered 25. 

\begin{table}
\resizebox{\linewidth}{!}{
\begin{filecontents*}{CSVFiles/CuratedR2ValuesHighwayTypes22relevanthighwaytypesFigure13CorresponidngTable.csv}
R2 Length,R2Composition,R2 Spatial,Threshold Value,Highway Type Length
0.781,0.885,0.885,-1.0,25
0.781,0.885,0.885,0.0,25
0.781,0.885,0.885,0.1,22
0.781,0.881,0.881,0.2,19
0.782,0.88,0.88,0.3,17
0.782,0.88,0.88,0.4,15
0.783,0.878,0.878,0.5,14
0.783,0.878,0.878,0.6,14
0.783,0.878,0.878,0.7,14
0.783,0.878,0.878,0.8,14
0.782,0.875,0.875,0.9,13
0.782,0.875,0.875,1.0,13
0.785,0.875,0.875,2.0,11
0.785,0.875,0.875,3.0,11
0.789,0.87,0.87,4.0,9
0.789,0.866,0.866,5.0,7
0.781,0.851,0.851,6.0,5
0.78,0.829,0.829,7.0,3
0.83,0.829,0.829,8.0,2
0.83,0.829,0.829,9.0,2
0.83,0.829,0.829,10.0,2
0.645,0.645,0.645,11.0,1
\end{filecontents*}
\pgfplotstabletypeset[
    multicolumn names=l, 
    col sep=comma, 
    string type, 
    header = has colnames,
    columns={R2 Length, R2Composition, R2 Spatial, Threshold Value, Highway Type Length},
    columns/R2 Length/.style={column type={S[round-precision=2, table-format=1.2, table-number-alignment=center]}, column name=$R^2$ Score Length Model},
    columns/R2Composition/.style={column type={S[round-precision=2, table-format=1.2, table-number-alignment=center]}, column name=$R^2$ Score Composition Model},
    columns/R2 Spatial/.style={column type={S[round-precision=2, table-format=1.2, table-number-alignment=center]}, column name=$R^2$ Score Spatial Model},
    columns/Threshold Value/.style={column type={S[round-precision=1, round-integer-to-decimal, table-format=-2.1, table-number-alignment=center]}, column name=Threshold Value},
    columns/Highway Type Length/.style={column type={S[round-precision=0, table-format=2.0, table-number-alignment=center]}, column name=No. of Road Types},
    every head row/.style={before row=\toprule, after row=\midrule},
    every last row/.style={after row=\bottomrule}
    ]{CSVFiles/CuratedR2ValuesHighwayTypes22relevanthighwaytypesFigure13CorresponidngTable.csv}}
    \smallskip
    \caption{{\bfseries Experiment results for model performance on reducing feature vector input road types, with road types with non-zero mutual information with all air pollutants.} The results from rerunning the experiment are detailed in Table \ref{tab:R2ValuesAll99HighwaysFeatureSelection} However, only road types with non-zero mutual information with all air pollutants are included in the initial set. } \label{tab:Table 14 R2 Values Highway Types 22 relevant highway types}
\end{table}

\begin{figure}[ht]
\begin{center}
\includegraphics[width=0.8\textwidth]{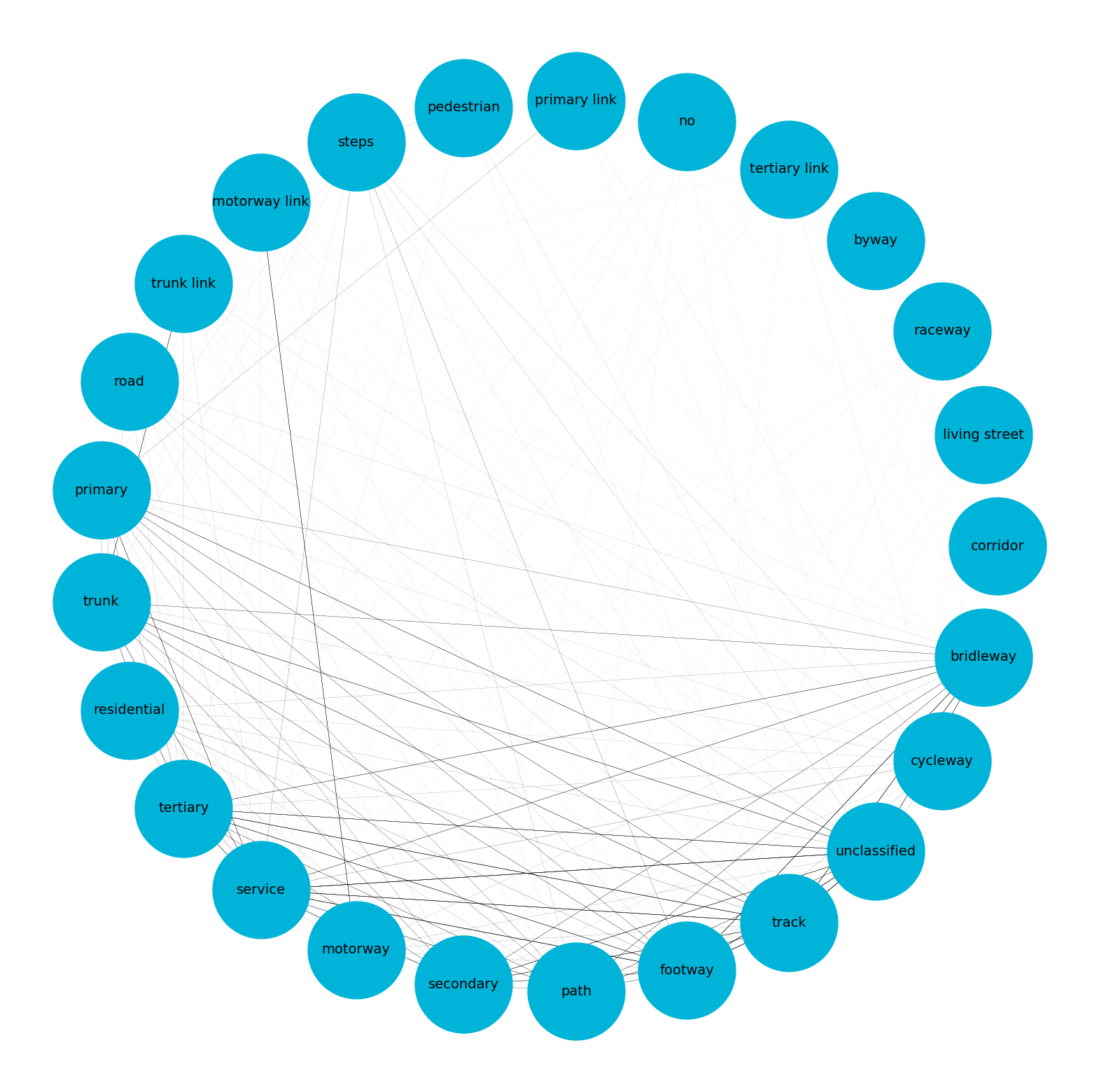}
\caption{{\bfseries Network model of the inter road mutual information.} The mutual information between the different road types was modelled as a network with the nodes representing road types and the edge between two nodes being weighted depending on the mutual information between the two. }
\label{fig:NetorkModelMutualInformation}
\end{center}
\end{figure}

\begin{figure}[ht]
\begin{center}
\text{Removal Process for Inter Road Mutual Information Network}\par\medskip
\includegraphics[width=0.15\textwidth]{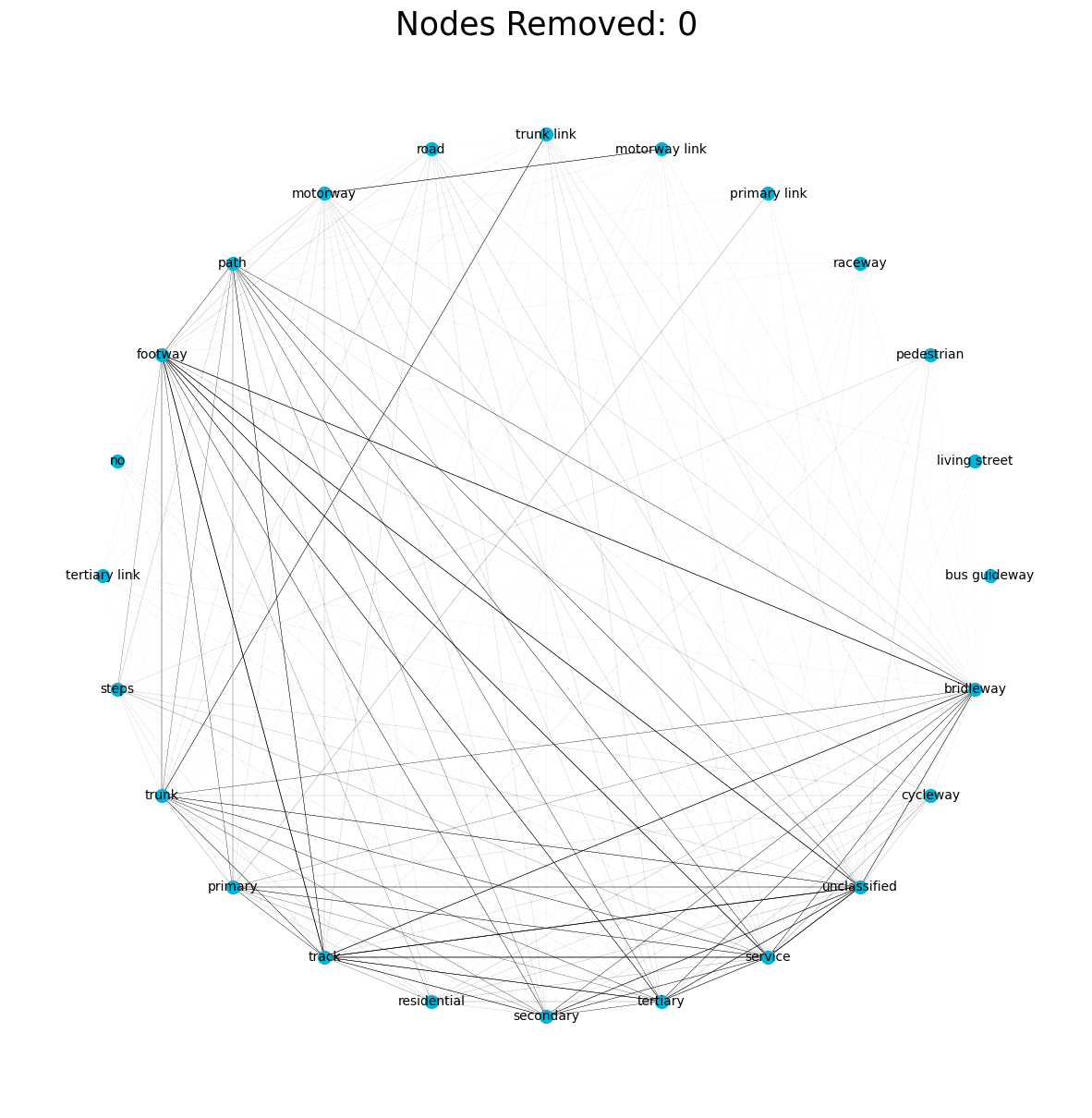}
\includegraphics[width=0.15\textwidth]{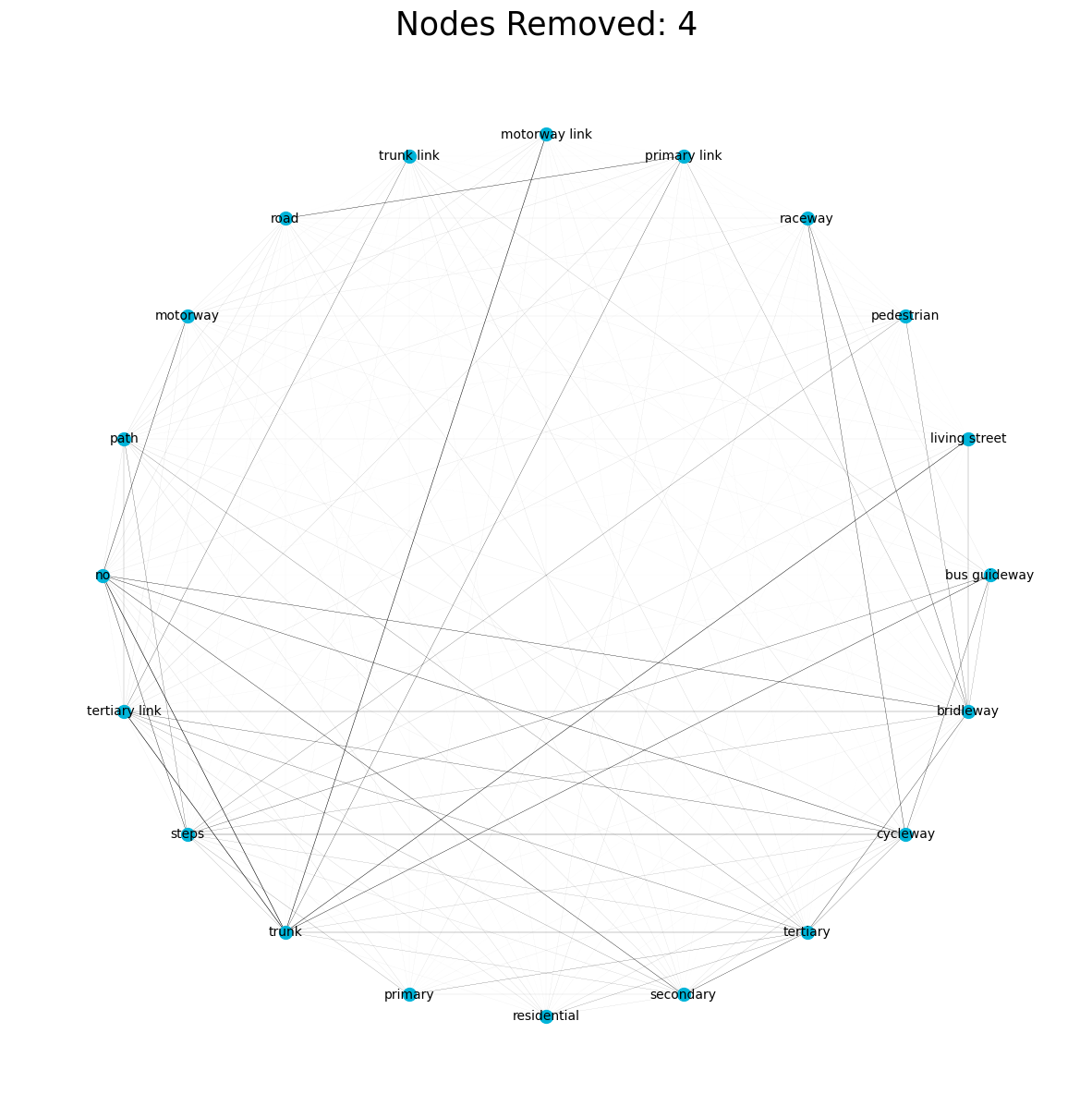}
\includegraphics[width=0.15\textwidth]{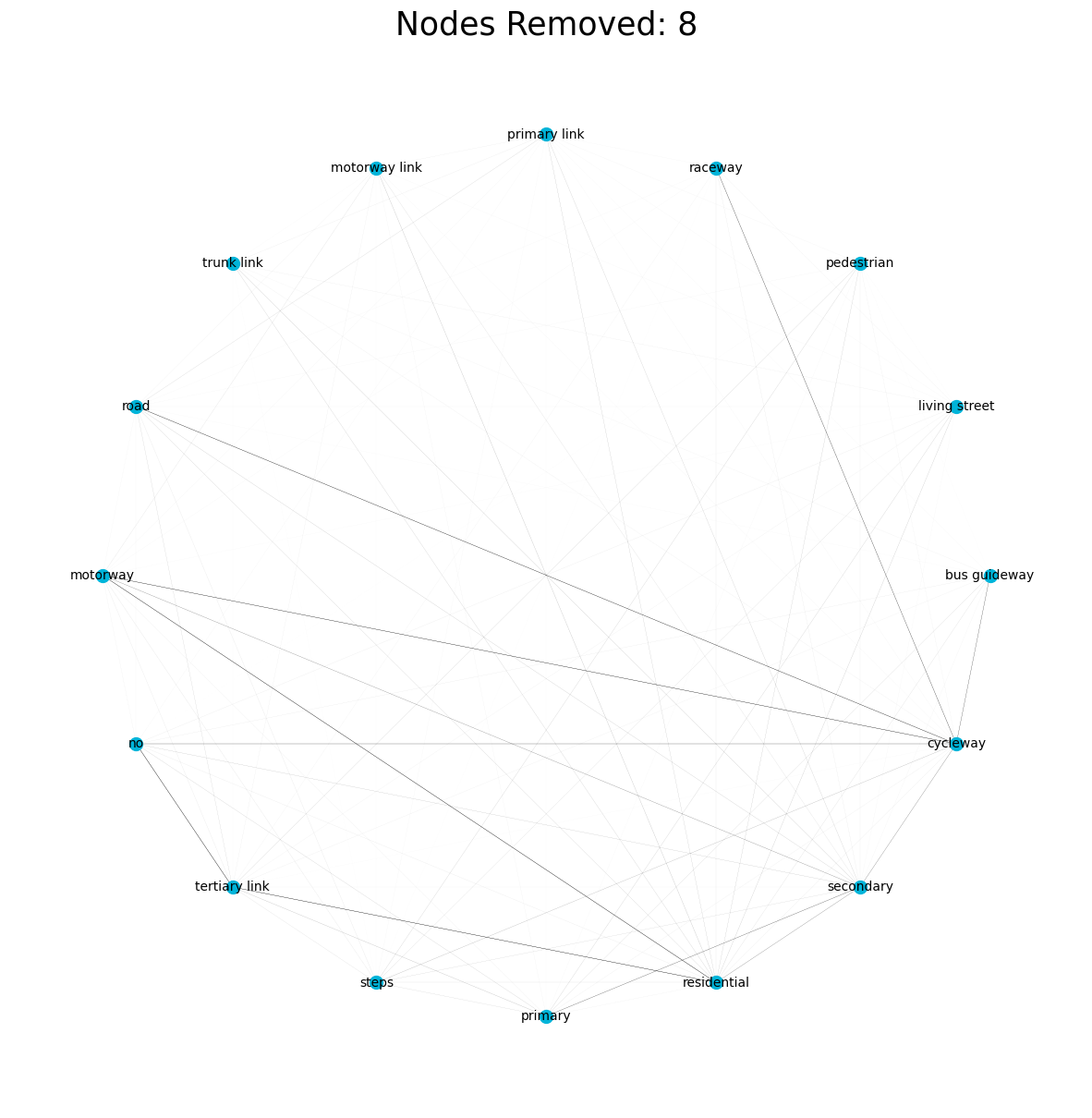}
\includegraphics[width=0.15\textwidth]{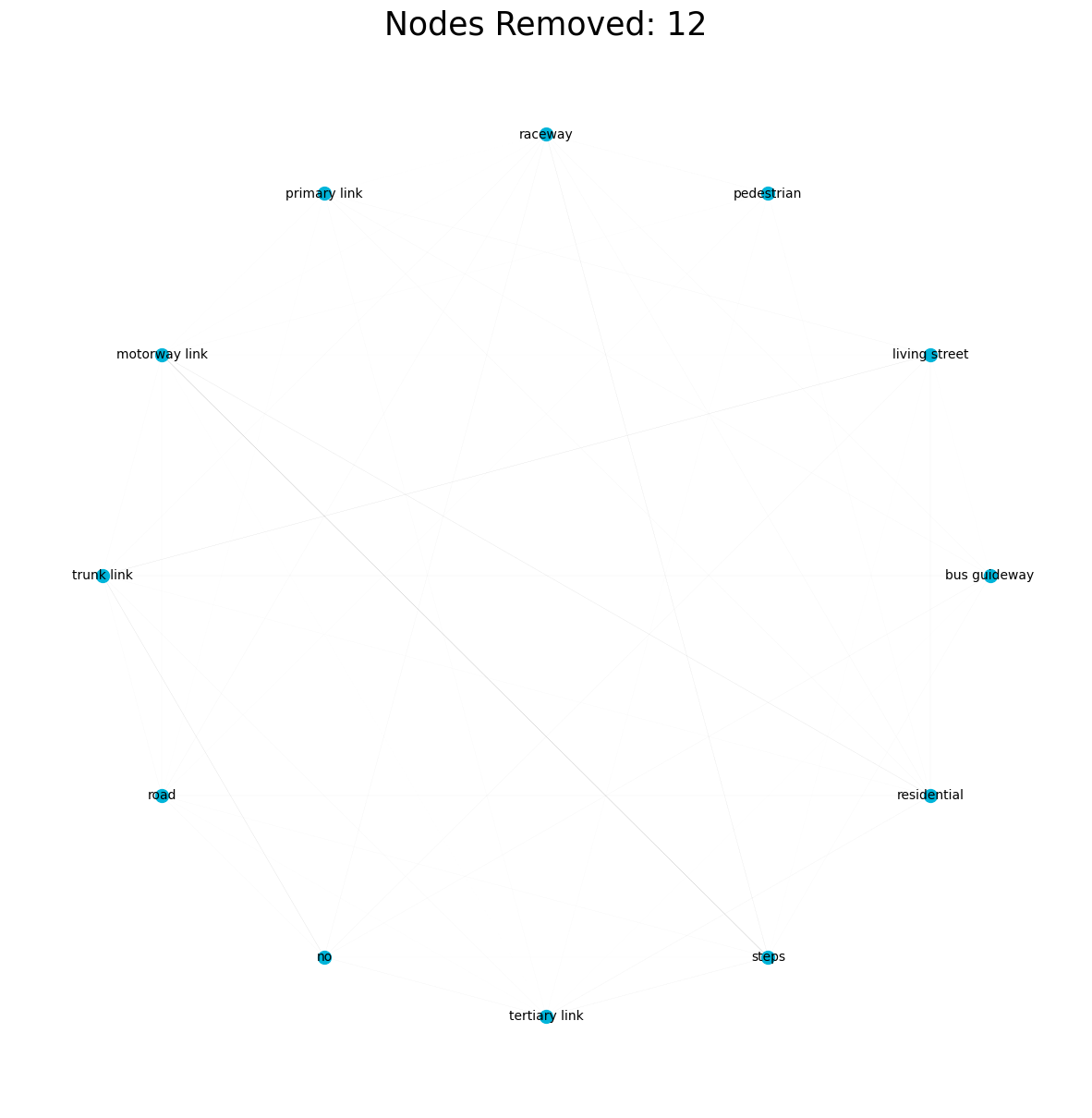}\\
\includegraphics[width=0.15\textwidth]{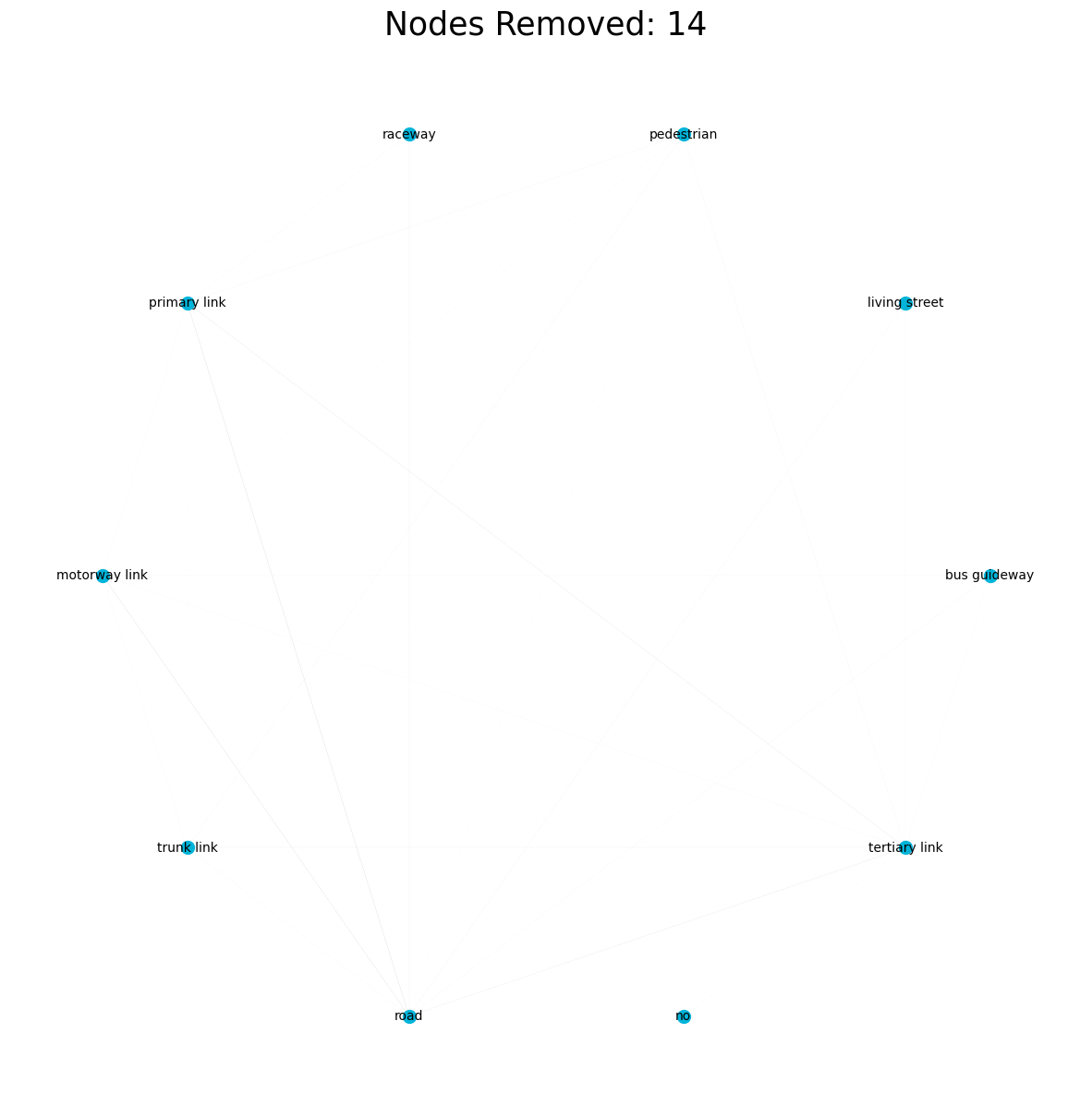}
\includegraphics[width=0.15\textwidth]{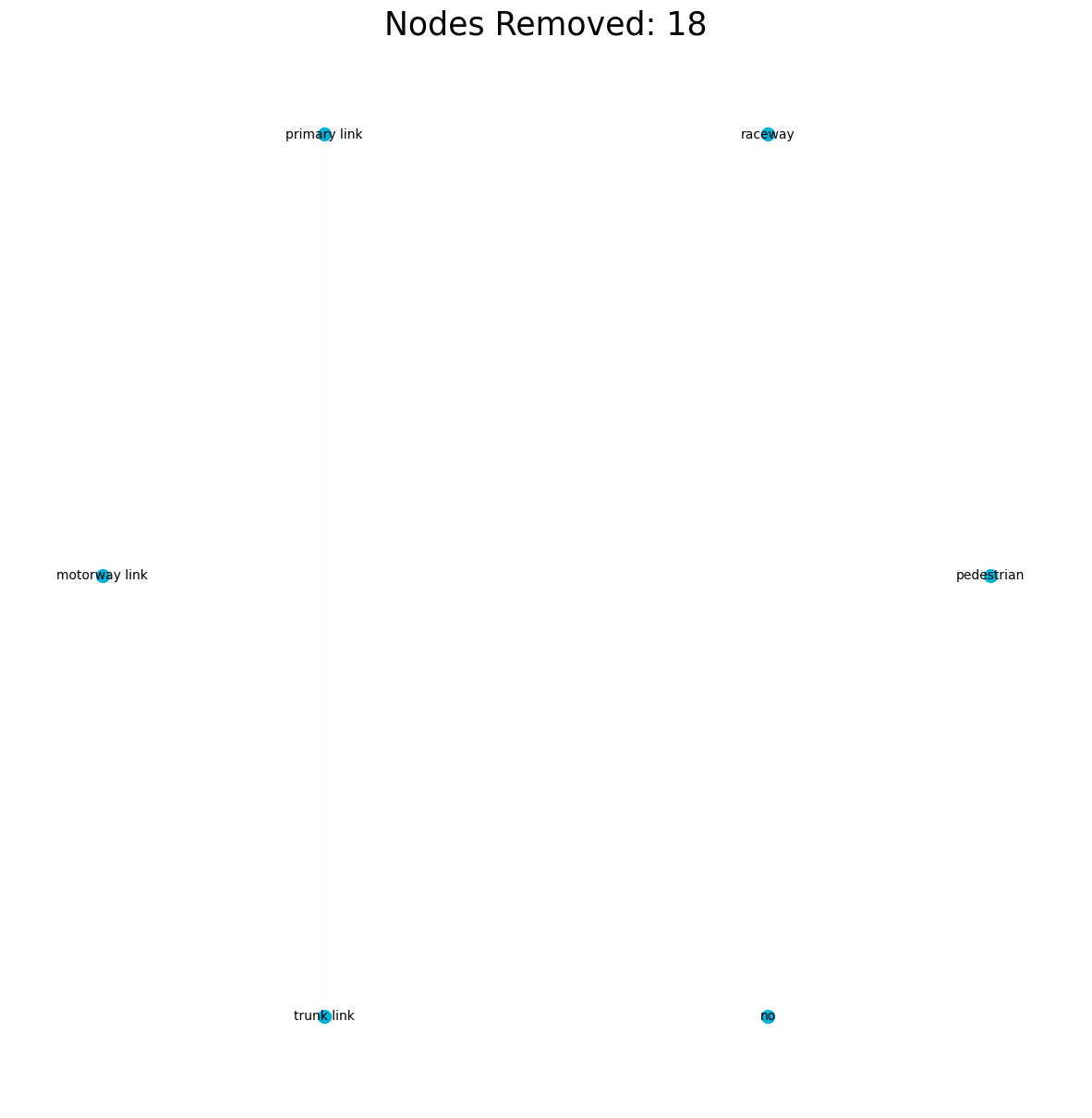}
\includegraphics[width=0.15\textwidth]{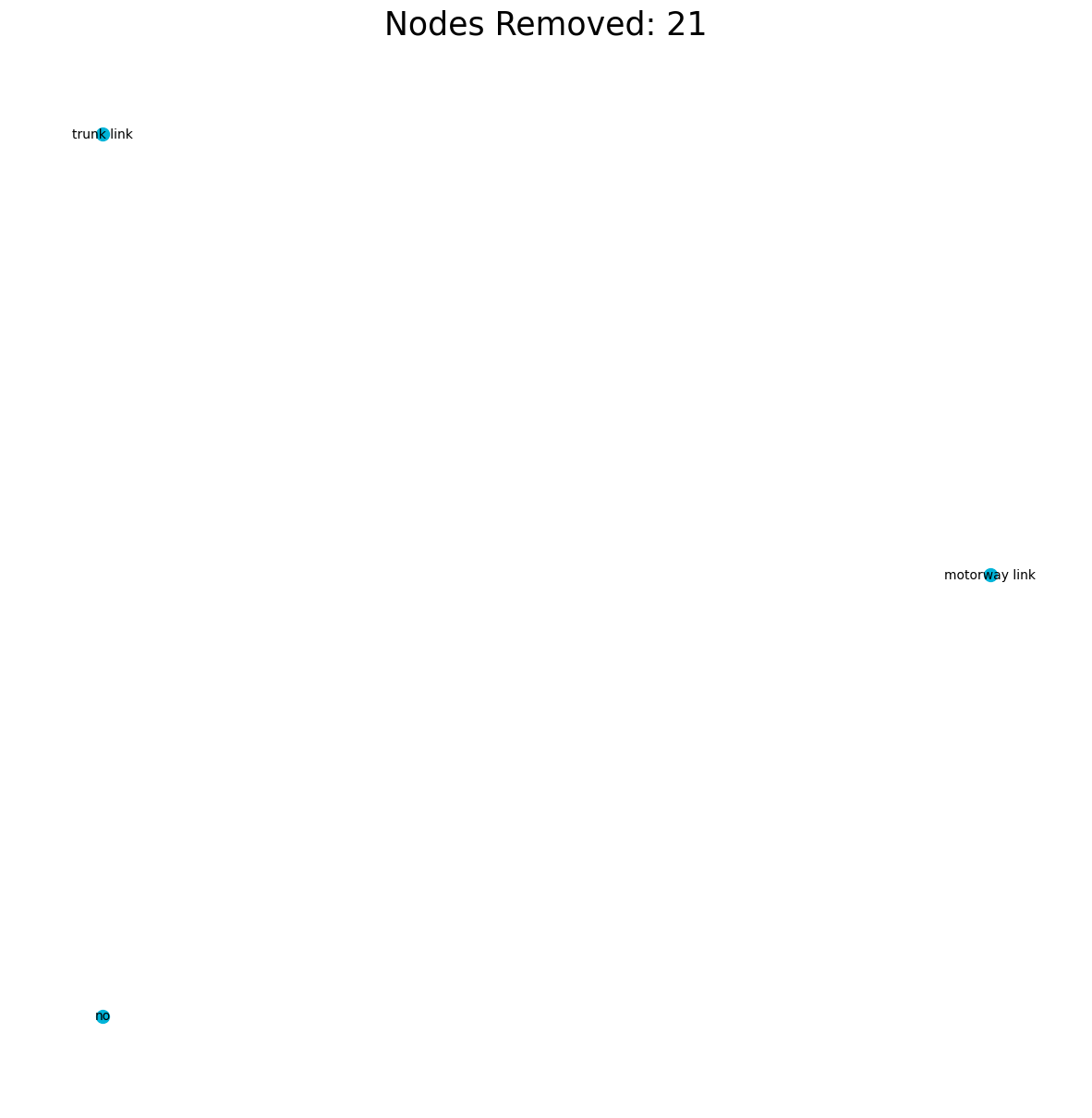}
\includegraphics[width=0.15\textwidth]{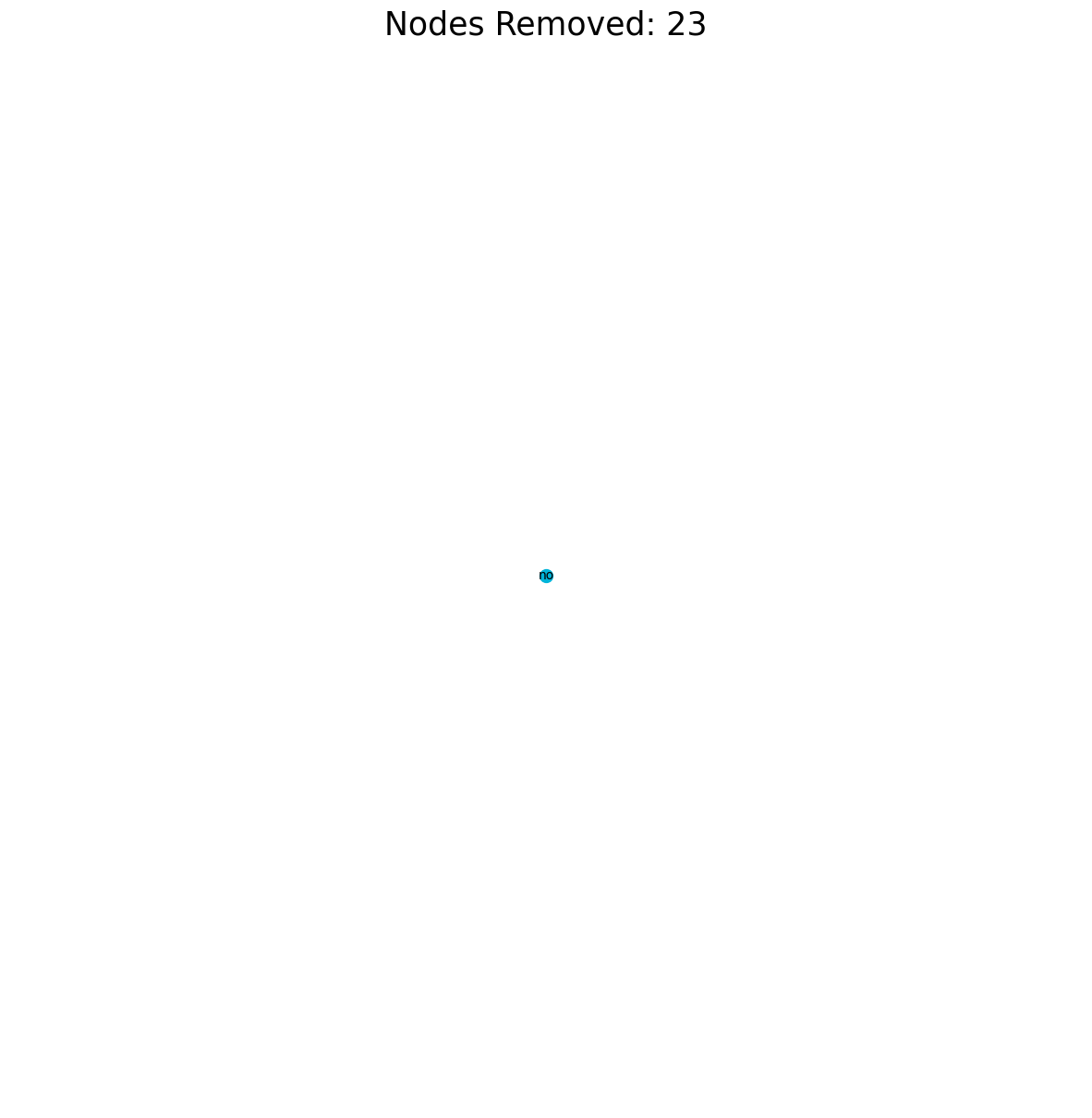}
\caption{{\bfseries Removal process for network model of the inter road mutual information.} Visualised are steps taken during the subsetting of the road types to determine which road types had the highest value of mutual information about all other road types to be included in the feature vector input into the models. }
\label{fig:NetworkModelRemoveTimesteps}
\end{center}
\end{figure}

Figure \ref{fig:NetorkModelMutualInformation} shows the full network representation of the 25 road types with non zero mutual information with all 12 pollutants, and Figure \ref{fig:NetworkModelRemoveTimesteps} the network state at key stages during the removal process for reducing the road type inclusion list, detailed in Table \ref{tab:Inter road mutual information tests}.

\begin{table}
\resizebox{\linewidth}{!}{
\begin{filecontents*}{CSVFiles/CuratedInterRoadMutualInformationResults.csv}
R2 Length,R2Composition,Inter Road Cumulative MutualInformation,Highway Type Length List,New Highway Type Included
0.645,0.645,3.003,1,track 
0.83,0.829,5.528,2,unclassified 
0.807,0.826,7.547,3,service 
0.764,0.826,9.153,4,footway 
0.781,0.851,10.202,5,tertiary 
0.787,0.854,11.042,6,trunk 
0.785,0.859,11.677,7,bridleway 
0.778,0.859,12.119,8,path 
0.778,0.859,12.475,9,motorway 
0.78,0.862,12.74,10,primary 
0.778,0.862,12.929,11,cycleway 
0.788,0.875,13.05,12,secondary 
0.787,0.875,13.094,13,steps 
0.782,0.875,13.134,14,residential 
0.782,0.875,13.151,15,tertiary link 
0.782,0.878,13.155,16,road 
0.782,0.878,13.155,17,corridor 
0.782,0.878,13.155,18,living street 
0.782,0.879,13.155,19,raceway 
0.782,0.878,13.155,20,byway 
0.782,0.882,13.155,21,no 
0.782,0.882,13.155,22,primary link 
0.782,0.882,13.155,23,pedestrian 
0.782,0.882,13.155,24,motorway link 
0.782,0.882,13.155,25,trunk link 
\end{filecontents*}
\pgfplotstabletypeset[
    multicolumn names=l, 
    col sep=comma, 
    string type, 
    header = has colnames, 
    columns={R2 Length, R2Composition, Inter Road Cumulative MutualInformation, Highway Type Length List, New Highway Type Included},
    columns/R2 Length/.style={column type={S[round-precision=2, table-format=1.2, table-number-alignment=center]}, column name=$R^2$ Length},
    columns/R2Composition/.style={column type={S[round-precision=2, table-format=1.2, table-number-alignment=center]}, column name=$R^2$ Composition},
    columns/Inter Road Cumulative MutualInformation/.style={column type={S[round-precision=2, table-format=2.2, table-number-alignment=center]}, column name=Inter Road Cumulative Mutual Information},
    columns/Highway Type Length List/.style={column type={S[round-precision=0, table-format=2.0, table-number-alignment=center]}, column name=No. of Road Types},
    columns/New Highway Type Included/.style={column type=l, column name=New Road Type Included},
    every head row/.style={before row=\toprule, after row=\midrule},
    every last row/.style={after row=\bottomrule}
    ]{CSVFiles/CuratedInterRoadMutualInformationResults.csv}}
    \smallskip
    \caption{{\bfseries Experiment results for the inter road mutual information subsetting tests.} Detailed is the model performance average for the Length and Composition models across all air pollutants with different road types subsets input into the feature vector. The New Road Type Included details the additional road type included over the previous row in the table input feature vector. } 
    \label{tab:Inter road mutual information tests}
\end{table}

\section{Missing Districts Tests Details}

\begin{figure}[ht]
\begin{center}
\text{Visualisation of Random State 12345 Test and Training Districts}\par\medskip
\includegraphics[width=0.35\textwidth]{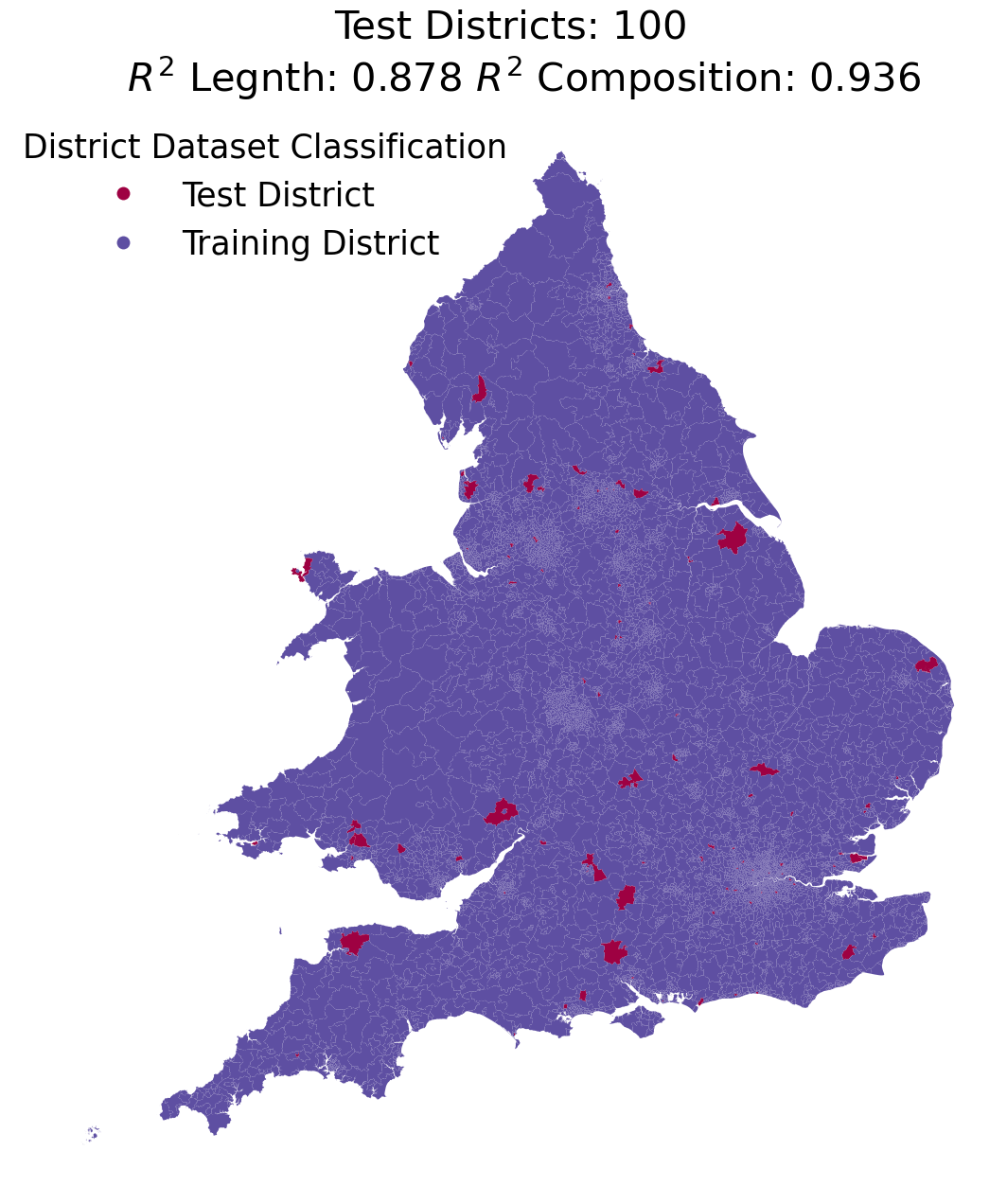}
\includegraphics[width=0.35\textwidth]{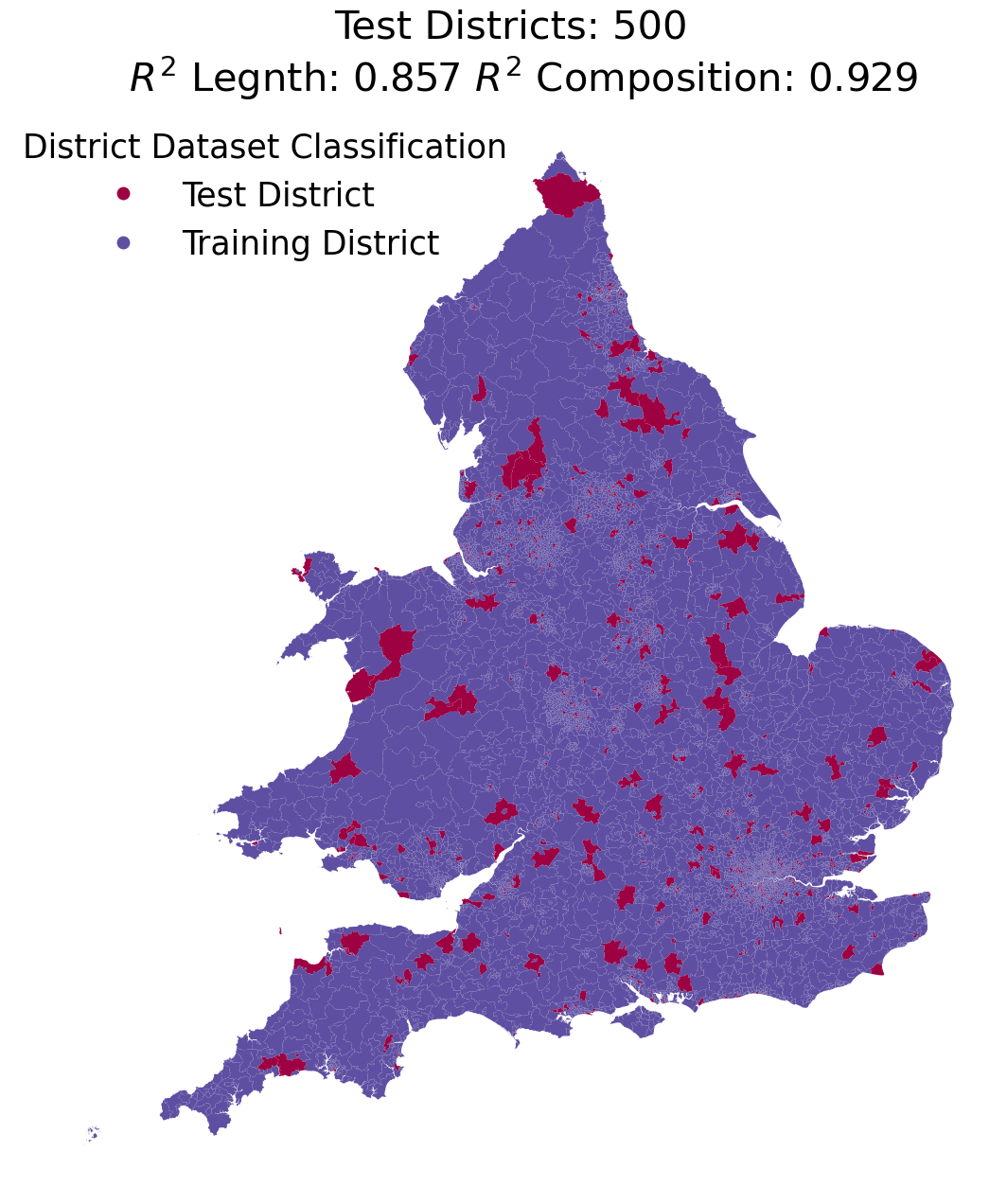}\\
\includegraphics[width=0.35\textwidth]{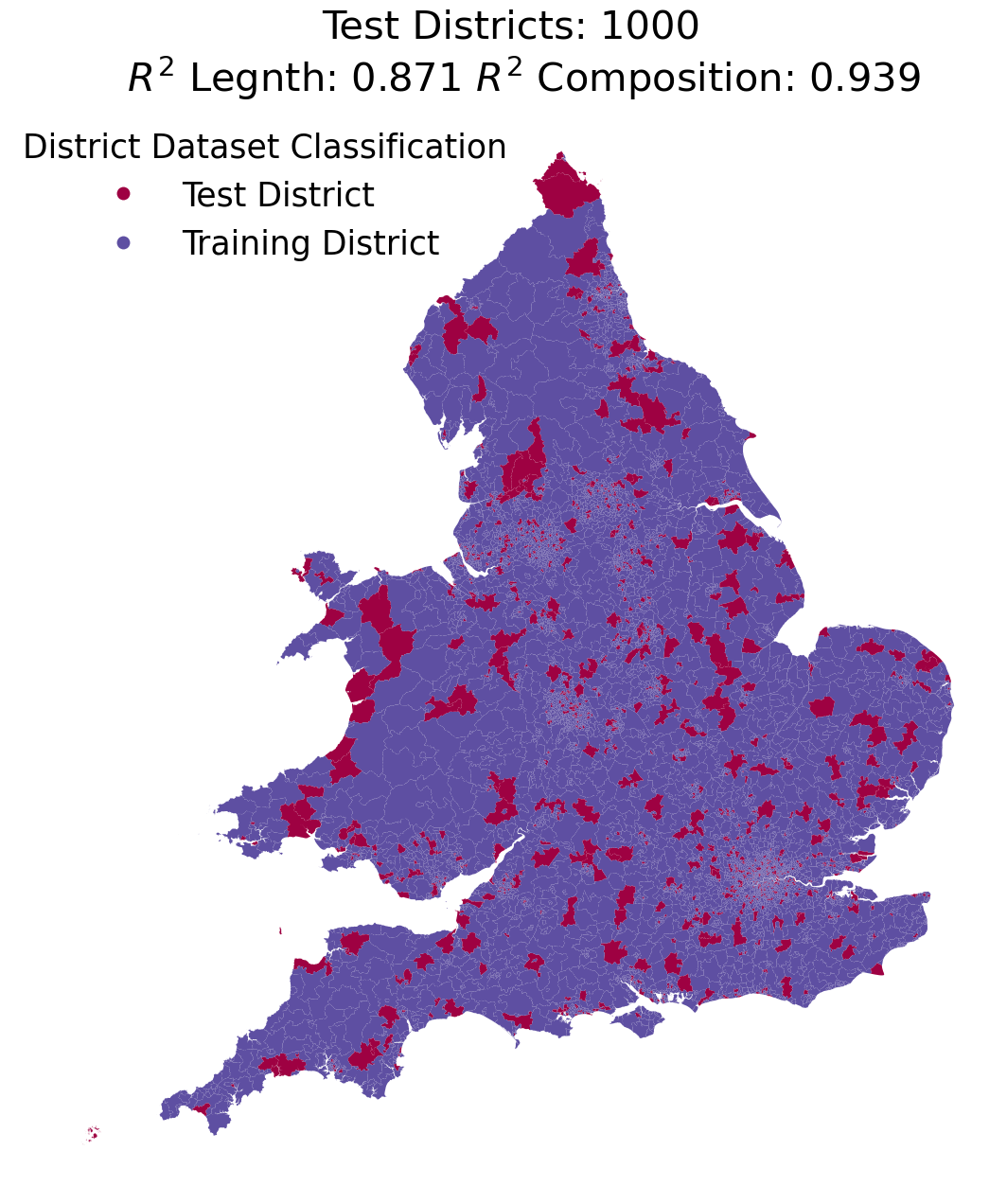}
\includegraphics[width=0.35\textwidth]{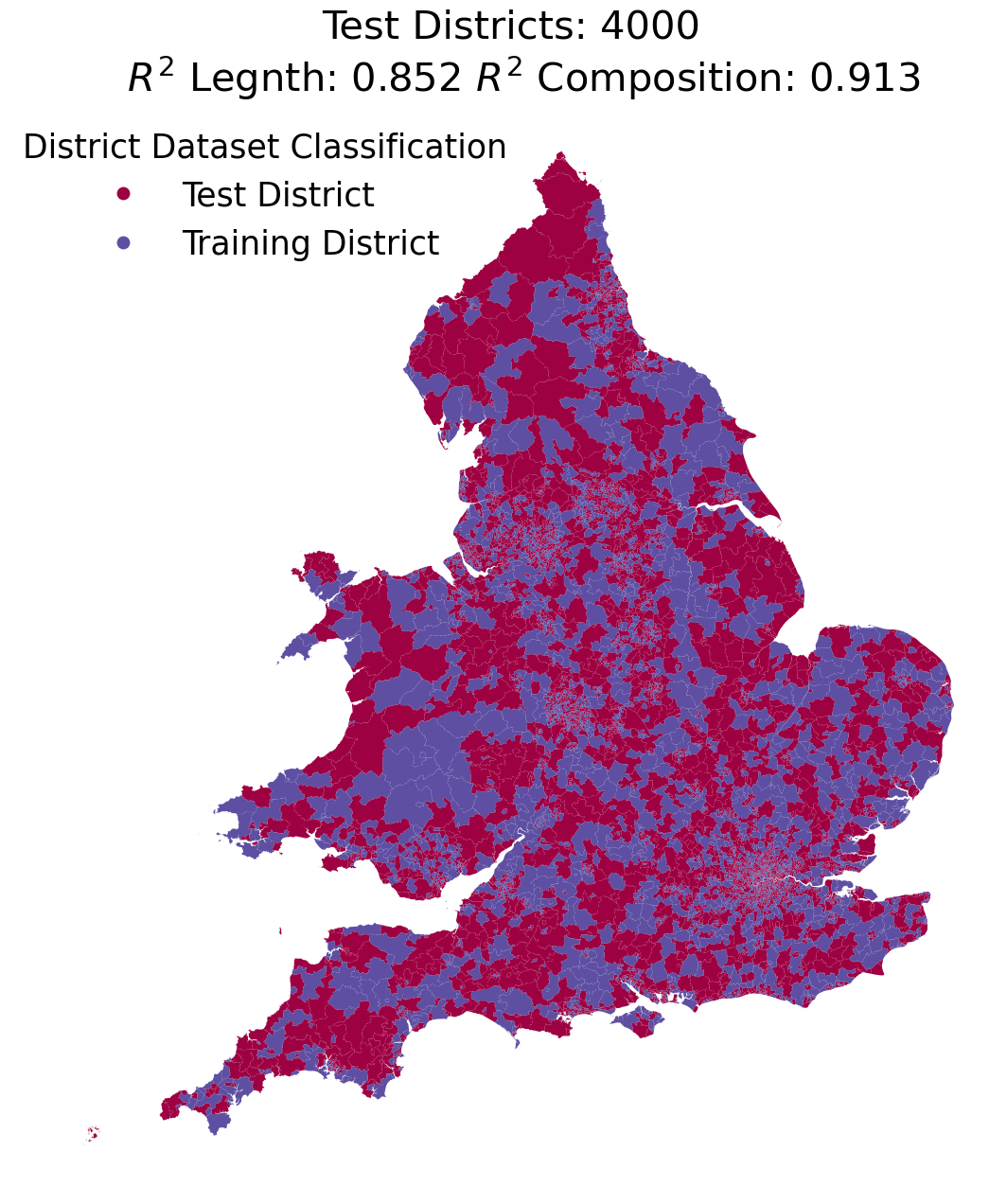}\\
\includegraphics[width=0.35\textwidth]{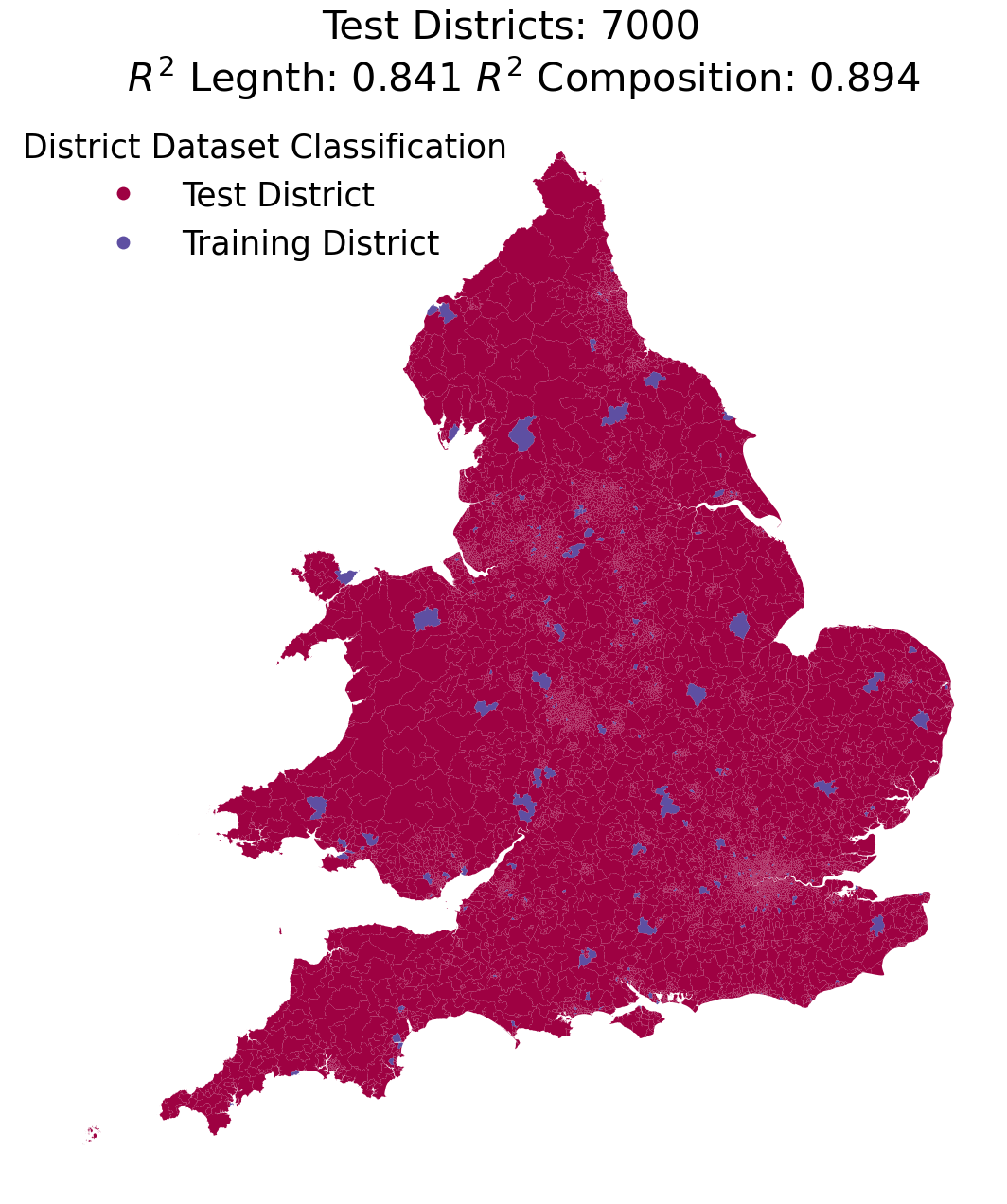}
\includegraphics[width=0.35\textwidth]{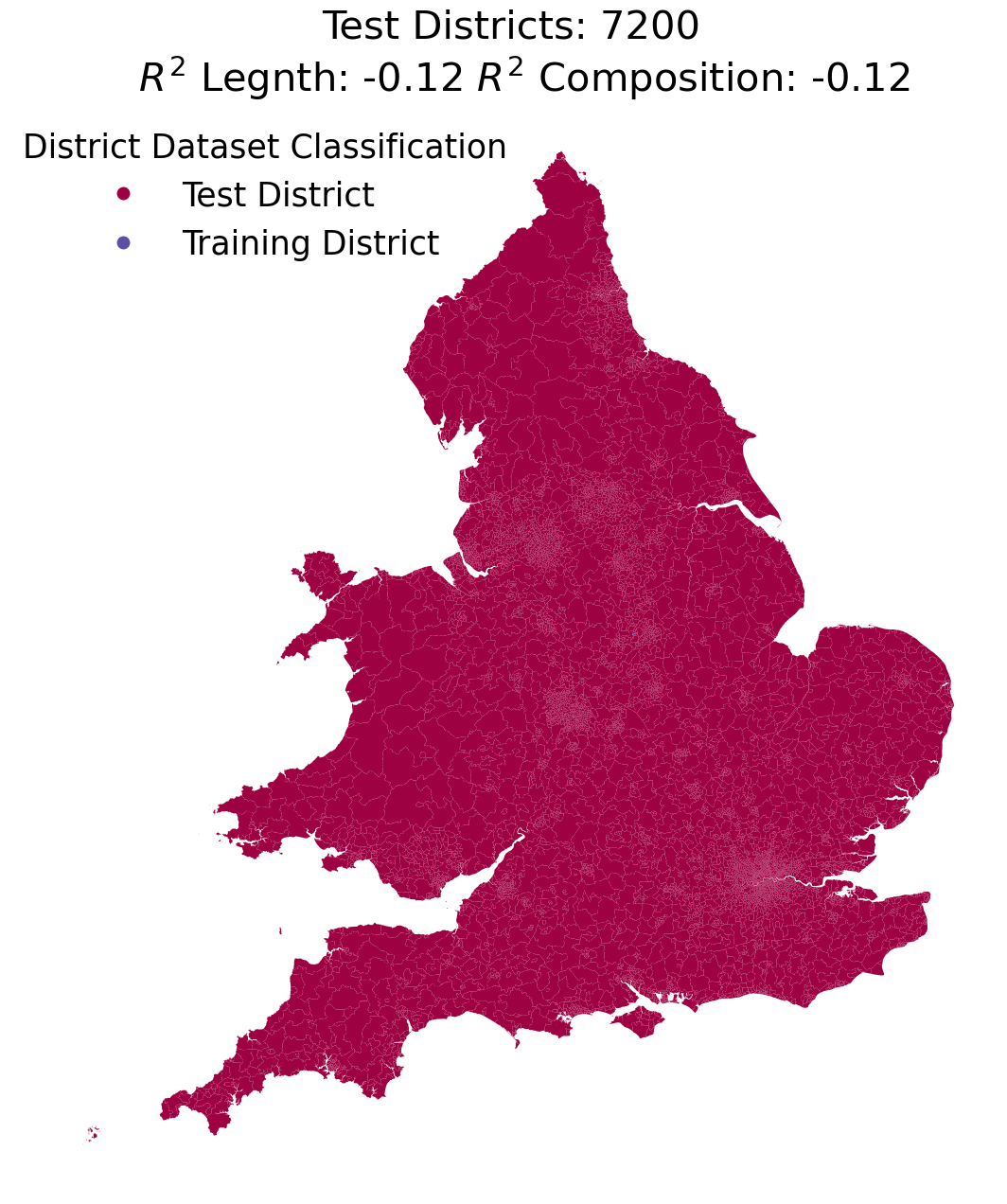}
\caption{{\bfseries Visualisation of changing MSOA test and training sets during the missing districts tests.} If a district is included within the test set during a previous experiment, e.g. test set size 100; then it will be included in the next test set, e.g. test set size 500. As the test set size increases (and training set size decreases), the model performance remains stable until eventually breaking down when the test set size encompasses over ~99.5\% of the districts.}
\label{fig:InterYearExperimentTrainingSetsVisualisation}
\end{center}
\end{figure}

Figure \ref{fig:InterYearExperimentTrainingSetsVisualisation} shows the visualisation of the MSOAs selected for inclusion in the training set during the missing districts tests. Table \ref{tab:Table 17 Population Density} shows the population density for a range of MSOAs. Table \ref{tab:Table 18 Inter year tests} details the results from the running of the length and composition model with a set of MSOAs restricted by population density distribution. 

\begin{table}
\resizebox{\linewidth}{!}{
\begin{filecontents*}{CSVFiles/CuratedMSOAPopulationDensitySorted.csv}
msoa11nm,Population Density,Population Density Normalized
Northumberland 019,5.614020926861652e-06,0.0195880692676168
Cornwall 058,0.0001674447056734,0.5842369552868992
Wakefield 007,0.0006826032327396,2.3816939016433283
Nuneaton and Bedworth 018,0.0015335269627431,5.350680512472427
Three Rivers 010,0.0025417715283788,8.868580543067846
Burnley 003,0.0035117516608292,12.252970852647884
Sedgemoor 009,0.0046096014584301,16.083515512350214
York 010,0.0067263513963433,23.46913892669779
Westminster 022,0.0286604098145742,100.0
\end{filecontents*}
\pgfplotstabletypeset[
    multicolumn names=l, 
    col sep=comma, 
    string type, 
    header = has colnames, 
    columns={msoa11nm, Population Density, Population Density Normalized},
    columns/msoa11nm/.style={column type=l, column name=MSOA Name},
    columns/Population Density/.style={column type={S[table-format=1.2e-1, round-precision=2, scientific-notation = true, round-integer-to-decimal, table-number-alignment=center]}, column name=Population Density (Individuals / m$^2$)},
    columns/Population Density Normalized/.style={column type={S[round-precision=2, round-integer-to-decimal, table-format=3.2, table-number-alignment=center]}, column name=Population Density Normalized (between 0-100)},
    every head row/.style={before row=\toprule, after row=\midrule},
    every last row/.style={after row=\bottomrule} 
    ]{CSVFiles/CuratedMSOAPopulationDensitySorted.csv}}
    \smallskip
    \caption{{\bfseries MSOA population density for every 900th MSOA.}  Shown are the MSOA population density values used in the study alongside a normalised score for the population density to act as an urbanisation metric in the study. } 
    \label{tab:Table 17 Population Density}
\end{table}

\begin{table}
\resizebox{\linewidth}{!}{
\begin{filecontents*}{CSVFiles/ResultsInterYearTests.csv}
,Test Region Size,Population Density Range Mean,R2 Length Mean,R2 Composition Mean
0,100,99.98041193073239,0.878561998367735,0.937966
1,500,99.98041193073239,0.8584575303270493,0.929488
2,1000,99.98041193073239,0.8721727484837706,0.940199
3,2000,99.97219776851142,0.8555816176636131,0.940494
4,3000,99.97021805297032,0.8530755719616859,0.915057
5,4000,91.73190753937853,0.8535623814127381,0.914723
6,5000,91.73190753937853,0.8414885924661156,0.907202
7,6000,91.73190753937853,0.8469324758957689,0.594034
8,6500,91.69692598407418,0.8470684153969098,-2.068818
9,7000,91.66055287397518,0.8420718258619121,0.897085
10,7025,91.66055287397518,0.8444981117340512,0.886647
11,7050,91.66055287397518,0.848000923510681,0.877503
12,7075,91.66055287397518,0.8481860654448644,0.6955
13,7100,72.45641210013484,0.8443905417173392,-8051239578094104.0
14,7125,72.39535860321976,0.8246933101672477,-263918633471537.9
15,7150,49.707523116638626,0.8173271199543329,0.729833
16,7175,49.707523116638626,0.845385275228534,0.605974
17,7190,48.77053702560652,0.649967384284791,0.797593
18,7195,48.67454100964591,0.778399800307453,0.823342
19,7196,48.67454100964591,0.7775146644406327,0.768328
20,7197,48.67454100964591,0.7838990737741306,0.649729
21,7198,19.6171640388845,0.8219871039295149,0.675514
22,7199,7.001794212473252,0.8166152937242189,0.603082
23,7200,0.0,-0.12007925099588514,-0.120079
\end{filecontents*}
\pgfplotstabletypeset[
    multicolumn names=l, 
    col sep=comma, 
    string type, 
    header = has colnames, 
    columns={Test Region Size, Population Density Range Mean, R2 Length Mean, R2 Composition Mean},
    columns/Test Region Size/.style={column type={S[round-precision=0, table-format=4.0, table-number-alignment=center]}, column name=Test Set Size},
    columns/Population Density Range Mean/.style={column type={S[round-precision=2, round-integer-to-decimal, table-format=2.2, table-number-alignment=center]}, column name=Population Density Range Mean (Individuals / m$^2$)},
    columns/R2 Length Mean/.style={column type={S[round-precision=2, table-format=-1.2, table-number-alignment=center]}, column name=$R^2$ Length Model Mean},
    columns/R2 Composition Mean/.style={column type={S[round-precision=2,scientific-notation = true, table-format=-1.2e-1, table-number-alignment=center]}, column name=$R^2$ Composition Model Mean},
    every head row/.style={before row=\toprule, after row=\midrule},
    every last row/.style={after row=\bottomrule}
    ]{CSVFiles/ResultsInterYearTests.csv}}
    \smallskip
    \caption{{\bfseries Length and composition model performance from the missing districts tests averaged across 16 random state tests.}  Shown are the performance of the length and composition model when changing the test set sizes alongside the associated population density range for the MSOAs included in the test set. Shown is how reducing the population density range does not affect the model performance of the length model, where the model only degrades when having a single MSOA in the training set. In contrast, the composition model is seen to be not as robust as the length model to the variations within the training set. } \label{tab:Table 18 Inter year tests}
\end{table}

\begin{table}
\resizebox{\linewidth}{!}{
\begin{filecontents*}{CSVFiles/popualtionDensityExperimentsResults.csv}
,Min and Max Urbanised MSOA In Training Set,R2 Length,Normalised Range of MSOA Population Density,Range of MSOA Population Density,Mean MSOA Population Density
0, Min: 0.0 Max: 1499.0,0.8422755124861023,1.5424793453788612,0.0004420809016907431,0.5699192910346527
1, Min: 125.0 Max: 1624.0,0.8441329061218298,1.713351635415779,0.0004910536002748717,0.7046163027441562
2, Min: 250.0 Max: 1749.0,0.845483190585183,2.0300834621146486,0.0005818302398199547,0.8608665932791458
3, Min: 375.0 Max: 1874.0,0.8432064612207538,2.3086576021877723,0.0006616707300023378,1.0418556331480806
4, Min: 500.0 Max: 1999.0,0.8444678125029994,2.61761834291082,0.0007502201444597074,1.248041708919249
5, Min: 625.0 Max: 2124.0,0.8440230902470386,2.9419807076087117,0.0008431837274663668,1.4797579936119407
6, Min: 750.0 Max: 2249.0,0.8445629512080134,3.342014655469625,0.0009578350963207248,1.742852895287027
7, Min: 875.0 Max: 2374.0,0.8443037460317694,3.6656072249252456,0.0010505780528562163,2.034754904386182
8, Min: 1000.0 Max: 2499.0,0.8456597156022453,3.9891623824828404,0.0011433102869884142,2.353985173989377
9, Min: 1125.0 Max: 2624.0,0.8485858178377348,4.334571267247977,0.0012423058888980528,2.702428450809058
10, Min: 1250.0 Max: 2749.0,0.8496132159216222,4.515064778145716,0.0012940360688100576,3.071841836480676
11, Min: 1375.0 Max: 2874.0,0.8508006929272948,4.758419898690624,0.0013637826436629797,3.4582425857203702
12, Min: 1500.0 Max: 2999.0,0.8447060875712662,4.912620350865119,0.0014079771251921164,3.8599421670245593
13, Min: 1625.0 Max: 3124.0,0.8513603058351453,5.155485378041872,0.001477583237277251,4.2810220055769905
14, Min: 1750.0 Max: 3249.0,0.8499216411691846,5.226269187207811,0.001497870167066575,4.714188789366606
15, Min: 1875.0 Max: 3374.0,0.8518342788385723,5.405852825079553,0.0015493395736405373,5.156270325654108
16, Min: 2000.0 Max: 3499.0,0.8505546552930411,5.548118589346295,0.0015901135247052213,5.610413974232163
17, Min: 2125.0 Max: 3624.0,0.8500974770998843,5.66940076977545,0.001624873494648269,6.077018435029041
18, Min: 2250.0 Max: 3749.0,0.8514161051853524,5.6445767565848355,0.0016177588307354147,6.5480552610344
19, Min: 2375.0 Max: 3874.0,0.8449329723606119,5.677751831537918,0.0016272669431732602,7.02035273791571
20, Min: 2500.0 Max: 3999.0,0.8436437719134812,5.6671887944449475,0.0016242395334535498,7.492623572946147
21, Min: 2625.0 Max: 4124.0,0.8220308070668844,5.609517362693431,0.0016077106647676325,7.961773661669881
22, Min: 2750.0 Max: 4249.0,0.8205065830606347,5.68537906732171,0.0016294529402064188,8.433539720674156
23, Min: 2875.0 Max: 4374.0,0.847788030892898,5.802523759698537,0.00166302708911764,8.912322779060462
24, Min: 3000.0 Max: 4499.0,0.8474760596190368,5.774066811212437,0.001654871211060801,9.396303022915244
25, Min: 3125.0 Max: 4624.0,0.8431473956560688,5.775045239421833,0.0016551516325953555,9.876915463331901
26, Min: 3250.0 Max: 4749.0,0.8210625778746959,5.832525517693483,0.0016716257159105686,10.360961955950955
27, Min: 3375.0 Max: 4874.0,0.8055112854399377,5.860183081718314,0.0016795524871048126,10.84982321826023
28, Min: 3500.0 Max: 4999.0,0.8045270229771034,5.876025673248128,0.0016840930387625064,11.341391857694255
29, Min: 3625.0 Max: 5124.0,0.8037031926661565,5.883193248672562,0.0016861472952529183,11.834257781479288
30, Min: 3750.0 Max: 5249.0,0.8004740605234896,5.992846432857052,0.001717574347214923,12.327926941135134
31, Min: 3875.0 Max: 5374.0,0.7996207552520542,6.0856437116585465,0.0017441704276162043,12.830548009787412
32, Min: 4000.0 Max: 5499.0,0.8134420012832341,6.277930379046815,0.0017992805745084698,13.344948238411144
33, Min: 4125.0 Max: 5624.0,0.7975764654284855,6.608189270951755,0.001893934126377496,13.879720303282188
34, Min: 4250.0 Max: 5749.0,0.793937848925113,6.901295345999612,0.001977939528677626,14.440916982173347
35, Min: 4375.0 Max: 5874.0,0.8356697911410295,7.083771354220325,0.0020302379004469577,15.02561199546334
36, Min: 4500.0 Max: 5999.0,0.8438578181078419,7.679641012119106,0.002201016586361451,15.639893650747732
37, Min: 4625.0 Max: 6124.0,0.8479289983437677,8.394493886416438,0.002405896349706329,16.306137833953883
38, Min: 4750.0 Max: 6249.0,0.8344911925577029,9.438616097409785,0.002705146054342015,17.053322646535708
39, Min: 4875.0 Max: 6374.0,0.8254430456002763,10.735613411388101,0.0030768707998122206,17.892075846800356
40, Min: 5000.0 Max: 6499.0,0.8167712112187863,12.789012169358038,0.003665383298973781,18.877286614400624
41, Min: 5125.0 Max: 6624.0,0.8138978935114464,15.736744745814748,0.0045102155356239815,20.063516967552516
42, Min: 5250.0 Max: 6749.0,0.8185990572419858,19.17915401352733,0.005496824139245291,21.51176205165391
43, Min: 5375.0 Max: 6874.0,0.8201169427579494,24.440978608522933,0.007004884631895091,23.33948291580329
44, Min: 5500.0 Max: 6999.0,0.8256630753484725,34.08321349147002,0.009768388664631554,25.733363246402384
45, Min: 5625.0 Max: 7124.0,0.8248128657911709,46.833434729730314,0.013422654323781834,29.068866213645755
\end{filecontents*}
\pgfplotstabletypeset[
    multicolumn names=l, 
    col sep=comma, 
    string type, 
    header = has colnames, 
    columns={Min and Max Urbanised MSOA In Training Set, R2 Length, Normalised Range of MSOA Population Density, Range of MSOA Population Density, Mean MSOA Population Density},
    columns/Min and Max Urbanised MSOA In Training Set/.style={column type=l, column name=Min and Max Urbanised MSOA In Training Set},
    columns/R2 Length/.style={column type={S[round-precision=2, table-format=4.0, table-number-alignment=center]}, column name=$R^2$ Score Length Model},
    columns/Normalised Range of MSOA Population Density/.style={column type={S[round-precision=2, round-integer-to-decimal, table-format=2.2, table-number-alignment=center]}, column name=Normalised Range of MSOA Population Density},
    columns/Range of MSOA Population Density/.style={column type={S[round-precision=5, table-format=-1.2, table-number-alignment=center]}, column name=Range of MSOA Population Density (Individuals / m$^2$)},
    columns/Mean MSOA Population Density/.style={column type={S[round-precision=2,scientific-notation = true, table-format=-1.2e-1, table-number-alignment=center]}, column name=Mean MSOA Population Density (Individuals / m$^2$)},
    every head row/.style={before row=\toprule, after row=\midrule},
    every last row/.style={after row=\bottomrule}
    ]{CSVFiles/popualtionDensityExperimentsResults.csv}}
    \smallskip
    \caption{{\bfseries Experiment results for a consistent training set size (1500) with changing mean urbanisation scores across included MSOAs.} Shown are the results from the experiment conducted to explore the effect of changing the mean urbanisation score in the training set on the performance of the length model. The training set moves from least (Min 0.0 Max: 1499.0) to most (Min 5625.0 Max: 7124.0) urbanised, where the performance of the length model remains consistent. Indicating there isn't a need to consider the distribution of how urban the MSOAs are within the training set. } \label{tab:trainingSetSizePopulationDensity}
\end{table}

